\documentclass{article}
\usepackage{graphicx} 
\usepackage[utf8]{inputenc}
\usepackage[left=3cm,right=3cm,top=3cm,bottom=3cm]{geometry}
\usepackage{amsmath}
\usepackage{amssymb}
\usepackage{esint}
\usepackage{enumitem}
\usepackage{multirow}
\usepackage{graphicx}
\usepackage{booktabs}
\usepackage{array}
\usepackage{verbatim}
\usepackage{hhline}
\usepackage{hyperref}
\usepackage{centernot}
\usepackage{amsthm}
\hypersetup{
  colorlinks   = true,    
  urlcolor     = blue,    
  linkcolor    = blue,    
  citecolor    = black     
}

\usepackage{cancel}

\newcommand{\tlam}{\tilde\lambda}
\newcommand{\MHVb}{\overline{\mathrm{MHV}}}
\newcommand{\Ncal}{\mathcal{N}}

\newcommand{\vts}[2]{v_{#1#2}}
\newcommand{\vta}[2]{v^{\star}_{#1#2}}
\newcommand{\bul}{\bullet}
\newcommand{\str}{\star}

\newcommand{\PTs}[1]{\mathrm{PT}^{\star}(#1)}
\newcommand{\PTa}[1]{\mathrm{PT}(#1)}
\newcommand{\Fsd}{\mathcal{F}[\phi]}
\newcommand{\Fasd}{\overline{\mathcal{F}}[\phi]}
\newcommand{\ab}[2]{\langle #1\,#2\rangle}
\newcommand{\sqb}[2]{[#1\,#2]}
\newcommand{\Tr}{\mathrm{Tr}}
\newcommand{\sgn}[1]{\mathrm{sgn}\left(#1\right)}

\renewcommand\Tr{\mathrm{Tr}}

\newcommand{\xdownarrow}[1]{%
  {\left\downarrow\vbox to #1{}\right.\kern-\nulldelimiterspace}
}

\title{Scattering amplitudes from geometry of kernel-free functionals on piecewise random null paths with memory}
\author{
Hiren Kakkad$^a$\thanks{E-mail: \texttt{hkakkad@shanghaitech.edu.cn}} \, and 
Piotr Kotko$^b$\thanks{E-mail: \texttt{pkotko@agh.edu.pl}}
\\ \\
$^a$ {\it ShanghaiTech University,} \\
{\it Center for Fundamental Physics, School of Physical Science and Technology,}\\ 
{\it 393 Middle Huaxia Road, Shanghai 201210, China,} \\  \\
$^b${\it AGH University Of Krakow,}\\ 
{\it Faculty of Physics and Applied Computer Science,} \\ 
{\it al. Mickiewicza 30, 30-059 Krak\'ow, Poland} 
}

\date{}

\begin{document}

\maketitle

\begin{abstract}
We introduce a simple functional of algebra-valued fields with definite Lorentz properties that is kernel-free yet it encodes the structure of tree-level scattering amplitudes. The fields are evaluated at the points of a piece-wise straight paths in the anti-self-dual plane of complexified Minkowski space, with the directions of consecutive segments correlated so that the path has a velocity memory. 
We show that momentum-space reduction of the functional generates the cyclic Parke--Taylor factor, off shell in terms of light-front bilinears and on shell in terms of spinor brackets. The same reduction, organized differently, can be naturally promoted to  twistor space without any twistor-space input: a closed cyclic chain of Penrose transforms on a twistor line, in which the Parke--Taylor denominators arise as boundary terms of the integrals along the paths. 
Finally we show that derivatives with respect to one of the path parameter generate numerators that together with the Parke-Taylor denominator build scattering amplitudes. Lorentz invariance fixes the spins of the inserted fields and the homogeneities of the corresponding twistor functions without any use of supersymmetry. The three-point outputs reproduce the kinematic forms of the light-cone cubic vertices of chiral higher-spin theory, and the derivative monomials of order four generate the components of the $\overline{\rm MHV}$ supermultiplet of $\mathcal N=4$ super Yang--Mills theory, off shell and on shell.  
\end{abstract}

\newpage

\tableofcontents

\newpage 

\section{Introduction}

Tree-level multi-gluon scattering was, for a long time, a showcase of the mismatch between the simplicity of a physical answer and the effort needed to obtain it. Individual Feynman diagrams proliferate factorially and conspire to produce results of striking compactness. The landmark example is the maximally-helicity-violating (MHV) amplitude conjectured by Parke and Taylor \cite{Parke:1986gb} and proven shortly afterwards by Berends and Giele through their off-shell recursion \cite{Berends:1987me}. In the conventions of this paper, where the distinguished sector is the conjugate one, the color-ordered $\MHVb$ amplitude for gluons $i$ and $j$ of positive helicity reads
\begin{equation}
  A\big(1,\dots,i^+,\dots,j^+,\dots,n\big)
  \;=\;
  \frac{\sqb{i}{j}^{\,4}}{\sqb{1}{2}\sqb{2}{3}\cdots\sqb{n}{1}}\,,
  \label{eq:PT_intro}
\end{equation}
with all remaining gluons of negative helicity. The supersymmetric Ward identities \cite{Grisaru:1976vm,Grisaru:1977px} and the early supersymmetry-based computations \cite{Parke:1985pn,Gunion:1985vca,Kunszt:1985mg} explained both the vanishing of the neighboring helicity sectors and the rigidity of the numerator. The technology of color ordering and helicity spinors that is essential in any discussion of \eqref{eq:PT_intro} is reviewed in \cite{Mangano:1990by,Elvang:2013cua,Travaglini:2022uwo}.

What makes \eqref{eq:PT_intro} more than a compact pocket formula is the ubiquity of its denominator. The cyclic chain, which we will refer to as the Parke--Taylor factor, is arguably the most universal rational structure in perturbative gauge theory. It carries the entire helicity dependence of Nair's superamplitude, in which the MHV sector of $\Ncal=4$ super Yang--Mills is generated by a current algebra on a degenerate twistor line \cite{Nair:1988bq}. It is the object on which the Kleiss--Kuijf \cite{Kleiss:1988ne} and Del Duca--Dixon--Maltoni \cite{DelDuca:1999rs} color structures act. It reappears, essentially unchanged, far outside its birthplace: as the universal integrand half of the Cachazo--He--Yuan formulas on the support of the scattering equations \cite{Cachazo:2013hca} and of the ambitwistor string \cite{Mason:2013sva}; as the MHV vertex of the Cachazo--Svrcek--Witten expansion \cite{Cachazo:2004kj} descending from the twistor string \cite{Witten:2003nn,Roiban:2004yf}; and, through on-shell recursion \cite{Britto:2004ap,Britto:2005fq}, as the seed from which all tree amplitudes can be reconstructed. Its gravitational cousins govern MHV graviton scattering \cite{Berends:1988zp,Mason:2008jy,Hodges:2012ym}. A structure this persistent invites a structural explanation.

One such explanation is dynamical and lives on the light front \cite{Dirac:1949cp}. In light-cone gauge Yang--Mills reduces to two physical fields, and Mansfield showed that a canonical field transformation, designed to absorb the non-MHV three-point coupling into the kinetic term, maps the light-cone action to an action whose vertices are precisely the off-shell continued MHV amplitudes \cite{Mansfield:2005yd}, a program anticipated in \cite{Gorsky:2005sf} and developed in detail in \cite{Ettle:2006bw}. In this MHV Lagrangian the Parke--Taylor factor appears off shell, written not in spinor brackets but in their light-front continuation, and the same continuation appears independently in the small-$x$ gluon cascade \cite{Motyka2009}. The quantum consistency of the framework is subtle: naive MHV rules are incomplete at one loop \cite{Brandhuber:2004yw,Brandhuber:2007vm}, and the missing pieces were traced to the S-matrix equivalence theorem evasions of the field transformation \cite{Ettle:2007qc}.

A second, closely related explanation is geometric and starts from the self-dual sector. Self-dual Yang--Mills is classically integrable, its perturbative expansion around it is organized by the Chalmers--Siegel action \cite{Chalmers:1996rq}, and its S-matrix is famously trivial at tree level while producing the rational one-loop all-plus amplitudes \cite{Bardeen:1995gk,Cangemi:1996rx,Bern:1993qk,Mahlon:1993fi,Bern:1996ja}. Classical solution-generating techniques in this sector, the perturbiners of Rosly and Selivanov \cite{Rosly:1996vr,Selivanov:1997aq}, package Berends--Giele currents into generating functionals of solutions, an idea whose spirit is close to the present work. The tree-level information hidden in the self-dual theory is not empty either: the form factor of the operator $\mathrm{Tr}\,F^2_{\mathrm{SD}}$ with all like-helicity gluons is a Parke--Taylor formula with a $q^4$ numerator \cite{Dixon:2004za}, rederivable directly from the Chalmers--Siegel embedding \cite{Boels:2013bi}, with higher-dimensional analogues such as $\mathrm{Tr}\,F^3$ studied in \cite{Brandhuber:2018kqb,AccettulliHuber:2019mxn} and the general form-factor technology developed in \cite{Brandhuber:2010ad,Brandhuber:2011tv}. More recently the self-dual sector has become the natural laboratory for the kinematic algebra underlying color-kinematics duality \cite{Monteiro2011,Bern:2008qj} and for celestial holography, where the $w_{1+\infty}$ symmetries of soft physics \cite{Guevara:2021abz,Strominger:2021mtt} find a top-down realization in twisted holographic duals of self-dual gauge theory \cite{Costello:2021bah,Costello:2022wso}. On the gravitational side the same sector is governed by Plebanski's heavenly equations \cite{Plebanski:1975wn} and their modern chiral formulations \cite{Krasnov:2016emc}.

The bridge between these two explanations, and the direct ancestor of this work, is built from Wilson lines. The solution of the  anti-self-dual Yang--Mills equations can be written in terms of straight infinite Wilson lines whose slopes scan the ruling directions of anti-self-dual, and Kotko and Stasto showed that Mansfield's field transformation is exactly such a Wilson line: the MHV action is Yang--Mills rewritten in terms of Wilson-line degrees of freedom \cite{Kotko2017}. This observation was developed into a systematic program: the geometry of the straight infinite Wilson lines in the self-dual and MHV theories was explored in \cite{Kakkad2020}, a new classical action with no triple-gluon vertices was obtained by a second Wilson-line transformation in \cite{Kakkad:2021uhv,Kakkad:2021dkw}, and its quantum completion, restoring the loop-level content that the tree-level transformations displace, was constructed in \cite{Kakkad_2022,Kakkad:2024ase,Kakkad:2023bnk}.
Wilson lines with prescribed slopes are also the gauge-invariant definition of off-shell amplitudes in high-energy factorization \cite{vanHameren:2012if,Kotko:2014aba}, which is the second place, besides the MHV Lagrangian \cite{Ettle:2006bw}, where an off-shell continuation of the spinor bracket is in daily use; and the general lesson that null polygonal Wilson loops know about amplitudes is by now classic \cite{Alday:2007hr}. The kinematic currency shared by this entire circle of ideas is what we call the $v$-symbols, $\vts{i}{j}=\varepsilon^+(p_i)\cdot p_j$: light-front bilinears defined for arbitrary off-shell momenta that reduce on shell to spinor brackets.

The same currency runs through a superficially distant subject. Metsaev's light-cone classification of cubic vertices for massless higher-spin fields \cite{Metsaev:1991mt,Metsaev:1991nb} produces, for each helicity triple, a unique kinematic structure built from exactly these bilinears, and the chiral closure of that classification defines chiral higher-spin theory \cite{Ponomarev:2016lrm}, which is in a precise sense the higher-spin extension of self-dual Yang--Mills \cite{Ponomarev:2017nrr}. The theory is one-loop finite \cite{Skvortsov:2018jea,Skvortsov:2020wtf,Skvortsov:2020gpn}, admits supersymmetric \cite{Metsaev:2019dqt,Metsaev:2019aig,Tsulaia:2022csz} and Chan--Paton extensions, and has been proposed as a model of flat-space holography. Its very existence illustrates how light-front kinematics sidesteps the covariant no-go theorems for interacting higher spins \cite{Weinberg:1964ew,Coleman:1967ad,Benincasa:2007xk,McGady:2013sga}, in the same technical way in which light-cone superspace once sidestepped covariant obstructions in $\Ncal=4$ super Yang--Mills \cite{Brink:1982wv,Mandelstam:1982cb}. Any object that naturally generates Metsaev's structures with their prescribed couplings is therefore of interest well beyond gauge theory.

Finally, all roads above intersect twistor space. The Penrose correspondence \cite{Penrose:1967wn,Penrose:1969ae}, with its prehistory in Bateman's integral representations \cite{Bateman:1904xu}, encodes free massless fields of helicity $h$ in functions of homogeneity $-2h-2$ on $\mathbb{CP}^3$, and Ward's construction geometrizes the self-dual sector \cite{Ward:1977ta}. 
In the modern amplitude literature twistor space is usually the input: one postulates a twistor string \cite{Witten:2003nn}, a twistor action \cite{Mason:2005zm,Boels:2006ir}, or an ambitwistor model \cite{Mason:2013sva}, and derives space-time physics from it. 
 One of the findings reported below is that the logic can be reversed. The same structures, the twistor line, the Parke--Taylor denominators on it, the reference twistor and the lines through it, \emph{emerge} from a remarkably simple position-space functional discussed in Section~\ref{sec:F_SD}, with no twistor-space input.

This work is built around a single question suggested by the ubiquity described above: is there one position-space object whose momentum-space reduction \emph{is} the Parke--Taylor factor, off shell and on shell, and whose natural deformations generate the numerators that turn the factor into amplitudes? We answer the question affirmatively. The object is a functional of a single algebra-valued field, defined by correlating field insertions along straight-line directions within anti-self-dual planes of complexified Minkowski space; it descends directly from the straight infinite Wilson lines of \cite{Kotko2017,Kakkad2020,Kakkad:2021uhv}.
The paths can be thought of as random piecewise trajectory with "memory", because a component of the direction of the $i+1$ step depends on the direction component of the $i$-th step.
Only two kinematic building blocks appear anywhere in this paper: off shell the $v$-symbols, on shell the spinor brackets.

Our results are threefold. First, we define the functional, provide its geometric interpretation, perform its momentum-space reduction, and show that it produces the cyclic off-shell Parke--Taylor factor built from $v$-symbols, which on shell becomes the spinor-bracket chain of \eqref{eq:PT_intro} stripped of its numerator. Second, we show that the same reduction, organized differently, makes twistor space emerge: keeping the integrals over the ruling directions and trading momentum components for incidence relations rewrites the functional as a closed cyclic chain of Penrose transforms on $\mathbb{CP}^3$.
Third, we extend the Parke-Taylor functional to a whole family of functionals by merely acting with derivatives of a path parameter on fields. The invariance under the boost and rotation symmetry of the construction forces the inserted fields up the light-cone spin ladder; at three points the raw outputs reproduce, exactly and with unit relative couplings, the kinematic forms of Metsaev's light-cone chiral vertices; and an amplitude extraction with half-integer clearing exponents is defined, together with precise claims about what it computes. On shell, the claims are then substantiated against known results: the extraction removes the light-front frame dependence uniquely beyond three points and delivers the $\MHVb$ amplitudes of $\Ncal\le4$ super Yang--Mills component by component through Nair's dictionary \cite{Nair:1988bq}; color structure and $U(1)$ decoupling \cite{Kleiss:1988ne} supply the non-adjacent helicity configurations; the collinear limits expose the underlying current algebra; and the spinor bracket itself acquires a derived off-shell continuation through the $v$-symbols, in exactly the normalized form used by the light-front MHV-action literature \cite{Ettle:2006bw}. In the twistor representation the same calculus acts by shifting the homogeneities of the slot functions, that is, the spins of the corresponding free fields.

The paper is organized as follows. Section~\ref{sec:Basics} collects the background material: complexified Minkowski space in double-null coordinates, self-dual and anti-self-dual planes and their ruling directions, helicity spinors and polarization vectors in our conventions, the $v$-symbols and their properties, the straight infinite Wilson lines that solve the  anti-self-dual Yang--Mills equations, and a short primer on twistor space. Section~\ref{sec:F_SD} introduces the Parke--Taylor functional on the anti-self-dual plane, carries out its momentum-space reduction to the off-shell Parke--Taylor factor, discusses the imposition of on-shellness, and derives the emergent twistor-space representation as a closed chain of Penrose transforms. Section~\ref{sec:amplitudes} discusses the amplitude landscape emerging when the functional is modified by derivatives of the fields with respect to the path parameter.

\section{Background material}
\label{sec:Basics}
In the following Section we set up out notation and gather essential prerequisites for the rest of the work.

\subsection{Complexified Minkowski space and its descendants}
\label{sub:complexified_Minkowski}

Extending the usual Minkowski space $\textbf{M}_4$ to the complex domain $\textbf{CM}_4$ allows not only to encapsulate  different types of the real inner product (Euclidean, hyperbolic and Minkowski), but can also be used to encode polarization states. This latter feature is usually utilized in the context of the twistor space, being especially useful in a description of \emph{on-shell} scattering amplitudes. However, part of our work concerns \emph{off-shell} currents, therefore $\textbf{CM}_4$ itself, as well as its subspaces, will play a central role here.

\paragraph{Double-null coordinates.} We introduce the following  coordinates in $\textbf{CM}_4$ 
\begin{gather} 
 x^{+}=\frac{1}{\sqrt{2}} \left(x^0+x^3\right) \,, \,\,\,\, \, x^{-}=\frac{1}{\sqrt{2}} \left(x^0-x^3\right) \,,\,\,\,\, x^{\bullet}=-\frac{1}{\sqrt{2}} \left(x^1+ix^2\right) \,,\,\,\,\,x^{\star}=-\frac{1}{\sqrt{2}} \left(x^1-ix^2\right) \,.
 \label{eq:double_null1}
\end{gather}
The appropriate basis four vectors are (in canonical coordinates)
\begin{gather}
\eta^\mu=\frac{1}{\sqrt{2}}\left(1,0,0,-1\right)\,,\,\,\,\,\widetilde{\eta}^\mu=\frac{1}{\sqrt{2}}\left(1,0,0,1\right)\, ,\,\,\,\, \varepsilon_{\perp}^{\pm\,\mu}=\frac{1}{\sqrt{2}}\left(0,1,\pm i,0\right)\,,
\label{eq:DNC_basis}
\end{gather}
so that 
\begin{gather}
x^{+}=x\cdot\eta\,,\,\,\, x^{-}=x\cdot\widetilde{\eta}\,,\,\,\, x^{\bullet}=x\cdot\varepsilon_{\bot}^{+}\,,\,\,\, x^{\star}=x\cdot\varepsilon_{\bot}^{-}\,
\end{gather}
with the dot being the standard Minkowski inner product with the $(+---)$ signature. If needed, the indices of the new coordinates can be lowered, for the transverse components by the cost of a minus sign. Because the basis vectors are all null, these coordinates are referred to as \emph{double null} coordinates.

Next, we define the two important null subspaces of the $\textbf{CM}_4$.

\paragraph{Anti-self-dual (ASD) plane.} We define a family of lines through $x\in \textbf{CM}_4$  
\begin{equation}
  \Pi^+(x)
  = \bigl\{\,x + s\,\varepsilon^+_\alpha\;:\;s,\alpha \in\mathbb{R}\,\bigr\}\,,
  \label{eq:SDplane}
\end{equation}
where 
\begin{equation}
    \varepsilon_{\alpha}^{+\, \mu} = \varepsilon_{\perp}^{+\, \mu }- \alpha \,\eta^{\mu} \,.
    \label{eq:epsilon_alpha}
\end{equation}
Above, $s$ and $\alpha$ are thought of as real parameters, but we will allow to deform their integrals to the complex domain.
The plane is totally null: it is spanned by the null vectors $\varepsilon^+_\perp$ and $\eta$, which are also mutually orthogonal, so the tangent bivector $w^{\mu\nu}=\varepsilon_\perp^{+\mu}\eta^{\nu}-\varepsilon_\perp^{+\nu}\eta^{\mu}$ is null, $w_{\mu\nu}w^{\mu\nu}=0$. To fix its duality we define $\star w^{\mu\nu}=\tfrac12\epsilon^{\mu\nu\rho\sigma}w_{\rho\sigma}$ with $\epsilon_{0123}=+1$. In Lorentzian signature $\star\star=-1$, so a bivector is self-dual if $\star w=i\,w$ and anti-self-dual if $\star w=-i\,w$. A direct computation gives $\star w=-i\,w$ for $\Pi^+(x)$, and we shall therefore call it the \emph{anti-self-dual} (ASD) plane\footnote{In \cite{Kotko2017,Kakkad2020} the same plane, and the sector of Yang--Mills theory whose Wilson lines lie on it, were called self-dual; this corresponds to the opposite orientation convention, $\epsilon_{0123}=-1$. We adopt $\epsilon_{0123}=+1$ because then the planes represented by single points of the twistor space of Section~\ref{sub:twistor_space} (the $\alpha$-planes) are the self-dual ones, as in the standard twistor literature \cite{Penrose:1985bww}. Accordingly, the names of the planes and of the corresponding sector of Yang--Mills theory used throughout this paper are the opposite of those of \cite{Kotko2017,Kakkad2020}; in the present convention the field strength of that sector, which has the non-vanishing component $F_{\mu\nu}\varepsilon_\perp^{+\mu}\eta^{\nu}=-\partial_-A^{\bullet}$ along $\Pi^+$ in the light-cone gauge, is anti-self-dual.}.
Similarly, we define the \emph{self-dual} (SD) plane through $x\in \textbf{CM}_4$ as
\begin{equation}
  \Pi^-(x)
  = \bigl\{\,x + s\,\varepsilon^-_\beta\;:\;s,\beta \in\mathbb{R}\,\bigr\}\,,
  \label{eq:ASDplane}
\end{equation}
where
\begin{equation}
    \varepsilon_{\beta}^{-\, \mu} = \varepsilon_{\perp}^{-\, \mu }- \beta \,\eta^{\mu} \,.
    \label{eq:epsilon_beta}
\end{equation}
Here any tangent bivector satisfies $\star w=+i\,w$.
In the terminology of \cite{Penrose:1985bww} the self-dual planes $\Pi^-$ are $\alpha$-planes and the anti-self-dual planes $\Pi^+$ are $\beta$-planes; their representation in twistor space is discussed in Section~\ref{sub:twistor_space}.

Sometimes we shall use the two-vector notation for objects within the (fixed) ASD plane $\Pi^+$
\begin{equation}
    \vec{x}\equiv (x^-,x^{\star}) \,.
\end{equation}
Consequently, the projection of $\varepsilon^+_{\alpha}$ on the ASD plane will be denoted as 
\begin{equation}
   \vec{\varepsilon}^+_{\alpha} = (-\alpha, -1) \,.
   \label{eq:epsilon_vec_SD}
\end{equation}

\paragraph{Helicity spinors.}
\label{sec:spinors}
Using the two dimensional representations of the $\mathbf{SL}(2,\mathbf{C})$, the on-shell massless four momentum $p^{\mu}$ can be represented as a bispinor 
$p_{\alpha\dot\alpha} = p^\mu(\sigma_\mu)_{\alpha\dot\alpha}$
with $\sigma^\mu = (\mathbf{1},\vec\sigma)$. In double-null coordinates it reads
\begin{equation}
  p_{\alpha\dot\alpha}
  = \sqrt{2}\begin{pmatrix}p^- & p^\str\\[2pt]
    p^\bul & p^+\end{pmatrix}.
  \label{eq:p_matrix_down}
\end{equation}
Since $p^2=0$, $p^-=p^{\bullet}p^{\star}/p^+$ and $\det p_{\alpha\dot\alpha} = 0$, the matrix has rank one and factors as
$p_{\alpha\dot\alpha} = \lambda_{\alpha}\tilde\lambda_{\dot\alpha}$. The little group transformations leave the $p_{\alpha\dot\alpha}$ unchanged, thus we have $\lambda\to t\lambda$,
$\tilde\lambda\to t^{-1}\tilde\lambda$ under a scaling $t$. Utilizing the scaling freedom we define the canonical \emph{minus-helicity} and \emph{plus-helicity} spinors for on-shell momentum $p$:
\begin{equation}
  \lambda_{\alpha}
  =2^{1/4} \begin{pmatrix}p^\str/\sqrt{p^+}\\[3pt]
    \sqrt{p^+}\end{pmatrix},
  \qquad
  \tilde\lambda_{\dot\alpha}
  = 2^{1/4}\begin{pmatrix}p^\bul/\sqrt{p^+}\\[3pt]
    \sqrt{p^+}\end{pmatrix}.
  \label{eq:spinor_standard}
\end{equation}

The upper index spinors are defined with the help of antisymmetric symbol $\epsilon^{\alpha\beta}$, $\epsilon^{12}=1=-\epsilon^{21}$ and $\epsilon_{12}=1=-\epsilon_{21}$,
\begin{equation}
    \lambda_{\alpha}=\epsilon_{\alpha\beta}\lambda^{\beta}\,,\qquad 
     \tilde\lambda_{\dot\alpha}=\epsilon_{\dot\alpha\dot\beta}\tilde\lambda^{\dot\beta}\,.
\end{equation}
Since $\epsilon^{\alpha\beta}\epsilon_{\beta\gamma} = -\delta^{\alpha}_{\gamma}$, the inverse relations read
\begin{equation}
    \lambda^{\alpha}=\lambda_{\beta}\,\epsilon^{\beta\alpha}\,,\qquad
    \tilde\lambda^{\dot\alpha}=\tilde\lambda_{\dot\beta}\,\epsilon^{\dot\beta\dot\alpha}\,,
\end{equation}
i.e.\ $\lambda^1=-\lambda_2$, $\lambda^2=\lambda_1$ (and likewise for dotted indices), with $\epsilon^{\dot1\dot2}=\epsilon_{\dot1\dot2}=1$.

\paragraph{Spinor products.}
For on-shell momenta $p_i$, $p_j$ and the corresponding spinors, the usual spinor brackets are defined as
\begin{equation}
  \ab{i}{j}
  = -\lambda_i^{\alpha}\lambda_{j,\alpha}
  = \lambda_{i,\alpha}\lambda_j^{\alpha}
  = -\epsilon^{\alpha\beta}\lambda_{i,\alpha}\lambda_{j,\beta}
  = \lambda_{i,2}\lambda_{j,1}-\lambda_{i,1}\lambda_{j,2}
  = \sqrt{2}\,\frac{p_j^\str\,p_i^+ - p_i^\str\,p_j^+}{\sqrt{p_i^+ p_j^+}}\,,
\end{equation}
\begin{equation}
  \sqb{i}{j}
  = \tilde\lambda_i^{\dot\alpha}\tilde\lambda_{j,\dot\alpha}
  = -\tilde\lambda_{i,\dot\alpha}\tilde\lambda_j^{\dot\alpha}
  = \epsilon^{\dot\alpha\dot\beta}\tilde\lambda_{i,\dot\alpha}\tilde\lambda_{j,\dot\beta}
  = \tilde\lambda_{i,\dot1}\tilde\lambda_{j,\dot2}-\tilde\lambda_{i,\dot2}\tilde\lambda_{j,\dot1}
  = \sqrt{2}\,\frac{p_i^\bul\,p_j^+ - p_j^\bul\,p_i^+}{\sqrt{p_i^+ p_j^+}}\,,
  \label{eq:brackets_def}
\end{equation}
so that $\ab{i}{j}\sqb{j}{i} = 2\,p_i\cdot p_j$.
Under the little group scaling
\begin{equation}
  \ab{i}{j}\to t_i\,t_j\,\ab{i}{j}\,,\qquad
  \sqb{i}{j}\to t_i^{-1}\,t_j^{-1}\,\sqb{i}{j}\,.
  \label{eq:LG_brackets}
\end{equation}

\paragraph{Polarization vectors.} 
We are using the following representations of a $+$ helicity polarization vector for on-shell momentum $p$:
\begin{equation}
    \varepsilon^{+\,\mu}(p,q) = \frac{\lambda_q^{\alpha}\sigma^{\mu}_{\alpha\dot\alpha}\tilde\lambda^{\dot\alpha}_p}{\sqrt{2}\ab{q}{p}}\,,
    \label{eq:pol_plus_def1}
\end{equation}
where $q$ is a reference momentum.
Similarly, for the $-$ helicity polarization vector we have
\begin{equation}
    \varepsilon^{-\,\mu}(p,q) = \frac{\lambda_p^{\alpha}\sigma^{\mu}_{\alpha\dot\alpha}\tilde\lambda^{\dot\alpha}_q}{\sqrt{2}\sqb{p}{q}}\,.
    \label{eq:pol_minus_def1}
\end{equation}
In 2D representation this reads:
\begin{equation}
    \varepsilon^{+}_{\alpha\dot\alpha}(p,q) = \sqrt{2}\frac{\lambda_{q,\alpha}\tilde\lambda_{p,\dot\alpha}}{\ab{q}{p}}\,, \qquad
    \varepsilon^{-}_{\alpha\dot\alpha}(p,q) = \sqrt{2}\frac{\lambda_{p,\alpha}\tilde\lambda_{q,\dot\alpha}}{\sqb{p}{q}}\, .
    \label{eq:pol_def2}
\end{equation}
Under the little group scaling chosen above $\varepsilon^+ \rightarrow  t^{-2} \varepsilon^+ $ and $\varepsilon^- \rightarrow  t^2 \varepsilon^- $.

\paragraph{Relation of $\varepsilon^{\pm}_{\alpha}$ to polarization vectors.} 
Notice, that the vectors \eqref{eq:epsilon_alpha}, \eqref{eq:epsilon_beta} are reminiscent of polarization vectors. Indeed, if, say $\alpha=p^{\bullet}/p^+$ in~\eqref{eq:epsilon_alpha}, the corresponding 
\begin{equation}
    \varepsilon^{+}_{p^{\bullet}/p^+}\equiv\varepsilon^{+}(p)= \varepsilon_{\perp}^{+\, \mu }- \frac{p^{\bullet}}{p^+} \,\eta^{\mu}
    \label{eq:polvec}
\end{equation} is the plus-helicity polarization vector for a particle with four momentum $p^{\mu}$ and reference vector $\eta$. As can be immediately checked it satisfies $\varepsilon^{+}(p)\cdot p =\varepsilon^{+}(p)\cdot\eta = 0$ and $\varepsilon^{+}(p)\cdot\varepsilon^{+}(p)=0$, $\varepsilon^{+}(p)\cdot \varepsilon^{-}(p) = -1$.
In 2D representation, directly from \eqref{eq:polvec} we get
\begin{equation}
  \varepsilon^+_{\alpha\dot\alpha}(p)
  = \sqrt{2}\begin{pmatrix} -p^{\bullet}/p^+ & -1\\[2pt]
    0 & 0 \end{pmatrix} \,.
  \label{eq:epsilon_plus_matrix}
\end{equation}
On the other hand, we recover the same matrix using \eqref{eq:pol_plus_def1} with the spinors 
\begin{equation}
    \lambda_{\eta,\alpha}=\tilde\lambda_{\eta,\dot\alpha}=2^{1/4}\begin{pmatrix}1\\[3pt]
    0\end{pmatrix} \,,
    \label{eq:lambda_eta_spinor}
\end{equation}
for the reference momentum $\eta^{\mu}$, and \eqref{eq:spinor_standard} for the momentum $p^{\mu}$ 
(Both spinors are written as columns, as in \eqref{eq:spinor_standard}; in \eqref{eq:pol_plus_def1} the dotted spinor is to be read as a row, i.e. $\varepsilon^{+}_{\alpha\dot\alpha}=\sqrt2\,\lambda_{q,\alpha}\tilde\lambda_{p,\dot\alpha}/\ab{q}{p}$ is the matrix $\sqrt2\,\lambda_q\tilde\lambda_p^{\,T}/\ab{q}{p}$.). 
Let us now comment on the little group scaling. In our formalism the information about polarization enters through vectors~\eqref{eq:polvec} or the corresponding matrix~\eqref{eq:epsilon_plus_matrix}, rather than through ~\eqref{eq:pol_plus_def1}. Since the matrix~\eqref{eq:epsilon_plus_matrix} has determinant zero, we could simply assume it is scale invariant and associate the scaling weights to the corresponding spinors. It is however more natural, to assume that~\eqref{eq:epsilon_plus_matrix} has the scaling of the plus helicity polarization vector~\eqref{eq:pol_plus_def1}. 

Similarly, we define the \emph{minus} helicity polarization vector
\begin{equation}
    \varepsilon^{-}(p)= \varepsilon_{\perp}^{-\, \mu }- \frac{p^{\star}}{p^+} \,\eta^{\mu}
    \label{eq:polvec_minus}
\end{equation}
Notice, that the polarization vectors $\varepsilon^{\pm}(p)$ \emph{can be defined regardless the momentum is on-shell or off-shell}.

\paragraph{The $v$-symbols.} 
It is convenient to introduce an inner product of the polarization vector~\eqref{eq:polvec},~\eqref{eq:polvec_minus}  corresponding to the four momentum $p_i$, with another four momentum $p_j$\footnote{We note the difference in the notation used throughout this work comparing to \cite{Kotko2017,Kakkad2020,Kakkad:2021uhv,Kakkad_2022,Kakkad:2024ase}. Our $\vts{i}{j}$ replaces $\widetilde{v}^*_{ij}$ and $\vta{i}{j}$ replaces $\widetilde{v}_{ij}$. } 
\begin{equation}
  \vts{i}{j}
  \;\equiv\; \varepsilon^{+}(p_i)\cdot p_j 
  = p_j^\bul - \frac{p^\bul_i}{p^+_i}\,p^+_j\,,
  \label{eq:vsymbol}
\end{equation}
\begin{equation}
  \vta{i}{j}
  \;\equiv\; \varepsilon^{-}(p_i)\cdot p_j 
  = p^\star_j - \frac{p^\star_i}{p^+_i}\,p^+_j\,.
  \label{eq:vsymbol_star}
\end{equation}
Above, the four momenta $p_i,p_j$ can be either on-shell or \emph{off-shell}. In the context of the latter, it is convenient to introduce 
\begin{equation}
    p_{i \dots j} \equiv p_i + p_{i+1} \dots + p_j \,, \quad j>i \,
\end{equation}
and define 
\begin{equation}
    v_{(i \dots j)(n\dots m)} \equiv  \varepsilon^+(p_{i\dots j})\cdot p_{n\dots m} 
\end{equation}
and similar for the $v^{\star}$.

\paragraph{Relation of $v$-symbols and spinor products.}

For on-shell $p_i$, $p_j$ we readily obtain the following relations
\begin{equation}
    \vts{i}{j} = -\frac{1}{\sqrt{2}}\sqrt{\frac{p_j^+}{p_i^+}}\,\sqb{i}{j}
    \label{eq:vij_to_spinor}
\end{equation}
and
\begin{equation}
    \vta{i}{j} = +\frac{1}{\sqrt{2}}\sqrt{\frac{p_j^+}{p_i^+}}\,\ab{i}{j} \,.
    \label{eq:vstarij_to_spinor}
\end{equation}
The second relation follows from the first under $p^\bul\leftrightarrow p^\str$, $\tlam\to\lambda$, up to an overall sign, which reflects the relative sign in the definitions~\eqref{eq:brackets_def} of the two brackets (equivalently, $\ab{i}{j}\sqb{j}{i}=+2\,p_i\cdot p_j$). A few further elementary properties of the $v$-symbols will be used repeatedly. Writing the definition over a common denominator,
\begin{equation}
  \vts{i}{j}=\frac{p^+_ip^\bul_j-p^+_jp^\bul_i}{p^+_i}\,,
  \qquad\text{so that}\qquad
  p^+_i\,\vts{i}{j}=-\,p^+_j\,\vts{j}{i}\,,
  \label{eq:v_lightfront}
\end{equation}
the numerator is antisymmetric under $i\leftrightarrow j$. On a three-point kinematics with momentum conservation this antisymmetric combination is the same for every pair,
\begin{equation}
  p^+_1\vts{1}{2}=p^+_2\vts{2}{3}=p^+_3\vts{3}{1}
  \qquad\big(\textstyle\sum_a p_a=0\,,\;n=3\big)\,,
  \label{eq:v_threept}
\end{equation}
an identity used heavily in Section~\ref{sec:amplitudes}.  
The vanishing locus $\vts{i}{j}=0$ is the light-front collinearity $p^\bul_i/p^+_i=p^\bul_j/p^+_j$, i.e.\ alignment of the polarization slopes in~\eqref{eq:polvec}. Finally, with the reference spinors $\lambda_\eta=\tlam_\eta=2^{1/4}(1,0)^T$ one checks
\begin{equation}
  \ab{i}{\eta}=\sqb{\eta}{i}=\sqrt{2p^+_i}\,,
  \label{eq:eta_contraction}
\end{equation}
so the ubiquitous light-front factors $\sqrt{p^+}$ can be interpreted as reference-spinor contractions.

In the next two subsections we discuss two functionals on the anti-self-dual plane that encode the simplest objects related to scattering amplitudes of  minus helicity states. Similar structures could be of course defined on the self-dual plane.

\subsection{Wilson lines on anti-self-dual plane}
\label{sub:Wilson_lines_in_CM4}

Consider an  anti-self-dual plane $\Pi^+$ fixed by some $x^+$ and $x^{\bullet}$ and a straight segment of the Wilson line within~\cite{Kotko2017}:
\begin{equation}
    \mathcal{W}_{\alpha}^{+}\left(x^+,x^{\bullet};\vec{y},\vec{x}\right)=\mathcal{P} \exp{\left\{ig \int_0^1\!\!ds\, \hat{A}\left(z\left(s\right)\right)\cdot \varepsilon^{+}_{\alpha}\right\} } \,,
    \label{eq:WL_generic}
\end{equation}
where the path is defined as
\begin{equation}
    z^{\mu}(s)=x^{\mu} + s \varepsilon^{+\mu}_{\alpha} \,,
\end{equation}
with the end points: $x^{\mu}=z^{\mu}(0)=(x^+,\vec{x},x^{\bullet})$, with $\vec{x}=(x^-,x^{\star})$,  $z^{\mu}(1)=(x^+,\vec{y},x^{\bullet})$, with $\vec{y}=(y^-,y^{\star})$. Above $\hat{A}^{\mu}=t^a A^{\mu}_a$ is the $\mathrm{SU}(N_c)$ algebra-valued field, with $[t^a,t^b]=i\sqrt{2}f^{abc}t^c$ and $\Tr(t^at^b)=\delta^{ab}$. 
Notice, that in the light-cone gauge $A^+=0$ we have $A\cdot\varepsilon_{\alpha}^+=A^{\bullet}$.
We define an \emph{infinite} straight line on anti-self-dual (self-dual) plane passing through $x^{\mu}$ by allowing the affine parameter $s$ to take any real value \eqref{eq:SDplane} (similarly, \eqref{eq:ASDplane} for SD plane)
\begin{equation}
    \mathcal{W}_{\alpha}^{\pm}\left(x\right)=\mathcal{P} \exp{\left\{ig \int\!\!ds\, \hat{A}^{\bullet}\left(x+s\varepsilon^{\pm}_{\alpha}\right)\right\} } \,.
    \label{eq:WL_inf}
\end{equation}

The Wilson line~\eqref{eq:WL_generic} is, by definition, a Lorentz-invariant object.
For our purposes, it is however important to restate the transformation properties for fixed basis $\varepsilon_{\perp}^{\pm}$ and $\eta$, $\tilde\eta$. We see that in order to maintain the boost invariance and the rotational invariance on the X-Y plane, for which 
\begin{equation}
 x^+ \rightarrow \kappa x^+ \,\,\,, x^- \rightarrow \bar{\kappa} x^- \,, \quad x^{\star}\rightarrow \bar\gamma  x^{\star}\,\,\,,  x^{\bullet}\rightarrow \gamma x^{\bullet} \,,
 \label{eq:Lorentz_transf_x}
\end{equation}
 with $\kappa\bar\kappa = 1$ and $\gamma\bar\gamma = 1$, we should choose that the $s$ parameter transforms as the "star" coordinate while $\alpha$ transforms as $x^- /x^{\star}$
\begin{equation}
 s\rightarrow \bar\gamma s \,, \quad \alpha \rightarrow \bar\kappa\gamma \alpha \,.
 \label{eq:s_alpha_scaling}
\end{equation}
See Appendix~\ref{app:Lorentz} the short summary of the Lorentz transformations relevant to our work.

Consider next the Wilson line~\eqref{eq:WL_inf} integrated over the line direction parameter $\alpha$
\begin{equation}
    \mathcal{W}^{\pm}(x)=\int\!\!d\alpha\, \partial_{\mp}\mathcal{W}_{\alpha}^{\pm}(x) \,.
    \label{eq:alpha-integrated_WL}
\end{equation}
Notice it transforms as the "bullet" component of a vector field, thanks to the light cone derivatives.

The object above turns out to be of great interest. Below, we list a few situations it emerged quite naturally:
\begin{enumerate}[label=\roman*)]
    \item $\mathcal{W}^{+}$ has been  used to construct the classical anti-self-dual connection for scattering of minus helicity gluons (the generating functional for the Berends-Giele off-shell currents)~\cite{Kotko2017,Kakkad2020} -- see also the next subsection.
    \item In momentum space, $\mathcal{W}^{+}$ encodes the generating functional for the so called gluon wave functions~\cite{Kotko2017}. Within the light front quantization they represent amplitudes for an on-shell gluon to split into $n$ same helicity gluons \cite{Motyka2009}.
    \item Upon projecting onto the octet, $\mathcal{W}^{+}$ can be related to the canonical field transformation from the light cone quantized Yang-Mills theory to the MHV theory \cite{Kotko2017,Kakkad2020,Kakkad_2022}, as well as the more recent, classically equivalent, theory not containing triple gluon interactions  \cite{Kakkad:2021uhv}, and its one-loop quantum completion \cite{Kakkad:2024ase}.
\end{enumerate}

\subsection{Same-helicity scattering tree amplitudes}
\label{sub:self-dual}

The integrated Wilson line~\eqref{eq:alpha-integrated_WL} can be used to find the anti-self-dual connection $A^{\bullet}$ as a functional of some external auxiliary source.
Since the endpoints were moved to infinity and the dependence on the path integrated out, we can interpret the  quantity
\begin{equation}
    j^a(x)=\frac{1}{2\pi g}\Tr\left\{t^a\, W^+(x) \right\} \,,
    \label{eq:SD_j}
\end{equation}
as a $\mathrm{SU}(N_c)$ field in adjoint representation (it should be clear from the previous considerations, that $j$ transforms as the "bullet" component of a vector  field).
Inverting now the functional $j^a(x)=\mathcal{F}[A^{\bullet}](x)$ we can attempt to find the connection $A_a^{\bullet}[j]$.

This has been explicitly done in momentum space in~\cite{Kotko2017} with the external field $j^a$ on-shell. Introducing a position three vector $\mathbf{x}\equiv\left(x^{-},x^{\bullet},x^{\star}\right)$ and its momentum space conjugate $\mathbf{p}\equiv\left(p^{+},p^{\star},p^{\bullet}\right)$, Eq.~\eqref{eq:SD_j} can be Fourier transformed on a fixed light cone time $x^+$ as
\begin{equation}
    \tilde{j}_a\left[A^{\bullet}\right](x^+;\mathbf{P}) = \sum_{n=1}^{\infty} 
    \int \prod_{i=1}^n \left( d^3\mathbf{p}_i \tilde{A}^{\bullet}_{b_i}(x^+;\mathbf{p}_i)\right)  \, \widetilde{\Gamma}_n^{ab_1\dots b_n}(\mathbf{P};\mathbf{p}_1,\dots ,\mathbf{p}_n) \,,
    \label{eq:SD_j_mom}
\end{equation}
where
\begin{equation}
    \widetilde{\Gamma}^{ab_1\dots b_n}_n (\mathbf{P};\mathbf{p}_1,\dots,\mathbf{p}_n) = g^{n-1} \frac{\delta^3\left(\mathbf{p}_1+\dots+\mathbf{p}_n-\mathbf{P}\right)\,\Tr\!\left(t^at^{b_1}\dots t^{b_n}\right)}{v_{(1\cdots n)1} v_{(1\dots n)(12)} \dots v_{(1\dots n)(1 \dots n-1)}} \,.
    \label{eq:Gamma_n}
\end{equation}
Above we used the off-shell $v$-symbols defined earlier. 
Inserting now the following ansatz 
\begin{equation}
    \tilde{A}^{\bullet}_a\left[j \right](x^+;\mathbf{P}) = \sum_{n=1}^{\infty} 
    \int \prod_{i=1}^n \left( d^3\mathbf{p}_i \tilde{j}_{b_i}(x^+;\mathbf{p}_i)\right)  \, \widetilde{\Psi}_n^{ab_1\dots b_n}(\mathbf{P};\mathbf{p}_1,\dots ,\mathbf{p}_n) 
    \label{eq:SD_Abullet_mom}
\end{equation}
to~\eqref{eq:SD_j_mom} we find
\begin{equation}
    \widetilde{\Psi}_n^{ab_1\dots b_n}(\mathbf{P};\mathbf{p}_1,\dots,\mathbf{p}_n) =     
    - g^{n-1}  
    \frac{v_{1(1 \dots n)}}{v_{(1 \dots n)1}}
    \frac{\delta^3\left(\mathbf{p}_1+\dots+\mathbf{p}_n-\mathbf{P}\right)\,\Tr\!\left(t^at^{b_1}\dots t^{b_n}\right)}{v_{12}v_{23} \dots v_{(n-1)n}} \, ,
    \label{eq:Psi_n} 
\end{equation}
where the prefactor simplifies as
\begin{equation}
    \frac{v_{(1 \cdots n)1}}{v_{1(1 \cdots n)}} = -\frac{p^+_{1}}{p^+_{1\dots n}} \,.
\end{equation}
Notice the familiar denominator that resembles the Parke-Taylor form (without the cyclicity, however). The kernel~\eqref{eq:Psi_n} can be shown to be equivalent to the Berends-Giele off-shell currents for scattering of minus helicity gluons, see eg.~\cite{Monteiro2011}.

\subsection{Berends-Giele current as a functional on piecewise correlated random path}
\label{sub:BG_position_space}

\begin{figure}
\centering
 \includegraphics[width=10cm]{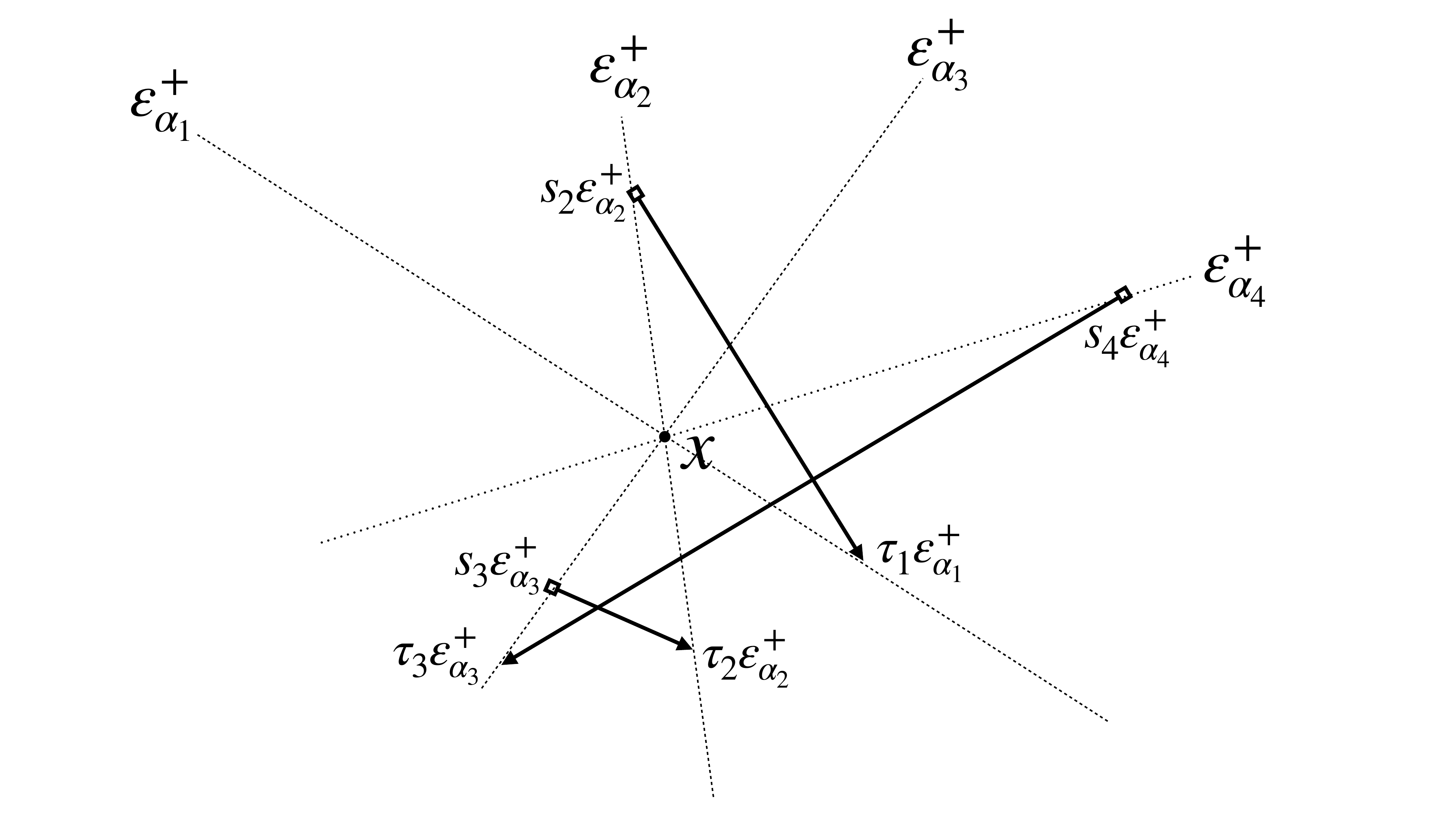} 
 \caption{Visualization of the points $\vec{y}_i$ on the ASD plane. The dashed lines represent direction vectors $\vec{\varepsilon}^+_{\alpha}$. The open circle (the nock of the vector) corresponds to $s_{i+1}\vec{\varepsilon}^+_{\alpha_{i+1}}$ while the head to $\tau_{i}\vec{\varepsilon}^+_{\alpha_{i}}$, with $\tau_i<0$. The vectors correspond thus to $\vec{y}_i$. \label{fig:vec_def} }
\end{figure}

While the momentum space representation of the anti-self-dual Berends-Giele currents is well known, the majority of the present work will be based on the corresponding \emph{position space} representation found in~\cite{Kotko2017}. The solution is written as a power expansion in the currents $\hat{j}\equiv j^a t^a$, where instead of a kernel the currents are being correlated by a direction vector on anti-self dual plane. More precisely, consider a set of points on ASD plane, $\vec{y}_i\equiv (y_i^-,y_i^{\star})$, such that
\begin{equation}
\begin{aligned}
 \vec{y}_1 & = \vec{x}+\tau_1 \vec{\varepsilon}^+_{\alpha_1} + s_2 \vec{\varepsilon}^+_{\alpha_2} \,, \\
  \vec{y}_2 & = \vec{x}+ \tau_2 \vec{\varepsilon}^+_{\alpha_2} + s_3 \vec{\varepsilon}^+_{\alpha_3} \,, \\
  \vdots \\
   \vec{y}_{n-1} & = \vec{x}+ \tau_{n-1} \vec{\varepsilon}^+_{\alpha_{n-1}} + s_n \vec{\varepsilon}^+_{\alpha_n} \,, 
\end{aligned}
\label{eq:piecewise_path}
\end{equation}
where $\tau_i<0$ and $s_i \in  \mathbf{R}$ are path parameters, see~Fig.~\ref{fig:vec_def}. 
The corresponding four-vectors $y^{\mu}=(y^+,y^-,y^{\bullet},y^{\star})$ have $y^+=x^+$ and $y^{\bullet}=x^{\bullet}$,
\begin{equation}
    y_{n-1}^{\mu}=(x^+,\,x^- - \tau_{n-1}\alpha_{n-1} -s_n\alpha_n,\, x^{\bullet},\,x^{\star} -\tau_{n-1} - s_n ) \,.
\end{equation}

\begin{figure}
\centering
 \includegraphics[width=11cm]{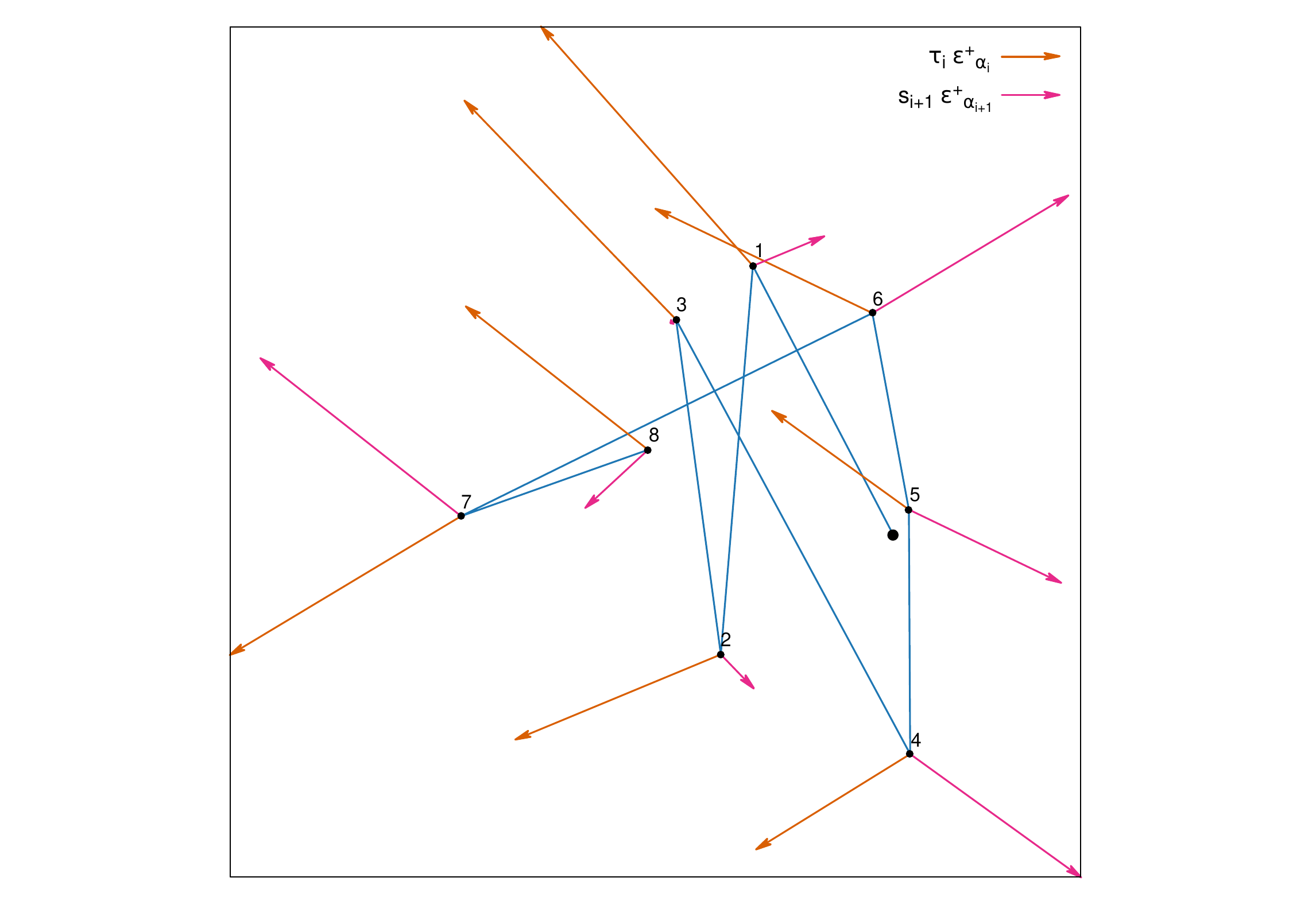}
 \caption{Visualization of the piecewise random path with memory $\{\vec{y}_i\}$ on ASD plane. The arrows at each point represent $s_{i+1}\vec{\varepsilon}^+_{\alpha_{i+1}}$ and $\tau_{i}\vec{\varepsilon}^+_{\alpha_{i}}$ (red and orange). Their sum is the position of the current vector from the origin (bold dot). Each pair of points share the same directions for the pair orange-red vectors, thus the path has the "velocity" memory. \label{fig:path} }
\end{figure}

In \cite{Kotko2017} attempts were made to interpret this set of points, but in the present work we explore the following new interpretation: for a given set of parameters $\{\tau_i,s_i,\alpha_i\}$ the set of points constitutes a piecewise trajectory with a "memory" of the previous step, due to correlated "velocities" (the direction vectors $\vec{\varepsilon}^+$), see~Fig.~\ref{fig:path}.

The position space solution corresponding to the Berends-Giele current can be written as a power expansion on these paths:
\begin{multline}
A^{\bullet}_a[j](x)= \sum_{n=1}^{\infty} \frac{g^{n-1}}{(2\pi)^n} \,\partial_-\Tr \Big\{
t^a
\int ds_{1}\,d\alpha_{1}\,\,\hat{j}\left(x+s_{1}\varepsilon_{\alpha_{1}}^{+}\right) \\
\prod_{i=1}^{n-1}\,  \int ds_{i+1}\,d\alpha_{i+1}\int_{-\infty}^{0}d\tau_{i}\,\,\partial_{-}\hat{j}\left(x+\tau_{i}\varepsilon_{\alpha_{i}}^{+}+s_{i+1}\varepsilon_{\alpha_{i+1}}^{+}\right) \Big\}\,.
\label{eq:SD_sol_pos}
\end{multline}
Above, the derivatives act on the $x$ coordinate. Since the current $j$ transforms as the "bullet" vector field component, while $\tau$ has the same scaling as $s$, we see that the above formula preserves the required Lorentz symmetry. 

It is remarkable, that a class of solutions to the classical ASD equation can be written in such a simple form.  Notice, that there is no integral kernel. The whole information on the field interactions is hidden in the cross correlation of the points via the common direction of one of the direction vectors. In Section~
\ref{sec:F_SD} we will explicitly see how the Parke-Taylor form emerges in the momentum space.

We can also easily rewrite the functional, still in position space, in an alternative form, where the field correlations are hidden in a kernel. We just change the variables so that each field is independent and compute the Jacobian. This gives 
\begin{equation}
A^{\bullet}_a[j](x)= \sum_{n=1}^{\infty} \frac{g^{n-1}}{(2\pi)^n}
 \int d^2y_1\dots d^2y_n\,\, \partial_- K_n(x;\vec{y}_1,\dots,\vec{y}_n) \,\, \Tr \left\{t^a \partial_{-}\hat{j}\left(y_1\right)\dots \partial_{-}\hat{j}\left(y_n\right) \right\}\,,
\label{eq:SD_sol_pos_with_kernel}
\end{equation}
where $y_i^{\mu} = (x^+,x^{\bullet},\vec{y}_i)$ is the space time point confined to the ASD plane and
\begin{equation}
  K_n(x;\vec{y}_1,\dots,\vec{y}_n) = \int_{\mathcal{V}_x}\! d\alpha_1 \dots d\alpha_n \prod_i \frac{1}{|\alpha_{i+1}-\alpha_i|}
  \,\,  \,,
\end{equation}
where $\mathcal{V}_x$ is rather-complicated $x$-dependent integration region in $\alpha_i$ space.
The kernel  acts thus as a volume of parameter space that can generate the points~\eqref{eq:piecewise_path}. Although this result is interesting on its own, the version with the cross correlated fields is much simpler and will be used throughout the work. 


In the main part of this work, we shall generalize the above functional to accommodate on-shell amplitudes and show its twistor space incarnation.

\subsection{Twistor space}
\label{sub:twistor_space}

For later use we collect the minimal twistor theory background. The detailed derivation of how twistors \emph{emerge} from the functional discussed above in the on-shell limit and  by imposing cyclicity of the trajectory  is given in Section~\ref{eq:F_in_twistor_space}. 

\paragraph{Twistors. }
For our purposes, the twistor space is defined as the projective complex space $\mathbb{CP}^3$ with homogeneous coordinates $Z^I=(Z^1,Z^2,Z^3,Z^4)$ defined up to $Z\sim cZ$, $c\in\mathbb{C}^{*}$. It is often convenient to map  $Z^I=(\pi_{\dot\alpha},\omega^{\alpha})$. 

\paragraph{Classification of null planes in $\mathbf{CM}_4$.} 
We have introduced two null planes $\Pi^+(x)$ (ASD plane) and $\Pi^-(x)$ (SD plane). In order to map these to twistor space, let us consider a variation of the coordinates within the planes. First consider $\xi\in\Pi^+(x)$. Varying $\xi^-$ and $\xi^{\star}$ we have
\begin{equation}
    \delta \xi_{\alpha\dot\alpha} = 
    \begin{pmatrix}
     \delta \xi^- & 0\\[2pt]
    0 & 0
    \end{pmatrix} =  
    \begin{pmatrix}
     1 \\[2pt]
     0 
    \end{pmatrix}
    \begin{pmatrix}
     \delta\xi^- & 0 
    \end{pmatrix}  
\quad \text{and} \quad
    \delta \xi_{\alpha\dot\alpha} = 
    \begin{pmatrix}
     0 & \delta \xi^{\star}\\[2pt]
    0 & 0
    \end{pmatrix} 
    = 
    \begin{pmatrix}
     1 \\[2pt]
     0 
    \end{pmatrix}
    \begin{pmatrix}
     0 & \delta\xi^{\star} 
    \end{pmatrix} \,,
\end{equation}
so they share the same un-dotted spinor $\lambda_{\eta,\alpha}$~\eqref{eq:lambda_eta_spinor} which is held fixed. In the twistor space nomenclature, such plane is called \emph{$\beta$-plane}.
On the other hand, for  $\xi\in\Pi^-(x)$ we can write
\begin{equation}
    \delta \xi_{\alpha\dot\alpha} = 
    \begin{pmatrix}
     \delta \xi^- & 0\\[2pt]
    0 & 0
    \end{pmatrix} =  
    \begin{pmatrix}
     \delta\xi^- \\[2pt]
     0 
    \end{pmatrix}
    \begin{pmatrix}
     1 & 0 
    \end{pmatrix}  
\quad \text{and} \quad
    \delta \xi_{\alpha\dot\alpha} = 
    \begin{pmatrix}
     0 & 0 \\[2pt]
    \delta \xi^{\bullet} & 0
    \end{pmatrix} 
    = 
    \begin{pmatrix}
     0  \\[2pt]
     \delta\xi^{\bullet} 
    \end{pmatrix}
    \begin{pmatrix}
     1 & 0 
    \end{pmatrix}   \,,
\end{equation}
so the points within the plane share the dotted spinor $\lambda_{\eta,\dot\alpha}$ of~\eqref{eq:lambda_eta_spinor}. In the twistor space terminology these are called \emph{$\alpha$-planes}.

A point $x\in\mathbf{CM}_4$ corresponds to the line (a $\mathbb{CP}^1$), which we denote by $L_x$, of twistors obeying the incidence relation $\omega^{\alpha}=x^{\alpha\dot\alpha}\pi_{\dot\alpha}$ (the precise normalization used in this work is dictated by the emergent incidence relations, see Eq.~\eqref{eq:incidence} in the next Section.). 

Conversely, a fixed twistor $Z$ is incident with a two-parameter family of points, $x^{\alpha\dot\alpha}=x_0^{\alpha\dot\alpha}+\mu^{\alpha}\pi^{\dot\alpha}$ with $\mu^\alpha\in\mathbb{C}^2$: a totally null plane with a fixed dotted spinor, i.e. a self-dual plane of the type of $\Pi^-(x)$ (an $\alpha$-plane). The anti-self-dual planes $\Pi^+(x)$, which have a fixed undotted spinor ($\beta$-planes), are not represented by points: each of them is represented by a dual twistor $W$, equivalently by the projective plane $\{Z:\,W\cdot Z=0\}\simeq\mathbb{CP}^2$ of all twistors whose planes meet it. We shall use the following consequences in Section~\ref{sub:twistor_geometry}: (i) two points $x,\xi$ are null separated if and only if $L_x$ and $L_\xi$ intersect, the common point being the twistor of the unique $\alpha$-plane containing the null line through $x$ and $\xi$; (ii) a null line is therefore represented by a single point of $\mathbb{CP}^3$, common to the lines $L_\xi$ of all its points; in particular the line $x+s\,\varepsilon^+_\alpha\subset\Pi^+(x)$ is represented by the point $P=(1,\alpha,\,x^{\bullet}-\alpha x^{+},\,x^{-}-\alpha x^{\star})$ of $L_x$, cf.~eq.~\eqref{eq:Pk} below; (iii) the lines $L_\xi$ of all points $\xi$ of a $\beta$-plane lie in the plane $W\cdot Z=0$ representing it.

The Penrose transform \cite{Penrose:1969ae,Bateman:1904xu} represents on-shell massless fields of helicity $h$ by (cohomology classes of) functions on twistor space homogeneous of degree $-2h-2$: contour integrals of such functions along the line of $x$, against the projective measure $\pi_{\dot\alpha}\,d\pi^{\dot\alpha}$ of weight $+2$, solve the free field equations. Two facts are used below: (i) the weight of the measure must be balanced by the weights of the rest of the integrand for the integral to be projectively well defined, and (ii) changing the homogeneity of the twistor function changes the helicity of the space-time field.

\section{Parke-Taylor functional on Anti-self-Dual plane}
\label{sec:F_SD}

It turns out that the position space solution to the anti-self-dual Yang-Mills equation~\eqref{eq:SD_sol_pos} can be turn into a powerful functional that describes various scattering amplitudes.

In this section we start with defining the functional that generates the Parke-Taylor cyclic denominator.

Let us \textit{introduce} the following functional of some algebra-valued field $\hat{\phi}$ 
\begin{equation}
\mathcal{F}[\phi]= \int d^4 x\,\sum_{n=3}^{\infty}  \,\Tr \left\{\prod_{i \in \mathbb{Z}_n}\,  \int ds_{i}\,d\alpha_{i}\int_{-\infty}^{0}d\tau_{i}\,\,\partial_{-}\hat{\phi}\left(x+\tau_{i}\varepsilon_{\alpha_{i}}^{+}+s_{i+1}\varepsilon_{\alpha_{i+1}}^{+}\right) \right\}\,.
\label{eq:F_functional}
\end{equation}
 The product is defined $\mathrm{mod}\,\,n$ i.e. $n+1 \equiv 1$.
In order to maintain the Lorentz invariance of the functional, the field has to have the following scaling properties:
\begin{equation}
    \hat\phi (\kappa x^+,\bar\kappa x^-,\gamma x^{\bullet},\bar\gamma x^{\star}) = \gamma\,\hat\phi (x^+,x^-,x^{\bullet}, x^{\star}) \, ,
    \label{eq:phi_homo_star}
\end{equation}
where the meaning of the scaling factors was explained below eq.~\eqref{eq:WL_inf}: the field transforms like the coordinate $x^{\bullet}$ in \eqref{eq:Lorentz_transf_x}, and the measure $ds\,d\alpha\,d\tau$ of each slot carries the weight $\bar\gamma\cdot\bar\kappa\gamma\cdot\bar\gamma=\bar\kappa\bar\gamma$, which together with the weight $\kappa$ of $\partial_-$ and $\gamma$ of the field gives one.
Thus, the $\phi$ field transforms as a "bullet" (spin-1) component of a vector field. For now, we shall, however, refrain from introducing explicit spin indicator.
The functional is Lorentz invariant with respect to general complex boosts and rotations around $Z$-axis.

\begin{figure}
\centering
 \includegraphics[width=11cm]{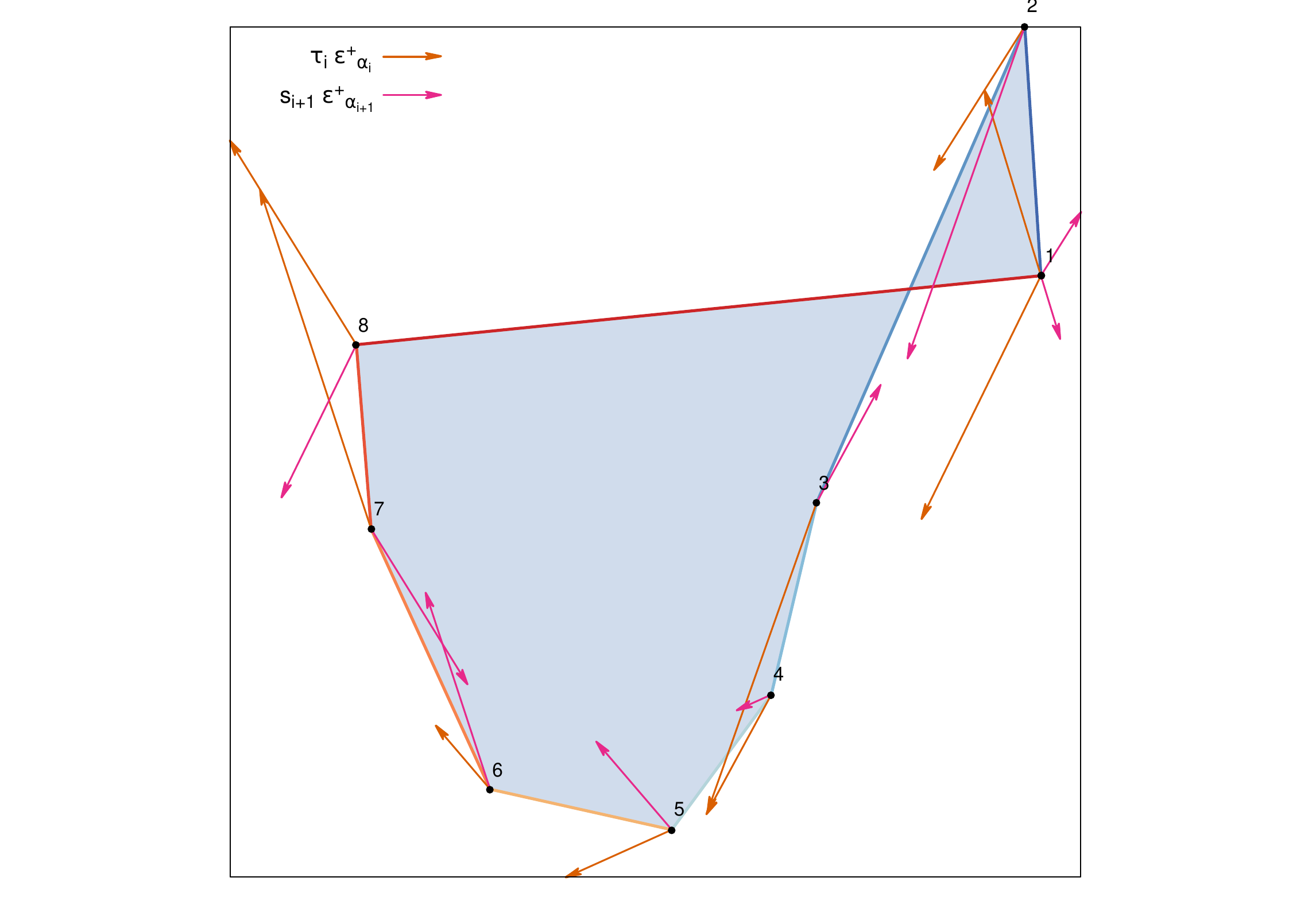}
 \caption{Visualization of the piecewise closed random path with memory on ASD plane, corresponding to the Parke-Taylor functional. The definition of arrows are same as in Fig.~\ref{fig:path}. The area enclosed is shaded for better visualization. \label{fig:closed_path} }
\end{figure}

There are couple of differences with the ASD solution~\eqref{eq:SD_sol_pos}: (i) there is an integration over $x$, so that it is a usual functional of the field $\phi$, (ii) the sum starts at $n=3$, so that there are at least three fields, (iii) there is a cyclicity property inferred, so that the \emph{last} field is shifted along the same $\varepsilon^+$ as the \emph{first} field, (iv) there is no projection onto an octet, so that we form a functional, (v) there is no global minus-component derivative.  
One should remember, that owing to the non-commutative nature of the generators, defining an appropriate ordering is necessary. This is achieved in Eq.~\eqref{eq:F_functional} as follows:  as $i$ increases, the fields with higher values of $i$ in the argument appear on the right. An example of a random closed path on which the fields are evaluated is shown in Fig.~\ref{fig:closed_path}.

We can also factor out the trace of the generators as shown below
\begin{equation}
\mathcal{F}[\phi]=\int d^4x\,\sum_{n=3}^{\infty} \,\mathrm{Tr}\left(t^{b_1}\dots t^{b_n}\right) \prod_{i \in \mathbb{Z}_n}\,  \int ds_{i}\,d\alpha_{i}\int_{-\infty}^{0}d\tau_{i}\,\,\partial_{-}{\phi}_{b_i}\left(x+\tau_{i}\varepsilon_{\alpha_{i}}^{+}+s_{i+1}\varepsilon_{\alpha_{i+1}}^{+}\right)\,.
\label{eq:F_funct_tr}
\end{equation}

Notice that the sum in the functional above starts with $n=3$, reminiscent of an action with a triple vertex. While the term corresponding to $n=1$ would vanish anyway, owing to $\Tr(t^a)=0$, the term for $n=2$ would become divergent (see also the discussion below).


\subsection{Momentum space}
\label{subsec:F_mom}

Let us express the function Eq.~\eqref{eq:F_funct_tr} in the momentum space. 

\paragraph{Step 1: Fourier transform.} 
Our conventions for the Fourier transform of the fields read
\begin{equation}
\phi_{b_i}\left(x\right)=\int \frac{d^{4}p}{(2\pi)^4}\, e^{-ix\cdot p}\widetilde{\phi}_{b_i}\left(p\right) = \int \frac{d^{4}p}{(2\pi)^4}\,e^{-i \left( x^{+}p^{-}+x^{-}p^{+}-x^{\bullet}p^{\star}-x^{\star}p^{\bullet}\right)} \widetilde{\phi}_{b_i}\left(p\right)\,.
\label{eq:FT_def}
\end{equation}

\paragraph{Step 2: $x$-integral.}
Each field insertion at $y_k = x + \tau_k\varepsilon^+_{\alpha_k}
+ s_{k+1}\varepsilon^+_{\alpha_{k+1}}$ gives a factor
$(-ip_k^+)$ because of the action of $\partial_-$
 and a phase that factorises as
\begin{equation}
  e^{-ip_k\cdot y_k}
  = e^{-ip_k\cdot x}
    \cdot e^{-i\tau_k(\varepsilon^+_{\alpha_k}\cdot p_k)}
    \cdot e^{-is_{k+1}(\varepsilon^+_{\alpha_{k+1}}\cdot p_k)}\,.
  \label{eq:phase_split}
\end{equation}
Integrating $\prod_k e^{-ip_k\cdot x}$ over $x$ gives the momentum conservation delta
$(2\pi)^4\delta^4(\sum_k p_k)$.

\paragraph{Step 3: $s_k$-integrals.}
The factor $e^{-is_k(\varepsilon^+_{\alpha_k}\cdot p_{k-1})}$
depends on 
$\varepsilon^+_{\alpha_k}\cdot p_{k-1}
= p_{k-1}^\bul - \alpha_k p_{k-1}^+$.
Integrating over $s_k$ gives
\begin{equation}
    \int\! ds_k \, e^{-is_k\,\, \varepsilon^+_{\alpha_k}\cdot p_{k-1}} = 2\pi \, \delta(p_{k-1}^\bul - \alpha_k p_{k-1}^+) \, ,
    \label{eq:alpha_delta}
\end{equation}
which fixes $\alpha_k$ to be equal to
\begin{equation}
  \alpha_{p_{k-1}} \equiv \frac{p_{k-1}^\bul}{p_{k-1}^+}\,.
  \label{eq:alpha_fixed}
\end{equation}
The $s_k$-integral selects the slope on the SD plane
of leg $k$ to make the corresponding direction transverse to the momentum of the leg $k-1$. This positive-helicity transversality condition is
imposed here geometrically by the auxiliary integration over the $s_k$ variables.

\paragraph{Step 4: $\alpha_k$-integrals.}
The $\alpha_k$-integral over the delta from Step~3
gives the Jacobian
\begin{equation}
  \int d\alpha_k\;
  \delta\!\left(p_{k-1}^\bul - \alpha_k p_{k-1}^+\right)
  =  \int d\alpha_k\ \delta\left(\alpha_k - \alpha_{p_{k-1}}\right) \frac{\sgn{p_{k-1}^+}}{p_{k-1}^+}\,,
  \label{eq:alpha_jacobian}
\end{equation}
providing a factor $\sgn{p_{k-1}^+}/p_{k-1}^+$ per leg.
Combined with the $(-ip_{k-1}^+)$ factor from $\partial_-$, the net factor per leg is $-i\sgn{p_{k-1}^+}$.

\paragraph{Step 5: Introducing the $v$-symbols.}
With $\alpha_k = p_{k-1}^\bul/p_{k-1}^+$ fixed,
the coefficient of the remaining $\tau_k$-phase for leg $k$ is
$-i\tau_k(\varepsilon^+_{\alpha_k}\cdot p_k)$.
We use~\eqref{eq:vsymbol}
\begin{equation}
  \varepsilon^+_{\alpha_k}\cdot p_k
  \;\Big|_{\alpha_k\,=\,p_{k-1}^\bul/p_{k-1}^+}
  = p_k^\bul - \frac{p_{k-1}^\bul}{p_{k-1}^+}\,p_k^+
  = \vts{(k-1)}{k}\,.
  \label{eq:tau_is_vtilde}
\end{equation}
The $\tau_k$-phase is therefore exactly
$-i\tau_k\vts{(k-1)}{k}$. 

\paragraph{Step 6: $\tau_k$-integrals.}
Understanding the $\tau$ integrals in the distributional sense 
\begin{equation}
  \int_{-\infty}^0 d\tau_k\;
  e^{-i\left(\tau_k\,\vts{(k-1)}{k}\,+i\,\epsilon\right)}
  = \frac{i}{\vts{(k-1)}{k}+i\epsilon}
  \,.
  \label{eq:halfline}
\end{equation}
In what follows we will treat the $i\epsilon$ prescription implicit.

\paragraph{Result: the $n$-point term.}
Collecting all the steps:
\begin{equation}
  \Fsd\Big|_n
  = (2\pi)^{n+4}\,\Tr\!\left(t^{b_1}\cdots t^{b_n}\right)
    \prod_{k \in \mathbb{Z}_n}\int\frac{d^4p_k}{(2\pi)^4}\;
    \delta^4\!\!\left(\textstyle\sum_k p_k\right)
    \frac{\sgn{p_k^+}}{\vts{(k-1)}{k}}\;
    \widetilde\phi_{b_k}(p_k)\,,
  \label{eq:Fsd_mom}
\end{equation}
where we identify $\vts{0}{1}\equiv \vts{n}{1}$ due to the imposed cyclicity property in our definition of the functional. 
The cyclic product in the denominator is the 
\emph{off-shell Parke-Taylor factor}
\begin{equation}
  \PTa{12\ldots n}
  = \vts{1}{2}\vts{2}{3}\dots \vts{(n-1)}{n}\vts{n}{1}\,.
  \label{eq:PT}
\end{equation}

The \emph{ASD functional} $\Fasd$ is defined by
replacing $\varepsilon^+_\alpha\to\varepsilon^-_\beta$
and, correspondingly, the direction parameters $\alpha_k$ by the $\beta_k$ of eq.~\eqref{eq:epsilon_beta}, with the reference field now transforming as the ``star'' component. Its corresponding
 PT factor is $\PTs{1 \ldots n}=\prod_k\vta{(k-1)}{k}$.

\subsection{Imposing on-shellness}
\label{subsec:ONshell_F}

So far the discussion has been fully off-shell. Now we consider imposing the on-shell condition on each of the fields in the functional defined in Eq.~\eqref{eq:F_functional}. 

It is natural to impose on-shellness in momentum space, by introducing a Dirac delta $(2\pi)\delta(p_i^2)\Theta(p_i^0) $, with $p_i^2 =2 p_i^- p_i^{+} - 2 p_i^{\bullet} p_i^{\star}$, for each of the external fields. Doing so, the phase space measure is
\begin{equation}
  \frac{d^4p_k}{(2\pi)^4}\cdot2\pi\delta(p_k^2) \Theta(p_i^0)
  = \frac{dp_k^+\,dp_k^\bul\,dp_k^\str}{(2\pi)^3\cdot2 p_k^+} \equiv \frac{d^3\mathbf{p}_k }{(2\pi)^3\cdot2 p_k^+} \, ,
  \label{eq:onshell_measure}
\end{equation}
where, as before, $\mathbf{p}\equiv(p^+,p^{\star},p^{\bullet})$. Note that for a massless momentum $\Theta(p^0)$ implies $p^+\ge0$, so the factors $\sgn{p_k^+}$ of \eqref{eq:Fsd_mom} are trivial on shell.
For the functional we get
\begin{multline}
\mathcal{F}[\phi]\Big|^{\mathrm{on-shell}}_n=    (-i)^{n}\,\Tr\left(t^{b_1} t^{b_2} \dots t^{b_n}\right) \int d^4 x\,\prod_{k \in \mathbb{Z}_n}\int ds_{k}\,d\alpha_{k} \int_{-\infty}^{0}d\tau_{k}\int \frac{d^3 \mathbf{p}_k \,}{2(2\pi)^3}\,\\
e^{-i \left[ x^{+}\frac{p_k^{\bullet} p_k^{\star}}{p_k^{+}}+(x^{-}- \tau_k \alpha_k - s_{k+1} \alpha_{k+1} )p_k^{+}-x^{\bullet}p_k^{\star}-(x^{\star}-\tau_k - s_{k+1}) p_k^{\bullet}\right] } \,\widetilde{\phi}_{b_k}\left(\mathbf{p}_k\right)\,,
\end{multline}
where the on-shell field is denoted by $\phi\left(p^-=p^{\bullet}p^{\star}/p^+,p^+,p^{\star},p^{\bullet}\right)\equiv\phi(\mathbf{p})$.

As before, the $x$ integration gives the overall momentum conservation delta, while the $s_k$ integrals give the delta functions~\eqref{eq:alpha_delta} that can be integrated fixing the $\alpha_k$~\eqref{eq:alpha_fixed}. Since $p_k$ are now on-shell we can use~\eqref{eq:vij_to_spinor} to replace the off-shell Parke-Taylor factor~\eqref{eq:PT} by
\begin{equation}
     \PTa{12\ldots n}
  = \left(\frac{-1}{\sqrt{2}}\right)^n \sqb{1}{2}\sqb{2}{3}\dots \sqb{(n-1)}{n}\sqb{n}{1} \quad (\textit{on-shell}) \,.
  \label{eq:PT_on-shell}
\end{equation}
Moreover, since $p_{k,\alpha\dot\alpha}=\lambda_k\tilde\lambda_k$ following the standard procedure we can replace $\delta^4(\sum_k p_i)$ by 
\begin{equation}
 \delta^4\Big(\sum_k \lambda_k \tilde\lambda_k \Big)
 = \sqb{i}{j}^2\, \delta^2\Big(\sum_k \lambda_k \sqb{k}{i}\Big)\delta^2\Big(\sum_l \lambda_l \sqb{l}{j}\Big) \, ,
 \label{eq:delta_spinors}
\end{equation}
where $i,j$ are two arbitrary legs. Finally, the phase space can also be written in terms of helicity spinors as
\begin{equation}
  \frac{dp_k^+\,dp_k^\bul\,dp_k^\str}{(2\pi)^3\cdot2 p_k^+}
  = \frac{d^2\lambda_k\,d^2\tilde\lambda_k}
         {2(2\pi)^3\,\mathrm{vol}\,GL(1)} 
  \equiv d\Lambda_k\,.
  \label{eq:LIPS_spinor}
\end{equation}
Putting everything together we get
\begin{equation}
  \Fsd\Big|_n^{\rm on-shell}
  = (-2\pi\sqrt2)^n (2\pi)^4\,\Tr\!\left(t^{b_1}\cdots t^{b_n}\right)
    \prod_{k \in \mathbb{Z}_n}\int d\Lambda_k\;
    \delta^4\!\!\left(\textstyle\sum_k
    \lambda_{k}\tilde\lambda_{k}\right)
    \frac{1}{\sqb{(k-1)}{k}}\;
    \widetilde\phi_{b_k}(\lambda_k,\tilde\lambda_k)\,,
  \label{eq:Fsd_final}
\end{equation}
where we have kept the same symbol for the on-shell field expressed in the spinor variables.

\subsection{Twistor space formulation}
\label{eq:F_in_twistor_space}

Although, in principle, we could "simply" take the Witten's half Fourier transform with respect to a spinor variable and attempt to write the functional in the twistor space,  we want to demonstrate how  -- quite  naturally -- the twistor space representation emerges, upon slight modification of the order of integration we did in the last section. Actually, it is quite amazing that merely restricting the arguments of the on-shell fields to wander across the plane in $\mathbf{CM}_4$ begs for the twistor space.

Initial steps are similar to the previous computation. (i) First, we put the fields on-shell by integrating the momentum "minus" components leaving the on-shell measure~\eqref{eq:onshell_measure}. (ii) Next, we integrate the  affine coordinates $s_k$ giving the delta functions ~\eqref{eq:alpha_delta}. (iii) Crucial: instead of integrating the directions $\alpha_k$, we integrate out $p_k^{\bullet}$'s. After integrating also the $\tau_k$'s this leaves 
\begin{multline}
\mathcal{F}[\phi]\Big|^{\mathrm{on-shell}}_n=     \Tr\left(t^{b_1} t^{b_2} \dots t^{b_n}\right) \int d^4 x\,\prod_{k \in \mathbb{Z}_n}  \int d\alpha_{k} 
\int \frac{dp^+_k dp^{\star}_k }{2(2\pi)^2\, p_k^+} \,\, \\
 e^{i p^{\star}_k ( x^{\bullet} - \alpha_{k+1}x^+) } 
  e^{-i p^+_k ( x^- - \alpha_{k+1}x^{\star}) }
  \,\, \frac{1}{\alpha_{k+1} - \alpha_k} \,\,\widetilde{\phi}_{b_k}\left(\alpha_{k+1}p_k^{\star},p_k^+,p_k^{\star},\alpha_{k+1}p_k^+\right)
 \, .
\label{eq:F-twistor1}
\end{multline}
Notice, there is no conservation delta (the $x$-integration is intact). We remind, that the order of the field arguments is $(p^-,p^+,p^{\star},p^{\bullet})$, but due to the integration the "minus" and "bullet" components are now expresses through $p^{\star}$ and $p^+$, multiplied by $\alpha$.

Since the field depends now, formally, only on $\alpha_{k+1}$, $p^+_k$ and $p^{\star}_k$, let us define a new function
\begin{equation}
  \tilde{f}_{b}\left( \alpha,p^{\star},p^+ \right) \equiv \frac{1}{2p^+} 
  \widetilde{\phi}_{b}\left(\alpha p^{\star},p^+,p^{\star},\alpha p^+\right) \, 
  \label{eq:twistor_f_def_1}
\end{equation}
and its formal inverse-Fourier transform
\begin{equation}
  f_{b}(\alpha,u,w) = \int\! \frac{dp^+ dp^{\star} }{(2\pi)^2}\, 
  e^{ip^{\star} u} e^{-ip^+ w} \,
  \tilde{f}_{b}\left( \alpha,p^{\star},p^+ \right) \,,
\end{equation}
where $u$,$w$ are, in general, complexes, and the integral is over contours that assure the existance of the transform\footnote{We could have chosen to work with a split signature that would render the "bullet" and "star" coordinates real, which would make our transforms ordinary FT. However, we choose to assume existance of a suitable contours instead.}.
With these definitions, we can write the $n$-th term of the functional, at fixed $x$, as
\begin{equation}
\mathcal{F}[\phi]\Big|^{\mathrm{on-shell}}_n (x) =    \,\Tr\left(t^{b_1} t^{b_2} \dots t^{b_n}\right) \,\prod_{k \in \mathbb{Z}_n}  \int\! d\alpha_{k}
  \,\,  
  \frac{f_{b_k}\left( \alpha_{k+1} , x^{\bullet} - \alpha_{k+1} x^+, x^- - \alpha_{k+1} x^{\star} \right)}{\alpha_{k+1} - \alpha_k}
 \, .
\label{eq:F-twistor2}
\end{equation}
We already see, that the two last arguments combine the coordinates in the Penrose-transform fashion \cite{Bateman:1904xu, Penrose:1969ae}. Indeed, the function $f$ looks like evaluated on a line in 3D complex space.

In order to actually formulate the above expression in the twistor space, we have to promote the function $f$ to a function on $\mathbb{CP}_3$ with a definite homogenity properties. To this end, we \emph{define}
\begin{equation}
 \mathfrak{f}_b \left( Z^1,Z^2,Z^3,Z^4 \right) =  \int\! \frac{dp^+ dp^{\star} }{2(2\pi)^2 p^+}\, 
  e^{ip^{\star} Z^3} e^{-ip^+ Z^4} \,
  \widetilde{\phi}_{b}\left( Z^1 Z^2 p^{\star},p^+,p^{\star}, Z^1 Z^2 p^+\right) \,,
  \label{eq:twistor_f_def}
\end{equation}
with 
\begin{equation}
 f_b(\alpha, u ,w ) =  \mathfrak{f}_b \left( 1, \alpha, u, w \right) \, ,
 \label{eq:twistor_f_dehomo}
\end{equation}
that is, the function $f$ is just a dehomogenization of $\mathfrak{f}$. Indeed, the function $\mathfrak{f}$ turns out to be \emph{homogenious of degree zero}.
For compactness, we will write
\begin{equation}
  \mathfrak{f}_b \left( Z^1,Z^2,Z^3,Z^4 \right) \equiv  \mathfrak{f}_b \left( Z^I \right) \,, \quad I=1,\dots,4 \, .
\end{equation}
The mapping of the fields to twistor functions proposed above may not be unique, but the precise form of the twistor function is not essential to our applications. What is important, is that we were able to map each field to a single twistor function operating on a copy of the twistor space. If not for this requirement, the functions could depend on the ratios $Z^1_{k}/Z^1_{l}$ without changing the homogenity and leading to the same $f$ upon dehomogenization. This possibility is actually used later to construct an alternative twistor space representation, which is useful in our geometric considerations.

\paragraph{Proof of the homogenity of $\mathfrak{f}(Z^I)$.} The proof is by a direct computation. We have
\begin{equation}
 \mathfrak{f}_b \left( \lambda Z^I  \right) =  \int\! \frac{dp^+ dp^{\star} }{2(2\pi)^2 p^+}\, 
  e^{ip^{\star} \lambda Z^3} e^{-ip^+ \lambda Z^4} \,
  \widetilde{\phi}_{b}\left( \lambda^2 Z^1 Z^2 p^{\star},p^+,p^{\star}, \lambda^2 Z^1 Z^2 p^+\right) \,,
\end{equation}
with $\lambda$ a complex number.
Changing variables $p^+\to\lambda^{-1}p^+$, $p^{\star}\to\lambda^{-1}p^{\star}$ in the Fourier transform we get
\begin{equation}
 \mathfrak{f}_b \left( \lambda Z^I \right) = \lambda^{-1} \int\! \frac{dp^+ dp^{\star} }{2(2\pi)^2 p^+}\, 
  e^{ip^{\star} Z^3} e^{-ip^+ Z^4} \,
  \widetilde{\phi}_{b}\left( \lambda Z^1 Z^2 p^{\star}, \lambda^{-1} p^+, \lambda^{-1} p^{\star}, \lambda Z^1 Z^2 p^+\right) \,.
\end{equation}
We recognize, that the scaling of the arguments of the field is a Lorentz transformation over the complexified Lorentz group: $(p^-,p^+,p^{\star},p^{\bullet})\to(\lambda p^-,\lambda^{-1}p^+,\lambda^{-1}p^{\star},\lambda p^{\bullet})$ is the transformation \eqref{eq:Lorentz_transf_x} with $\kappa=\bar\lambda$, $\gamma=\lambda$ (the components of $p$ transform like those of $x$, so that $x\cdot p$ is invariant); it is the combined scaling (iii) of Appendix~\ref{app:Lorentz}. The Fourier transform of the field $\tilde{\phi}$ has the same scaling as~\eqref{eq:phi_homo_star} (but notice the different order of the components):
\begin{equation}
  \widetilde{\phi}_{b}\left(\lambda Z^1 Z^2 p^{\star},\, \lambda^{-1} p^+,\, \lambda^{-1} p^{\star},\, \lambda Z^1 Z^2 p^+\right) = 
   \lambda\, \widetilde{\phi}_{b}\left(Z^1 Z^2 p^{\star},\,  p^+,\,  p^{\star},\, Z^1 Z^2 p^+\right)\,,
\end{equation}
 Thus
 \begin{equation}
 \mathfrak{f}_b \left( \lambda Z^I  \right) = 
 \mathfrak{f}_b \left( Z^I \right) 
  \,,
\end{equation}
as required.
\\

Due to~\eqref{eq:twistor_f_dehomo} we identify $\alpha = Z^2/Z^1$, $u=Z^3/Z^1$, $w=Z^4/Z^1$ and write the functional in terms of the homogenious function as
\begin{multline}
\mathcal{F}[\phi]\Big|^{\mathrm{on-shell}}_n (x) =     \,\Tr\left(t^{b_1} t^{b_2} \dots t^{b_n}\right) 
\\
\,\prod_{k \in \mathbb{Z}_n} 
 \int\! d\alpha_{k}
  \,\,  
  \frac{ \mathfrak{f}_{b_k}\left( Z_{k+1}^1 , Z^2_{k+1}, Z^1_{k+1} x^{\bullet} - Z^2_{k+1}  x^+, Z^1_{k+1} x^- - Z^2_{k+1}  x^{\star} \right)}{\alpha_{k+1} - \alpha_k}
 \, ,
\label{eq:F-twistor3}
\end{multline}
where
\begin{equation}
 d\alpha_k = d \left( \frac{Z_k^2}{Z_k^1} \right) = (Z^1_k)^{-2} \left( Z^1_k dZ^2_k - Z^2_k dZ^1_k \right) \,.
 \label{eq:alpha_measure_twistor}
\end{equation}
Introducing now the twistor $Z^I$ as 
\begin{equation}
 Z^I = \left( \pi_{\dot\alpha} , \omega^{\alpha} \right) \,,\qquad 
 \pi_{\dot\alpha} = \begin{pmatrix} Z^2 \\ Z^1\end{pmatrix}\,,\qquad 
 \omega^{\alpha} = \begin{pmatrix} -Z^3 \\ Z^4\end{pmatrix}\,,
 \label{eq:twistor_spinors}
\end{equation}
where the ordering of the components is chosen so that the incidence relation below takes its standard bispinor form.
We can write
\begin{equation}
 d\alpha_k = (Z^1_k)^{-2} \pi_{k,\dot\alpha} d\pi_k^{\dot\alpha} \,
 \label{eq:alpha_measure_twistor_pi}
\end{equation}
and 
\begin{equation}
 \prod_{k \in \mathbb{Z}_n} \frac{1}{\alpha_{k+1} - \alpha_k} = \prod_{k \in \mathbb{Z}_n} (Z^1_k)^{2} \frac{1}{\pi_{k,\dot\alpha} \pi_{k+1}^{\dot\alpha} } \,,
\end{equation}
where we have used the cyclicity. The \emph{incidence relations} simply follow from the arguments of $\mathfrak{f}$ in~\eqref{eq:F-twistor3}. With $x^{\alpha\dot\alpha}$ obtained from \eqref{eq:p_matrix_down} by raising both indices, we have
\begin{equation}
 \omega^{\alpha} = \frac{1}{\sqrt2}\,x^{\alpha\dot\alpha}\pi_{\dot\alpha}\,,\qquad
 \frac{1}{\sqrt2}\,x^{\alpha\dot\alpha}=\begin{pmatrix} x^{+} & -x^{\bullet} \\ -x^{\star} & x^{-} \end{pmatrix}\,,
 \qquad\text{i.e.}
 \label{eq:incidence}
\end{equation}
\begin{equation}
\begin{pmatrix} 
Z^3 \\ Z^4 
\end{pmatrix} 
= 
\begin{pmatrix} 
x^{\bullet} & -x^+ \\ 
x^- & -x^{\star} 
\end{pmatrix} 
\begin{pmatrix} 
Z^1 \\ Z^2
\end{pmatrix} \, .
\end{equation}
Since each $\mathfrak{f}_{b_k}$ is evaluated on the incidence relation of $x$, every factor is integrated over the line $L_x$ of Section~\ref{sub:twistor_space}. Denoting by $\Gamma_k\subset L_x$ the contour of real slopes $\alpha_k\in\mathbb{R}$ of the $k$-th factor (see also the end of Section~\ref{sub:twistor_geometry}), we can finally write
\begin{equation}
\mathcal{F}[\phi]\Big|^{\mathrm{on-shell}}_n (x) =  \,\Tr\left(t^{b_1} t^{b_2} \dots t^{b_n}\right) \,\prod_{k \in \mathbb{Z}_n}  \int_{\Gamma_k} \!  \pi_{k,\dot\alpha} d\pi_k^{\dot\alpha}  
  \,\,  
  \frac{ \mathfrak{f}_{b_k}\left( Z^I_{k+1}\right) }{\pi_{k,\dot\alpha} \pi_{k+1}^{\dot\alpha} }
 \, ,
\label{eq:F-twistor4}
\end{equation}
where the projective weights $Z_k^1$ perfectly cancelled between the measure and the cyclic denominator. Since the function $\mathfrak{f}$ is of homogenity zero, the expression is projectively well defined.

What we have achieved is a twistor space representation of the $n$-th term in the expansion of the functional (here at fixed $x$). We saw that keeping the integrals over the directions of the $\varepsilon_{\alpha}$, $d\alpha_k$, and integrating the "bullet" momentum components instead, introduces the incidence relations, i.e. emergent twistors.

Before closing this section, let us discuss a few additional points.

\paragraph{Momentum vs twistor representations.} Although the twistor and momentum space representations of the functional look structurally very similar, the products of $\pi_{k,\dot\alpha}$ spinors are not the spinor brackets of~\eqref{eq:Fsd_final}. Further, in the latter we have external fields with well defined Lorentz properties (there corresponding to spin -1), here we have twistor homogenious functions. Finally, \eqref{eq:F-twistor4} is the integrand at fixed $x$: the integration over $x$, which in momentum space produced the conservation delta, has not been performed. In twistor space it becomes an integral over the position of the line $L_x$, and momentum conservation is encoded in the way the incidence relations change under a translation of $x$.

\paragraph{Subtlety regarding the space-time derivatives.} Our functional contains the $\partial_-$ derivatives acting on each space-time field to assure the boost invariance. When building the twistor function~\eqref{eq:twistor_f_def} we have already acted with the derivatives on the Fourier transforms of the fields to obtain the $p_k^+$ factors, which in turn got cancelled with the phase space factors (the factors $1/p_k^+$ in~\eqref{eq:twistor_f_def} come from the $\tau$ integration). 
In order to promote the functional to the twistor space we had to impose the projective character of the field arguments -- doing so made the $x^-$ coordinate multiplied by the $Z_k^1$ twistor coordinate, see~\eqref{eq:F-twistor3}. Since the twistor functions depend on $x$ through the incidence relations one expects that we can formulate the functional with the derivatives acting on some other version of the twistor functions. The difficulty arises because now the functions have projective character -- in contrast to the considerations from the beginning of this subsection,  where we executed derivatives on dehomogenized functions. In what follows we will demonstrate, that an equivalent functional, with explicit derivatives, can be written as
\begin{multline}
\mathcal{F}[\phi]\Big|^{\mathrm{on-shell}}_n (x) =  \,\Tr\left(t^{b_1} t^{b_2} \dots t^{b_n}\right) \,\prod_{k \in \mathbb{Z}_n}  \int_{\Gamma_k} \!  \pi_{k,\dot\alpha} d\pi_k^{\dot\alpha} 
 \int_{-\infty}^0\! d\tau_k  \\
  \,\,  
   \partial_-\mathfrak{g}_{b_k}\left( Z^1_{k+1}, Z^2_{k+1}, Z^1_{k+1} x^{\bullet} - Z^2_{k+1}  x^+, Z^1_{k+1} x^- - Z^2_{k+1}  x^{\star} + \tau_k\left(Z_{k+1}^2-\frac{Z_{k+1}^1}{Z_k^1}\,Z_{k}^2\right) \right) 
 \, ,
\label{eq:F-twistor-alt1}
\end{multline}
where $\mathfrak{g}$ is defined as
\begin{equation}
 \mathfrak{g}_b \left( Z^I \right) =  \int\! \frac{dp^+ dp^{\star} }{2(2\pi)^2 p^+}\, \left(\frac{1}{Z^1}\right)^2 
  e^{ip^{\star} Z^3} e^{-ip^+ Z^4} \,
  \widetilde{\phi}_{b}\left( Z^1 Z^2 p^{\star},p^+,p^{\star}, Z^1 Z^2 p^+\right) = \frac{\mathfrak{f}_b(Z^I)}{(Z^1)^2}\,
  \label{eq:twistor_g_def}
\end{equation}
and has homogenity (-2) as is easily verified. To check~\eqref{eq:F-twistor-alt1} against~\eqref{eq:F-twistor4}, notice first that $\partial_-$ on $k$-th function gives $Z_{k+1}^1 p_k^+$. The $\tau_k$ integral, on the other hand, give $Z_k^1/\pi_{k,\dot\alpha}\pi_{k+1}^{\dot\alpha}p_k^+$. Due to cyclic nature of the product we thus have $(Z_1^1 Z_2^1 \dots Z_n^1)^2$ in the numerator that is exactly canceled by the $(Z_k^1)^2$ factors in~\eqref{eq:twistor_g_def}. The integrals over $dp^{\star}_k dp_k^+$ can be put back into $\mathfrak{f}$ form, retrieving~~\eqref{eq:F-twistor4}.

The formula with explicit derivatives~\eqref{eq:F-twistor-alt1} will be extremely useful in  Section~\ref{sec:amplitudes}, where we will generalize the functional to different spins.


\subsection{Geometry of the functional in twistor space}
\label{sub:twistor_geometry}
 
The alternative twistor space representation~\eqref{eq:F-twistor-alt1} can be used to reexamine the geometry of the functional. 
The result of this subsection is the following. At fixed $x$ the integrand of the $n$-th term is supported on a single plane $\Pi_x\subset\mathbb{CP}^3$. The $k$-th factor of the integrand is the twistor function $\mathfrak g_{b_k}$ evaluated at a twistor $Z_{k+1}(\tau_k)$, and the integral over $\tau_k$ is an integral along a straight line of $\Pi_x$ through a fixed point $Z_*$. The Parke--Taylor factor $1/(\pi_{k,\dot\alpha}\pi_{k+1}^{\dot\alpha})$ is the boundary term of this integral at the endpoint $\tau_k=0$, which is a point of the line $L_x$. The integration paths of consecutive factors are connected at points of $L_x$ and together form a closed polygon in $\Pi_x$ with alternate vertices on $L_x$. We establish this in four steps. First we show that the fourth argument of $\mathfrak g_{b_k}$ in \eqref{eq:F-twistor-alt1} is the incidence relation of a point displaced from $x$ inside the ASD plane. Second we identify the plane $\Pi_x$ and the point $Z_*$ and describe the integration path of a single factor. Third we perform the $\tau_k$ integral and obtain the Parke--Taylor factor as a boundary term. Fourth we combine the $n$ factors into the polygon. Throughout we use the twistor $Z^I=(\pi_{\dot\alpha},\omega^{\alpha})$, the coordinates $\alpha=Z^2/Z^1$, $u=Z^3/Z^1$, $w=Z^4/Z^1$ and the convention of \eqref{eq:F-twistor4} that the twistor function $\mathfrak g_{b_k}$ is evaluated at the twistor $Z_{k+1}$.

\paragraph{
The function $\mathfrak{g}$ localises on incidence relations
\label{subsec:g_incidence}}
Let us notice that the fourth argument of $\mathfrak{g}_{b_k}$ in \eqref{eq:F-twistor-alt1} (corresponding to general $Z^4$ twistor components) is not part of the usual incidence relation of the point $x$. In this paragraph we show that it is the incidence relation of another point.

We recall the standard facts \cite{Penrose:1985bww}, see Section~\ref{sub:twistor_space}: a point $\xi$ of complexified Minkowski space is represented by the line $L_\xi\subset\mathbb{CP}^3$ of twistors satisfying its incidence relation; a null line in spacetime is represented by a single point of $\mathbb{CP}^3$, the point common to the lines $L_\xi$ of all its points $\xi$  (the twistor of the $\alpha$-plane containing the null line); two lines $L_\xi$, $L_{\xi'}$ intersect if and only if $\xi$ and $\xi'$ are null separated.

The argument of $\mathfrak{g}_{b_k}$ in \eqref{eq:F-twistor-alt1} is
\begin{align}
 Z_{k+1}(\tau_k)&=\big(Z^1_{k+1},\,Z^2_{k+1},\,Z^1_{k+1}x^{\bullet}-Z^2_{k+1}x^{+},\,Z^1_{k+1}x^{-}-Z^2_{k+1}x^{\star}+\tau_k\,c_k\big)\,,
 \label{eq:Zshift}\\
 c_k&\equiv Z^2_{k+1}-\frac{Z^1_{k+1}}{Z^1_k}\,Z^2_k=\frac{\pi_{k,\dot\alpha}\pi_{k+1}^{\dot\alpha}}{Z^1_k}\,.
 \label{eq:ck}
\end{align}
In the chart $Z^1_{k+1}=1$ one has $c_k=\alpha_{k+1}-\alpha_k$, and the fourth argument in \eqref{eq:Zshift} rearranges as
\begin{equation}
 x^{-}-\alpha_{k+1}x^{\star}+\tau_k\,(\alpha_{k+1}-\alpha_k)
 =\big(x^{-}-\tau_k\alpha_k\big)-\alpha_{k+1}\big(x^{\star}-\tau_k\big)
 =\xi_k^{-}-\alpha_{k+1}\,\xi_k^{\star}\,,
 \label{eq:fourth_slot}
\end{equation}
with, by \eqref{eq:epsilon_vec_SD},
\begin{equation}
 \vec{\xi}_k=\vec{x}+\tau_k\,\vec{\varepsilon}^{\,+}_{\alpha_k}\,,
 \qquad\text{or}\qquad
 \xi^{\mu}_k=x^{\mu}+\tau_k\,\varepsilon^{+\mu}_{\alpha_k}\,.
 \label{eq:xik}
\end{equation}
The third argument in \eqref{eq:Zshift}, which involves only $x^{+}$ and $x^{\bullet}$, is untouched because $\varepsilon^{+}_{\alpha}$ has neither a plus nor a bullet component. Hence $\mathfrak{g}_{b_k}$ is evaluated on the incidence relation of $\xi_k$,
\begin{equation}
 \begin{pmatrix} Z^3_{k+1}\\ Z^4_{k+1}\end{pmatrix}
 =\begin{pmatrix} \xi_k^{\bullet} & -\xi_k^{+}\\ \xi_k^{-} & -\xi_k^{\star}\end{pmatrix}
 \begin{pmatrix} Z^1_{k+1}\\ Z^2_{k+1}\end{pmatrix}\,, \quad \mathrm{where}\quad \xi_k^{\bullet} = x^\bullet;\,\, \xi_k^{+} = x^+\,,
 \label{eq:incidence_xik}
\end{equation}
that is, on the line $L_{\xi_k}$ of the displaced point $\xi_k$. The parameter $\tau_k$ moves the point at which the field $\phi_{b_k}$ is evaluated from $x$ to $\xi_k$ along the null line $x+s\,\varepsilon^{+}_{\alpha_k}$. 
In twistor space it moves the line on which the twistor of the field sits from $L_x$ to $L_{\xi_k}$ replaces the line $L_x$, on which the argument of the twistor function $\mathfrak g_{b_k}$ lies at $\tau_k=0$, by the line $L_{\xi_k}$.
Two more facts will be used below. Since $\xi_k-x$ is a null vector the lines $L_{\xi_k}$ and $L_x$ meet in one point. And $Z_k$ enters \eqref{eq:Zshift} only through the slope $\alpha_k=Z^2_k/Z^1_k$ of $\varepsilon^{+}_{\alpha_k}$ while its components $Z^3_k$ and $Z^4_k$ do not enter.
  
\paragraph{The plane $\Pi_x$ and the point $Z_*$}
\label{subsec:zstar}
 
Homogeneously, \eqref{eq:Zshift} reads
\begin{equation}
 Z_{k+1}(\tau_k)=Z_{k+1}(0)+\tau_k\,c_k\,Z_*\,,\qquad Z_*\equiv(0,0,0,1)\,,
 \label{eq:Zshift_hom}
\end{equation}
where $Z_{k+1}(0)$ is the twistor space point corresponding to $L_x$ with dotted spinor $\pi_{k+1}$. The twistor $Z_*$ has $\pi_*=0$ and $\omega^{\alpha}_*=(0,1)^{T}$, the undotted spinor of $\eta$ with its index raised; it is the point of the line at infinity $\{\pi_{\dot\alpha}=0\}$ associated with $\eta$, and the reference twistor of the light-cone gauge. Its role follows from the incidence relation alone: for a displacement $\delta \xi$ inside $\Pi^{+}(x)$, i.e. $\delta \xi^{+}=\delta \xi^{\bullet}=0$, the third argument in \eqref{eq:Zshift} does not change, while
\begin{equation}
 \delta Z^4=Z^1\,\delta \xi^{-}-Z^2\,\delta \xi^{\star}\,.
 \label{eq:deltaZ4}
\end{equation}
Every direction of the ASD plane moves a twistor along the same direction $Z_*$, with a coefficient that is the antisymmetric contraction of the dotted spinor of $\delta \xi$ with $\pi$ (all directions of $\Pi^{+}(x)$ share the undotted spinor of $\eta$). 

Every displacement inside $\Pi^{+}(x)$ changes only the fourth component of a twistor and it does so along $Z_*$. Hence the lines $L_\xi$ of all points $\xi\in\Pi^{+}(x)$ lie in one plane of twistor space. It is the plane spanned by the line $L_x$ and the point $Z_*$,
\begin{equation}
 \Pi_x\equiv\big\{Z\in\mathbb{CP}^3:\;W_x\cdot Z=0\big\}\simeq\mathbb{CP}^2\,,
 \qquad W_x\equiv(x^{\bullet},\,-x^{+},\,-1,\,0)\,,
 \label{eq:Pix}
\end{equation}
which we also denote by $\langle L_x,Z_*\rangle$. The dual twistor $W_x$ is the one that represents the ASD plane $\Pi^+(x)$ in the sense of Section~\ref{sub:twistor_space}. On $\Pi_x$ the coordinate $u=x^{\bullet}-\alpha x^{+}$ is fixed by $\alpha$. Hence $(\alpha,w)$ is a chart on $\Pi_x$ and all the twistors $Z_{k+1}(\tau_k)$ of the $n$-th term at fixed $x$ lie in this plane. In this chart $L_x$ is the line $w=x^{-}-\alpha x^{\star}$ and the lines through $Z_*$ are the vertical lines $\alpha=\mathrm{const}$.

We now consider a single factor of the product in \eqref{eq:F-twistor-alt1}. Fix $k$ and consider the twistor function $\mathfrak{g}_{b_k}$, which is evaluated at the twistor $Z_{k+1}(\tau_k)$ of \eqref{eq:Zshift}; the space-time field $\phi_{b_k}$ enters only through the Fourier transform in the definition \eqref{eq:twistor_g_def} of this function.
Two points of the line $L_x$ play a role. The first one is
\begin{equation}
 P_k\equiv Z_k(0)=\big(Z^1_k,\;Z^2_k,\;Z^1_kx^{\bullet}-Z^2_kx^{+},\;Z^1_kx^{-}-Z^2_kx^{\star}\big)\in L_x\,,
 \label{eq:Pk}
\end{equation}
the point of $L_x$ with slope $\alpha_k$. By the facts recalled above it represents the null line $x+s\,\varepsilon^{+}_{\alpha_k}$ through $x$. The second one is $P_{k+1}$, the point of $L_x$ with slope $\alpha_{k+1}$.

The displaced point $\xi_k=x+\tau_k\varepsilon^{+}_{\alpha_k}$ lies on the null line $x+s\,\varepsilon^{+}_{\alpha_k}$ for every value of $\tau_k$. Therefore the line $L_{\xi_k}$ passes through $P_k$ for every value of $\tau_k$. One can see this directly in the chart: the line $L_{\xi_k}$ is given by $w=x^{-}-\alpha x^{\star}+\tau_k(\alpha-\alpha_k)$ and at $\alpha=\alpha_k$ the dependence on $\tau_k$ drops out. So the only thing that happens to $L_{\xi_k}$ when $\tau_k$ changes is a rotation about the fixed point $P_k$. At $\tau_k=0$ the line $L_{\xi_k}$ coincides with $L_x$. As $\tau_k$ decreases to $-\infty$ it turns towards the vertical line $\alpha=\alpha_k$ through $P_k$ and $Z_*$.

The twistor function $\mathfrak g_{b_k}$ is evaluated at the point of $L_{\xi_k}$ with slope $\alpha_{k+1}$, i.e. at $Z_{k+1}(\tau_k)$. At $\tau_k=0$ this point is $P_{k+1}$ on $L_x$.
When $\tau_k$ decreases the point leaves $L_x$ and moves along the vertical line $\alpha=\alpha_{k+1}$ through $P_{k+1}$ and $Z_*$. Its coordinate is $w=x^{-}-\alpha_{k+1}x^{\star}+\tau_k(\alpha_{k+1}-\alpha_k)$ and it moves towards $Z_*$ in the direction fixed by the sign of $\alpha_{k+1}-\alpha_k$. 
The integral over $\tau_k$ in \eqref{eq:F-twistor-alt1} is therefore an integral of the twistor function $\partial_-\mathfrak g_{b_k}$ along the vertical segment from $P_{k+1}$ towards $Z_*$. It is evaluated in the next paragraph.

This picture respects the projective nature of the twistor variables. The coefficient $c_k$ in \eqref{eq:Zshift} is linear in $Z_{k+1}$ and depends on $Z_k$ only through the ratio $\alpha_k=Z^2_k/Z^1_k$. Hence $Z_{k+1}(\tau_k)\to\lambda Z_{k+1}(\tau_k)$ under $Z_{k+1}\to\lambda Z_{k+1}$ and $Z_{k+1}(\tau_k)$ does not change under $Z_k\to\mu Z_k$. The shift by $\tau_k c_k Z_*$ preserves the homogeneity of the integrand in both twistors and \eqref{eq:F-twistor-alt1} is well defined on projective twistor space.
 
\paragraph{The $\tau$-integral.} We evaluate the integral along the vertical segment found above without using the explicit relation between the twistor functions and the fields.
Since $x^{-}$ enters the incidence relation only through $Z^4=Z^1x^{-}-Z^2x^{\star}$,
\begin{equation}
 \partial_-\,\mathfrak{g}_{b_k}\big(Z_{k+1}(\tau_k)\big)\equiv\frac{\partial}{\partial x^{-}}\,\mathfrak{g}_{b_k}\big(Z_{k+1}(\tau_k)\big)
 =Z^1_{k+1}\,\frac{\partial\mathfrak{g}_{b_k}}{\partial Z^4}\Big|_{Z_{k+1}(\tau_k)}\,.
 \label{eq:dminus_twistor}
\end{equation}
By \eqref{eq:Zshift_hom} and \eqref{eq:dminus_twistor},
\begin{align}
 \frac{d}{d\tau_k}\,\mathfrak{g}_{b_k}\big(Z_{k+1}(\tau_k)\big)&=c_k\,\frac{\partial\mathfrak{g}_{b_k}}{\partial Z^4}\Big|_{Z_{k+1}(\tau_k)}\,,\nonumber\\
 \partial_-\mathfrak{g}_{b_k}\big(Z_{k+1}(\tau_k)\big)&=\frac{Z^1_{k+1}}{c_k}\,\frac{d}{d\tau_k}\,\mathfrak{g}_{b_k}\big(Z_{k+1}(\tau_k)\big)\,.
 \label{eq:chain}
\end{align}
Therefore, the integrand of \eqref{eq:F-twistor-alt1} is a total derivative:
\begin{equation}
 \int_{-\infty}^{0}d\tau_k\;\partial_-\mathfrak{g}_{b_k}\big(Z_{k+1}(\tau_k)\big)
 =\frac{Z^1_kZ^1_{k+1}}{\pi_{k,\dot\alpha}\pi_{k+1}^{\dot\alpha}}
 \big[\mathfrak{g}_{b_k}\big(Z_{k+1}(0)\big)-\lim_{\tau_k\to-\infty}\mathfrak{g}_{b_k}\big(Z_{k+1}(\tau_k)\big)\big]\,.
 \label{eq:FTC}
\end{equation}
 The prefactor $Z^1_{k+1}/c_k=1/(\alpha_{k+1}-\alpha_k)$ diverges when $\alpha_{k+1}=\alpha_k$, i.e. when the line along which $\phi_{b_k}$ is integrated is parallel to the line $x+s\,\varepsilon^{+}_{\alpha_k}$ along which $\xi_k$ moves; then, by \eqref{eq:fourth_slot}, the argument of $\mathfrak{g}_{b_k}$ does not depend on $\tau_k$ at all.
The boundary term is the field at $\tau_k\to-\infty$, where the twistor has reached $Z_*$; in momentum space it is $e^{-ip_k^{+}\tau_kc_k}$ at $\tau_k\to-\infty$, and the $i\epsilon$ prescription of \eqref{eq:halfline} is the statement that it vanishes. 
With $\mathfrak{g}_{b_k}\big(P_{k+1}\big)=(Z^1_{k+1})^{-2}\mathfrak{f}_{b_k}\big(P_{k+1}\big)$ the first term above is
\begin{equation}
 \frac{Z^1_kZ^1_{k+1}}{\pi_{k,\dot\alpha}\pi_{k+1}^{\dot\alpha}}\big(Z^1_{k+1}\big)^{-2}\mathfrak{f}_{b_k}\big(P_{k+1}\big)
 =\frac{Z^1_k}{Z^1_{k+1}}\,\frac{\mathfrak{f}_{b_k}\big(P_{k+1}\big)}{\pi_{k,\dot\alpha}\pi_{k+1}^{\dot\alpha}}\,.
 \label{eq:leftover}
\end{equation}
Because $\prod_kZ^1_k/Z^1_{k+1}=1$  \eqref{eq:F-twistor-alt1} reduces to \eqref{eq:F-twistor4} with the same prefactor, as stated. 
In words, the factor $1/(\pi_{k,\dot\alpha}\pi_{k+1}^{\dot\alpha})$ of \eqref{eq:F-twistor4}, i.e. the $k$-th factor of the Parke--Taylor denominator, arises as the boundary term of the integral of $\partial_-\mathfrak g_{b_k}$ along the vertical line through $Z_*$, evaluated at the endpoint $\tau_k=0$ of the integration path, which is the point $P_{k+1}$ where this line meets $L_x$.
Before the $\tau_k$-integration the integrand has no poles; the contour of the $\pi_k$-integrals in \eqref{eq:F-twistor-alt1} is therefore the real slope $\alpha_k\in\mathbb{R}$, the real slice $\Gamma_k\simeq\mathbb{RP}^1$ of $L_x$, and the prescription for the Parke-Taylor poles is inherited from the convergence of the $\tau_k$-integral.

\paragraph{The full functional as a closed polygon.} 
Since $Z_k(\tau_{k-1})$ lies on the line through $Z_*$ at $\alpha=\alpha_k$, $P_k$ is its projection from $Z_*$ onto $L_x$. The line $L_{\xi_k}$ thus joins $P_k$, the projection of the twistor of leg $k-1$, to $Z_{k+1}(\tau_k)$, the twistor of leg $k$, which in turn projects to $P_{k+1}$. Going once around the colour trace therefore closes a polygon,
\begin{equation}
 P_1\;\to\;Z_2\;\to\;P_2\;\to\;Z_3\;\to\;\cdots\;\to\;P_n\;\to\;Z_1\;\to\;P_1\,,
 \qquad Z_{k+1}\equiv Z_{k+1}(\tau_k)\,,
 \label{eq:polygon}
\end{equation}
whose edges $P_k\to Z_{k+1}$ lie on the lines $L_{\xi_k}$ and whose edges $Z_{k+1}\to P_{k+1}$ lie on the lines through $Z_*$. The $n$-th term of the functional at fixed $x$ is an integral over closed $2n$-gons in the plane $\Pi_x$ with vertices alternately on $L_x$ and on the lines through $Z_*$; $\partial_-$ differentiates and $\int d\tau_k$ integrates along the even edges, and the Parke-Taylor denominators are what these two operations leave behind.  At $\tau_k=0$ the polygon collapses onto $L_x$, which is \eqref{eq:F-twistor4}: $n$ points on one line weighted by the cyclic product of their brackets. In the twistor space of \cite{Witten:2003nn} such a line supports the MHV amplitude; our twistor space is the conjugate one and the line $L_x$ supports the $\overline{\rm MHV}$ amplitude. This completes the picture announced at the beginning of the subsection.
 
\begin{figure}[t]
 \centering
 \includegraphics[width=\textwidth]{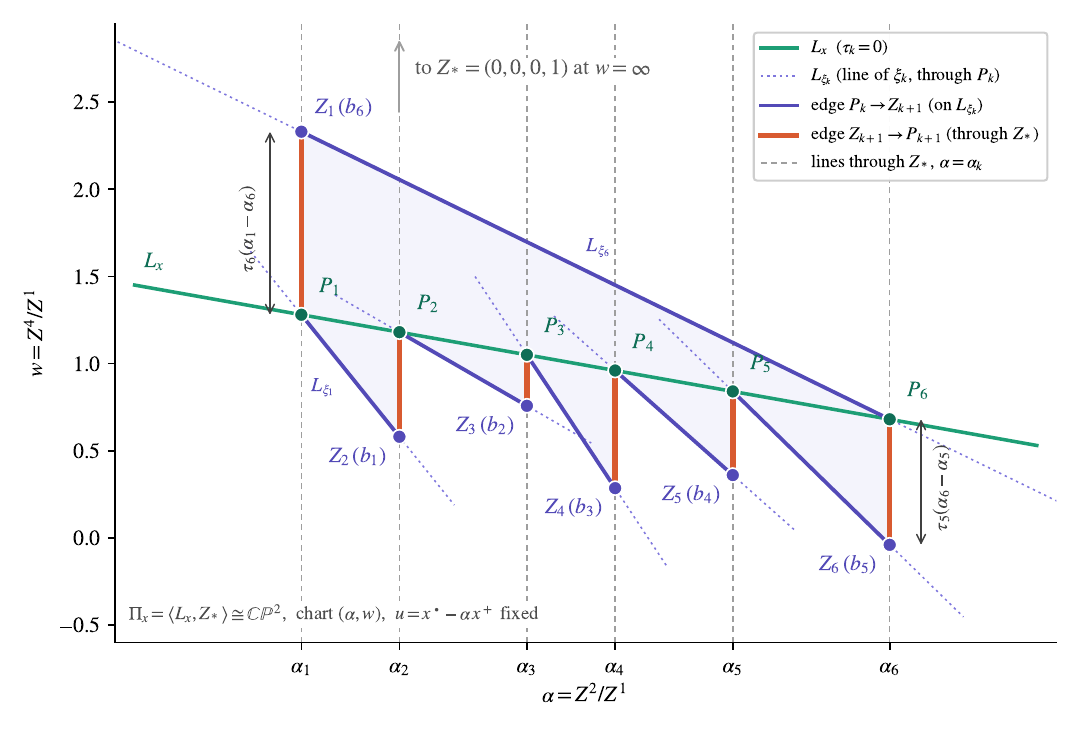}
 \caption{The six-point configuration with $x^{-}=1$, $x^{\star}=0.2$, $\alpha_k=(-1.4,\,-0.9,\,-0.25,\,0.2,\,0.8,\,1.6)$, $
 \tau_k=(-1.2,\,-0.45,\,-1.5,\,-0.8,\,-0.9,\,-0.35)$, in the plane $\Pi_x=\langle L_x,Z_*\rangle$, chart $(\alpha,w)$. Teal: $L_x$ with the points $P_k$ at $\alpha=\alpha_k$, representing the lines $x+s\,\varepsilon^{+}_{\alpha_k}$. Dashed: the lines through $Z_*$, the point at infinity of the $w$-axis. Dotted purple: the lines $L_{\xi_k}$ of the points $\xi_k=x+\tau_k\varepsilon^{+}_{\alpha_k}$, each through $P_k$. Purple dots: the twistors $Z_{k+1}(\tau_k)$ carrying $\phi_{b_k}$, where $L_{\xi_k}$ meets the vertical at $\alpha_{k+1}$. The closed $12$-gon $P_1\to Z_2\to P_2\to\cdots\to P_6\to Z_1\to P_1$ has its odd edges (solid purple) on the lines $L_{\xi_k}$ and its even edges (thick orange) on the lines through $Z_*$, the segments swept by the $\tau_k$-integrals. Five vertices lie below $L_x$ and $Z_1$ above it, since $\alpha_1-\alpha_6<0$. At $\tau_k=0$ the polygon collapses onto $L_x$.}
 \label{fig:polygon6}
\end{figure}

\begin{figure}
 \centering
 \includegraphics[width=0.8\textwidth]{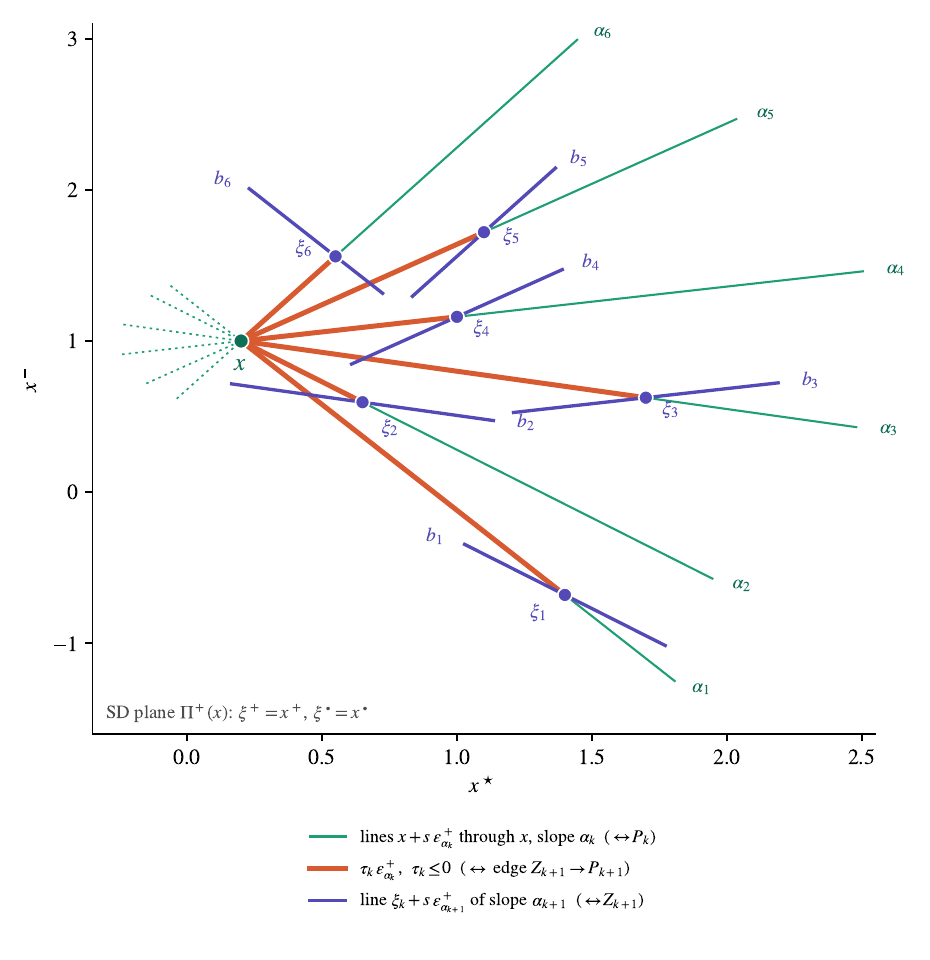}
 \caption{The same configuration on the ASD plane $\Pi^{+}(x)$ in the coordinates $(x^{\star},x^{-})$, in which $\xi+s\,\varepsilon^{+}_{\alpha}$ is the line of slope $\alpha$ through $\xi$. Teal: the lines $x+s\,\varepsilon^{+}_{\alpha_k}$ through $x$ (solid: $\tau\le0$). Orange: the displacements $\tau_k\varepsilon^{+}_{\alpha_k}$ to the points $\xi_k$ (purple dots). Purple: the lines $\xi_k+s\,\varepsilon^{+}_{\alpha_{k+1}}$ along which $\phi_{b_k}$ is integrated. Under the duality of Sec.~\ref{sec:dictionary}, each line through $x$ is a point $P_k$ of Fig.~\ref{fig:polygon6}, each point $\xi_k$ is a line $L_{\xi_k}$, and each purple line is a twistor $Z_{k+1}$.}
 \label{fig:sdplane6}
\end{figure}

Two remarks. With $\tau_k\le0$ the vertex $Z_{k+1}$ lies below $L_x$ (decreasing $w$) when $\alpha_{k+1}>\alpha_k$ and above it when $\alpha_{k+1}<\alpha_k$. Since the differences $\alpha_{k+1}-\alpha_k$ sum to zero around the cycle, both cases occur for every configuration of slopes, so that the polygon has vertices on both sides of $L_x$. Nothing in the construction orders the slopes: they are independent integration variables, and for a generic configuration the polygon is self-intersecting. It is free of self-intersections only for special orderings, for instance when the slopes increase monotonically along the trace except for the single step that closes the cycle, and the configurations of Figures~\ref{fig:polygon6} and \ref{fig:sdplane6} were chosen in this way for clarity only. The self-intersections have no consequence, because the integrals over the $\tau_k$ are taken independently for each factor.

For $n=2$ both coefficients $c_1$ and $c_2$ vanish on the same locus $\alpha_1=\alpha_2$, the points $P_1$ and $P_2$ coincide, and the two simple poles combine into the non-integrable double pole noted after \eqref{eq:F_funct_tr}; for $n\ge3$ the poles sit at distinct loci and are simple.

Figures~\ref{fig:polygon6} and \ref{fig:sdplane6} show the example configuration for $n=6$ drawn in the plane $\Pi_x$ in the chart $(\alpha,w)$; the coordinate $u=x^{\bullet}-\alpha x^{+}$ is not shown. Figure~\ref{fig:sdplane6} shows the same data on the ASD plane $\Pi^{+}(x)$ in the coordinates $(x^{\star},x^{-})$, in which, by \eqref{eq:epsilon_vec_SD}, the line $\xi+s\,\varepsilon^{+}_{\alpha}$ through any point $\xi$ has slope~$\alpha$.
 
\paragraph{A dictionary between the fields on ASD plane and the twistor plane.}\label{sec:dictionary}
The two pictures are related by projective duality: $\Pi_x$ is the dual of the compactified ASD plane $\Pi^{+}(x)$, points of one being lines of the other. In the chart $(\xi^{\star},\xi^{-})$ on $\Pi^{+}(x)$ and $(\alpha,w)$ on $\Pi_x$, the incidence relation \eqref{eq:incidence_xik} assigns to the point $\vec{\xi}$ the line
\begin{equation}
 L_\xi:\qquad w=\xi^{-}-\alpha\,\xi^{\star}\,,
 \label{eq:Lxi_line}
\end{equation}
and to the line of slope $\alpha_0$ through $\vec{\xi}$, i.e. $\{\xi':\;\xi'^{-}-\xi^{-}=\alpha_0(\xi'^{\star}-\xi^{\star})\}$, the point of $L_\xi$ at $\alpha=\alpha_0$: for $\xi'$ on that line, $\xi'^{-}-\alpha \xi'^{\star}=\xi^{-}-\alpha \xi^{\star}+(\alpha_0-\alpha)(\xi'^{\star}-\xi^{\star})$ is independent of $\xi'$ exactly at $\alpha=\alpha_0$. This is the Penrose correspondence between null lines and points restricted to $\Pi^{+}(x)$. Since $L_\xi$ becomes vertical when $\xi$ runs off to infinity along the direction of slope $\alpha_0$, the point at infinity of $\Pi^{+}(x)$ in that direction is represented by the line $\alpha=\alpha_0$ through $Z_*$, and the line at infinity of $\Pi^{+}(x)$ by the point $Z_*$ itself: this is why every displacement inside $\Pi^{+}(x)$ is a motion towards $Z_*$, eq.~\eqref{eq:deltaZ4}, and why $\eta$ and $\varepsilon^{+}_{\alpha_k}$, different directions in spacetime, act in twistor space along the same line and differ only by their coefficients. Table~\ref{tab:dictionary} collects the correspondences.

\begin{table}[t]
 \centering
 \renewcommand{\arraystretch}{1.25}
 \begin{tabular}{>{\raggedright\arraybackslash}p{0.46\textwidth} >{\raggedright\arraybackslash}p{0.48\textwidth}}
 \toprule
 \textbf{ASD plane $\Pi^{+}(x)$}, coordinates $(x^{\star},x^{-})$ & \textbf{twistor plane $\Pi_x=\langle L_x,Z_*\rangle$}, chart $(\alpha,w)$\\
 \midrule
 \multicolumn{2}{l}{\emph{Points and lines}}\\
 a point $\xi$ & the line $L_\xi$: $w=\xi^{-}-\alpha\,\xi^{\star}$\\
 the line of slope $\alpha_0$ through $\xi$ & the point of $L_\xi$ at $\alpha=\alpha_0$\\
 the line $x+s\,\varepsilon^{+}_{\alpha_k}$ through $x$ & the point $P_k$ of $L_x$\\
 moving $\xi$ along the line $x+s\,\varepsilon^{+}_{\alpha_k}$ & rotating $L_\xi$ about $P_k$\\
 all parallel lines of slope $\alpha_0$ (their common point at infinity) & the vertical line $\alpha=\alpha_0$ through $Z_*$\\
 the line at infinity of $\Pi^{+}(x)$ & the point $Z_*$\\
 \midrule
 \multicolumn{2}{l}{\emph{The reference direction $\eta$}}\\
 the direction $\eta$, i.e. slope $\alpha_0=\infty$ & the point $Z^1=0$ of $L_x$; it is the point excluded by the chart $\alpha=Z^2/Z^1$ and the point at which the factor $(Z^1)^{-2}$ of \eqref{eq:twistor_g_def} is singular\\
 the SD plane $\Pi^{-}(x)$ & the point $\alpha=\infty$ of $L_x$, i.e. the point of the previous row: $\Pi^{-}(x)$ is the $\alpha$-plane of the twistor on $L_x$ whose dotted spinor is that of $\eta$\\
 the derivative $\partial_-=\eta\cdot\partial$ & $Z^1\,\partial/\partial Z^4$, the derivative towards $Z_*$\\
 \midrule
 \multicolumn{2}{l}{\emph{The integrations in \eqref{eq:F-twistor-alt1}}}\\
 $\int ds_{k+1}$ along the line $\xi_k+s\,\varepsilon^{+}_{\alpha_{k+1}}$ & already absorbed: the whole line is the single point $Z_{k+1}$\\
 $\int_{-\infty}^{0}d\tau_k$ along $\varepsilon^{+}_{\alpha_k}$ & the integral along the vertical line $\alpha=\alpha_{k+1}$ from $Z_*$ ($\tau_k=-\infty$) to $P_{k+1}$ ($\tau_k=0$)\\
 real slopes $\alpha_k\in\mathbb{R}$ & the real contour $\Gamma_k\simeq\mathbb{RP}^1$ of $L_x$\\
 $\alpha_{k+1}=\alpha_k$: the two lines of the field $k$ coincide & $c_k=0$: the Parke-Taylor pole\\
 \bottomrule
 \end{tabular}
 \caption{Dictionary between the ASD plane $\Pi^{+}(x)$ and the twistor plane $\Pi_x$. The two planes are projectively dual: a point of one corresponds to a line of the other.}
 \label{tab:dictionary}
\end{table}

The vector $\eta$ enters twice: its undotted spinor gives the point $Z_*$ towards which all motions inside $\Pi^{+}(x)$ are directed, and its direction gives the point $\alpha=\infty$ of every line $L_x$, where the weight factor $(Z^1)^{-2}$ of $\mathfrak{g}$ has its double zero. The non-locality of the functional in spacetime, the integral over $\tau_k$, becomes an integral along the line joining $Z_{k+1}$ to $Z_*$, which is the form of the propagator in the axial gauge on twistor space, often called the CSW gauge, in which the component of the twistor-space connection along the lines through the reference twistor $Z_*$ is set to zero; it is the twistor-space counterpart of the light-cone gauge, with $Z_*$ in the role of the vector $\eta$.


\section{Scattering amplitudes}
\label{sec:amplitudes}

\subsection{General idea}

So far we have discussed only the functional which generates the Parke-Taylor cyclic denominator. In this Section we will explore  a simple modification, which will allow to generate a large class of scattering amplitudes in an entirely new way.

Intuitively, we want to generate various numerators, that ultimately should be expressible in terms of spinor products on-shell. 
If we restrict ourselves to purely "holomorphic" class of amplitudes, i.e. the ones containing only one type of spinor product (here $\sqb{i}{j}$), we see that acting with $\tau_k$ derivative generates
\begin{multline}
 \dots\,\, \partial_-\phi_{b_{k-1}}\left( x + \tau_{k-1} \varepsilon^+_{\alpha_{k-1}} + s_{k} \varepsilon^+_{\alpha_{k}} \right)   \boldsymbol{\partial_{\tau_k}} \partial_-\phi_{b_{k}}\left( x + \tau_k \varepsilon^+_{\alpha_k} + s_{k+1} \varepsilon^+_{\alpha_{k+1}} \right) \dots \\ 
 =  \dots\,\, \partial_-\phi_{b_{k-1}}\left( x + \tau_{k-1} \varepsilon^+_{\alpha_{k-1}} + s_{k} \varepsilon^+_{\alpha_{k}} \right)     \boldsymbol{\varepsilon_{\alpha_k}^{+\mu} \partial_{\mu}} \partial_-\phi_{b_{k}}\left( x + \tau_k \varepsilon^+_{\alpha_k} + s_{k+1} \varepsilon^+_{\alpha_{k+1}} \right) \dots \,,
 \label{eq:tau_derivat_1}
\end{multline}
 which in momentum space will generate the $v$-symbol $v_{(k-1)k}$ (or the corresponding spinor bracket $\sqb{k-1}{k}$ on-shell). As we have already seen in Subsection~\ref{subsec:F_mom}, this is a consequence of the interleaved paths in the functional: the $\alpha_k$ integral will fix the $\varepsilon^+_{\alpha_k}$ to become the polarization vector $\varepsilon^+(p_{k-1})$, with $p_{k-1}$ being the momentum of the neighboring field, while the $\partial_{\mu}$ field derivative above will produce the momentum $p_{k}$. Thus, the above will give a factor $\varepsilon^+(p_{k-1})\cdot p_{k} = v_{(k-1)k}$.

There is however an important issue to be addressed on general grounds, before going any further. Namely, introducing the $\tau_k$ derivatives spoils the Lorentz symmetry of the functional. 
Moreover, as we shall see, the $\tau_k$ derivative alone is not permitted for the same reason.

The solution is simple and natural: we will modify the Lorentz property of some of the fields to balance the action of the $\partial_{\tau_k}$ derivative. As we shall argue in details below, a single $\tau_k$ derivative requires two neighboring fields $k-1$, $k$ to behave like the  spin (-1/2) components of the "good" light-cone spinors, i.e. built from the components of the Dirac field that remain dynamical in the light-cone formulation \cite{Kogut:1969xa}, $\Psi_+=\Lambda_+\Psi$ in the notation of Appendix~\ref{app:Lorentz}  (see also item (ii) of Subsection~\ref{sub:tau_spin}). This is in fact not surprising in momentum or twistor space, as the single derivative produces a product of spinors corresponding to two adjacent fields. 

The rest of this Section is organized as follows. In subsection~\ref{sub:tau_spin} we will start again with the twistor space formulation, which will help to better understand what derivative operators are actually permitted and what are the allowed Lorentz properties of the affected fields. 
Next, in subsection~\ref{sub:generating_functional} we will define a family of functionals  and discuss the amplitudes they generate.

\subsection{Single \texorpdfstring{$\tau$}{tau} derivative and spin-1/2 pairing}
\label{sub:tau_spin}

We begin by considering a single $\tau_m$ derivative acting upon a field. As elucidated above, such derivative would generate $v_{(m-1)m}=-\sqb{m-1}{m}\sqrt{p_{m}^+/2p_{m-1}^+}$ in momentum space (for the time being we assume the on-shell condition, but many of the conclusions will also hold off-shell). 

In order to understand how to incorporate the $\tau$ derivative in a consistent way, let's start with  the twistor space representation.

\paragraph{Insight from the twistor space.} Consider the twistor space formula for the functional without any derivatives, Eq.~\eqref{eq:F-twistor4}. We immediately see, that for the twistor space formula to remain well defined, the only allowed monomial of legs $m-1$,$m$ in the numerator, leading in momentum space to $\sqb{m-1}{m}$, is 
\begin{equation}
\mathcal{F}^{(m)}[\phi]\Big|^{\mathrm{on-shell}}_n (x) =   \,\Tr\left(t^{b_1} t^{b_2} \dots t^{b_n}\right) \,\prod_{k \in \mathbb{Z}_n}  \int_{\Gamma_k} \!  \pi_{k,\dot\alpha} d\pi_k^{\dot\alpha}  
  \,\,  
  \pi_{m,\dot\beta} \pi_{m+1}^{\dot\beta}
  \frac{ \mathfrak{f}_{b_k}\left( Z^I_{k+1}\right) }{\pi_{k,\dot\alpha} \pi_{k+1}^{\dot\alpha} }
 \, ,
\label{eq:Fm-twistor1}
\end{equation}
where now the homogeneity of functions $\mathfrak{f}_{b_{m-1}}\left( Z^I_m\right)$ and $\mathfrak{f}_{b_{m}}\left( Z^I_{m+1}\right)$ must be -1\footnote{ Although well defined in twistor space at fixed $x$, this expression is not an amplitude. The numerator of an amplitude in this sector is a component of the fermionic delta function $\delta^8$ of the $\Ncal=4$ superamplitude, eq.~\eqref{eq:nair}, which expresses the conservation of supermomentum, i.e. the invariance under the fermionic translations, and every such component is of order four in the brackets $\pi_{k,\dot\alpha}\pi_{l}^{\dot\alpha}$.}. 
This formula tells us that: (i) the only truly dynamical factor in the numerator in momentum space is $\sqb{m-1}{m}$, (ii) any possible factors of $\sqrt{p^+}$ must be build in the mapping of the fields to the twistor function and those factors are related to properties under boosts, (iii) the factors $\sqrt{p^+}$ must be identical for both fields $m-1$, $m$. 

The above observations are sufficient to construct the correct differential operators, consistent with both the Lorentz symmetry in space-time and the twistor space formula.

\paragraph{Introducing scalar components of spin-1/2 field.} We postulate the following modification of~\eqref{eq:tau_derivat_1} to consistently generate $\sqb{k-1}{k}$ in the numerator
\begin{multline}
 \dots\,\, \partial_-\phi^{(-1)}_{b_{k-2}}\left( x + \tau_{k-2} \varepsilon^+_{\alpha_{k-2}} + s_{k-1} \varepsilon^+_{\alpha_{k-1}} \right) \\
 \boldsymbol{\partial_-}\partial_-\phi^{(-1/2)}_{b_{k-1}}\left( x + \tau_{k-1} \varepsilon^+_{\alpha_{k-1}} + s_{k} \varepsilon^+_{\alpha_{k}} \right)  
  \boldsymbol{\partial_{\tau_k}} 
  \partial_-\phi^{(-1/2)}_{b_{k}}\left( x + \tau_k \varepsilon^+_{\alpha_k} + s_{k+1} \varepsilon^+_{\alpha_{k+1}} \right) \\
\partial_-\phi^{(-1)}_{b_{k+1}}\left( x + \tau_{k+1} \varepsilon^+_{\alpha_{k+1}} + s_{k+2} \varepsilon^+_{\alpha_{k+2}} \right)   
 \dots \,.
 \label{eq:tau_derivat_2}
\end{multline}
Let us summarize the modifications:
\begin{enumerate}[label=(\roman*)]
 \item we introduced explicit spin labels for the fields; in particular we relabel $\phi \equiv \phi^{(-1)}$ to stress that old fields transformed as the "bullet" spin (-1) components.
 \item  the fields $k-1$, $k$ are promoted to the helicity $-1/2$ components of the Dirac field. More precisely $\phi^{(-\frac12)}$ is the good component $\Psi_+^{(-\frac12)}$ of Appendix~\ref{app:Lorentz} with one inverse light-cone derivative, $\phi^{(-\frac12)}=\partial_-^{-1}\Psi_+^{(-\frac12)}$, i.e. $\widetilde\phi^{(-\frac12)}=\widetilde\Psi_+^{(-\frac12)}/(-ip^+)$ in momentum space. Since $\Psi_+$ scales as $\sqrt\kappa$ under the boost and $\partial_-^{-1}$ as $\bar\kappa$ (see Appendix~\ref{app:Lorentz}) these fields transform as
 \begin{equation}
  \phi^{(-\frac{1}{2})} (\kappa x^+,\bar\kappa x^-, x^{\bullet}, x^{\star}) = \sqrt{\bar\kappa}\,\, \phi^{(-\frac{1}{2})} (x^+,x^-,x^{\bullet}, x^{\star}) \, ,
 \end{equation} with respect to the boosts, and
 \begin{equation}
  \phi^{(-\frac{1}{2})} ( x^+, x^-,\gamma x^{\bullet},\bar\gamma x^{\star}) = \sqrt{\gamma}\,\, \phi^{(-\frac{1}{2})} (x^+,x^-,x^{\bullet}, x^{\star}) \, ,
 \end{equation} 
 with respect to complex rotations.
 \item in order to assure the boost invariance the spin-1/2 field without the $\tau$ derivative (the $k-1$ one) is accompanied by the extra light-cone derivative $\partial_-$;
 \item the transformation of the $\tau$ derivative with respect to complex rotations is balanced by the combine spin-1/2 field transformations; 
 \item since the spin-1/2 fields carry adjoint SU($N$) indices, we naturally interpret these as \emph{gluinos}.
\end{enumerate}
Note that the overall Lorentz invariance of the functional constrains only the total weights of all fields; the assignment above is dictated by the stronger requirement that the momentum-space output be little-group covariant and boost invariant leg by leg, which is the position-space counterpart of the projective well-definedness of \eqref{eq:Fm-twistor1} in each twistor $Z_k$ separately.
In what follows, we will explicitly check the above construction both in the momentum and twistor spaces.

\paragraph{Momentum space.} Executing the $\tau$ derivative action on $m$-th field and extra $\partial_-$ acting on \mbox{$(m-1)$-th} field at \emph{Step 2} of Subsection~\ref{subsec:F_mom} we find 
\begin{multline}
 \mathcal{F}^{(m)}[\phi]\Big|_n
  =  \frac{(2\pi)^{n+4}}{\sqrt{2}}\,\Tr\!\left(t^{b_1}\cdots t^{b_n}\right)
    \prod_{k \in \mathbb{Z}_n}\int\frac{d^4p_k}{(2\pi)^4}\;   
    \frac{\sgn{p_k^+}}{\vts{(k-1)}{k}}\; 
     \delta^4\!\!\left(\textstyle\sum_k p_k\right) \sqb{m-1}{m}
     \\
       \widetilde\phi^{\,(-1)}_{b_1}(p_1)\dots \widetilde\phi^{\,(-1)}_{b_{m-2}}(p_{m-2})\,\, \sqrt{p_{m-1}^+}\, 
       \widetilde\phi^{\,(-\frac{1}{2})}_{b_{m-1}}(p_{m-1})\,\, \sqrt{p_m^+} \,
       \widetilde\phi^{\,(-\frac{1}{2})}_{b_{m}}(p_{m})
      \widetilde\phi^{\,(-1)}_{b_{m+1}}(p_{m+1}) \dots \widetilde\phi^{\,(-1)}_{b_n}(p_n) \,.
  \label{eq:Fsd_tau_mom1}
\end{multline}
where $\sqb{m-1}{m}\equiv-\sqrt2\sqrt{p^+_{m-1}/p^+_m}\;\vts{(m-1)}{m}$ is the off-shell continuation of the spinor bracket, eq.~\eqref{eq:offshell_bracket}, to which it reduces on shell.
We see that now the spin-1/2 fields are accompanied by $\sqrt{p^+}$ factors that are balanced by their Lorentz boost transformation. Thus, the construction is consistent in the momentum space.

\paragraph{Twistor space.}
We shall now prove that it is possible to lift the functional to the twistor space in a consistent way. First, let us define the twistor function corresponding to the spin-1/2 field as
\begin{equation}
 \mathfrak{g}^{(-\frac12)}_b \left( Z^I \right) =  \int\! \frac{dp^+ dp^{\star} }{2(2\pi)^2 }\,\frac{1}{Z^1}
  e^{ip^{\star} Z^3} e^{-ip^+ Z^4} \,
  \widetilde{\phi}^{(-\frac{1}{2})}_{b}\left( Z^1 Z^2 p^{\star},p^+,p^{\star}, Z^1 Z^2 p^+\right) \,,
  \label{eq:twistor_g_def_half_spin}
\end{equation}
Notice, there is no $1/p^+$ comparing to \eqref{eq:twistor_g_def}. 
Using the fact that the combined complex boost and rotation transform the -1/2 spin field, a little unexpectedly, as the "bullet" component (see Appendix~\ref{app:Lorentz})
 \begin{equation}
  \phi^{(-\frac{1}{2})} (\lambda x^+, \bar\lambda x^-,\bar\lambda x^{\bullet},\lambda x^{\star}) = \bar\lambda \,\, \phi^{(-\frac{1}{2})} (x^+,x^-,x^{\bullet}, x^{\star}) \, ,
 \end{equation}
we can readily prove that the twistor function~\eqref{eq:twistor_g_def_half_spin} is homogeneous of degree -2
\begin{equation}
\mathfrak{g}^{(-\frac12)}_b(\lambda Z^I ) = \lambda^{-2}  \mathfrak{g}^{(-\frac12)}_b(Z^I ) \,.
\end{equation}
With this definition, the on-shell functional, with a single $\tau$ derivative (and extra $\partial_-$ derivative acting on $m-1$ function, as discussed above) is
\begin{multline}
\mathcal{F}^{(m)}[\phi]\Big|^{\mathrm{on-shell}}_n (x) =  \,\Tr\left(t^{b_1} t^{b_2} \dots t^{b_n}\right) \,\prod_{k \in \mathbb{Z}_n}  \int_{\Gamma_k} \!  \pi_{k,\dot\alpha} d\pi_k^{\dot\alpha} 
 \int_{-\infty}^0\! d\tau_k  \\
  \,\, \mathcal{D}_m 
   \partial_-\mathfrak{g}^{(h_k)}_{b_k}\left( Z^1_{k+1}, Z^2_{k+1}, Z^1_{k+1} x^{\bullet} - Z^2_{k+1}  x^+, Z^1_{k+1} x^- - Z^2_{k+1}  x^{\star} + \tau_k\left(Z_{k+1}^2-\frac{Z_{k+1}^1}{Z_k^1}\,Z_{k}^2\right) \right) 
 \, ,
\label{eq:F-twistor-alt-half-spin}
\end{multline}
where $h_k=-\tfrac12$ for $k=m-1,m$ and $h_k=-1$ otherwise, with $\mathfrak g^{(-1)}\equiv\mathfrak g$ of \eqref{eq:twistor_g_def}, and
\begin{equation}
 \mathcal{D}_m = \partial^{(m-1)}_{-}  \partial_{\tau_m} \,,
 \label{eq:D_m}
\end{equation}
with the $\partial_-$ derivative acting on the $m-1$ function (as indicated by an upperscript). The above is a projectively well defined integral: both $\mathfrak g$ and $\mathfrak g^{(-\frac12)}$ have homogeneity $-2$, and the derivatives do not change the homogeneities. The only difference to the Parke-Taylor case~\eqref{eq:F-twistor-alt1} are the extra derivatives. If we wanted to go back to the position space, we will find that the  Lorentz weights of those extra derivatives are balanced in the definitions of the twistor functions for spin -1/2 fields. We also see that the derivatives correctly reproduce~\eqref{eq:Fm-twistor1}. Using~\eqref{eq:twistor_g_def_half_spin}, utilizing the derivatives and performing the $\tau$ integrals\footnote{The $\tau_m$-integral of the total derivative $\partial_{\tau_m}\partial_-\mathfrak g^{(-\frac12)}$ is the boundary term $\partial_-\mathfrak g^{(-\frac12)}_{b_m}(P_{m+1})$, the $\tau_{m-1}$-integral of $\partial_-\partial_-\mathfrak g^{(-\frac12)}$ gives, by \eqref{eq:FTC}, $Z^1_{m-1}Z^1_m/(\pi_{m-1,\dot\alpha}\pi_{m}^{\dot\alpha})$ times $\partial_-\mathfrak g^{(-\frac12)}_{b_{m-1}}(P_m)$, and the other links give $Z^1_kZ^1_{k+1}/(\pi_{k,\dot\alpha}\pi_{k+1}^{\dot\alpha})$ times $\mathfrak g_{b_k}(P_{k+1})$ as before; the powers of $Z^1_k$ cancel around the cycle.} we get
\begin{equation}
\mathcal{F}^{(m)}[\phi]\Big|^{\mathrm{on-shell}}_n (x) =  \,\Tr\left(t^{b_1} t^{b_2} \dots t^{b_n}\right) \,\prod_{k \in \mathbb{Z}_n}  \int_{\Gamma_k} \!  \pi_{k,\dot\alpha} d\pi_k^{\dot\alpha}  
  \,\,  
  \pi_{m,\dot\beta} \pi_{m+1}^{\dot\beta}
  \frac{ \mathfrak{f}^{(t_k)}_{b_k}\left( Z^I_{k+1}\right) }{\pi_{k,\dot\alpha} \pi_{k+1}^{\dot\alpha} }
 \, ,
\label{eq:Fm-twistor2}
\end{equation}
where $t_k=-2h_k-2$ is the homogeneity of the twistor function of the field of helicity $h_k$: $t_k=-1$ for $k=m-1,m$, with $\mathfrak f^{(0)}\equiv\mathfrak f$ otherwise, and
\begin{equation}
 \mathfrak{f}^{(-1)}_b\left(Z^I\right)\equiv (Z^1)^2\,\frac{\partial\,\mathfrak{g}^{(-\frac12)}_b}{\partial Z^4}
 = -i\,Z^1\!\int\! \frac{dp^+ dp^{\star} }{2(2\pi)^2}\, p^+\,
  e^{ip^{\star} Z^3} e^{-ip^+ Z^4} \,
  \widetilde{\phi}^{(-\frac{1}{2})}_{b}\left( Z^1 Z^2 p^{\star},p^+,p^{\star}, Z^1 Z^2 p^+\right) \,,
\end{equation}
which is homogeneous of degree $-1$, as required; indeed $\partial_-\mathfrak g^{(-\frac12)}=Z^1\partial\mathfrak g^{(-\frac12)}/\partial Z^4=\mathfrak f^{(-1)}/Z^1$, and the factors $1/Z^1_{m}$, $1/Z^1_{m+1}$ so produced are canceled by the factors $Z^1$ left over from the neighboring links.

\paragraph{Multiple $\tau$ derivatives and spin incrementation.}
Basing on the above considerations we expect that each action with $\tau$ derivative increases the spin of the fields $m-1,m$ by 1/2 unit. The nontrivial aspect is the right number of $\partial_-$ derivatives acting on the pair of fields, so that: (i) the functional remains boost invariant in position (or momentum) space, (ii) the numerator is a pure spinor product, without extra boost invariant $p^+$ component ratios, (iii) the ramaining $p^+$ factors can be assimilated in a definition of a twistor function with a proper homogenity.
One particular difficulty is related to rather unobvious boost transformations of the \emph{physical} degrees of freedom of fields we are using in our formalism, comparing to complex rotations. 
That is, the $\pm 1$ spin fields transform as scalars for the boost $\kappa x^+$, while $\pm 1/2$ fields get both a factor $\sqrt{\bar\kappa}$. For complex rotations $\gamma x^{\bullet}$, on the other hand, the field transformation is gradual 
\begin{equation}
 \phi^{(h)} ( x^+, x^-,\gamma x^{\bullet},\bar\gamma x^{\star}) = \left(\gamma\right)^{-h}\,\, \phi^{(h)} (x^+,x^-,x^{\bullet}, x^{\star}) \,, \quad  h=-1,-1/2,0,+1/2,+1 \,.
\end{equation}
Therefore, a scaling of subsequent $\tau$ derivatives is balanced by the increasing spin of the two neighbouring spin fields, but, in general, it is not sufficient to pair each $\tau$ derivative with a $\partial_-$. The exact chain of $\partial_-$ depends on the exact chain of $\tau$ derivatives under consideration, as we shall see more explicitly below. 

Before closing this section, let us comment, that 
the outputs of the functional are, by construction, functions on a single twistor line with the Parke--Taylor denominator, i.e. degree-one curves in the counting of \cite{Witten:2003nn}; such configurations are amplitudes only in the $\MHVb$ sector, $\sum_k (h_k+1)=4$ \cite{Nair:1988bq,Witten:2003nn}. Since each $\tau$ derivative raises $\sum_k h_k$ by one unit, we restrict ourselves to at most four $\tau$ derivatives acting within the functional.

\subsection{Generating functional}
\label{sub:generating_functional}

Our generalization of the Parke-Taylor functional still will be evaluated on the closed random paths with memory, see Fig.~\ref{fig:closed_path}.

Let us start with defining some terminology we shall use throughout.

\paragraph{Slots and links.} It will be useful to call the position $i=1,\dots,n$ of the field in~\eqref{eq:F_functional} a \emph{slot}. The \emph{link} $k$ is the pair of consecutive slots $(k-1,k)$ together with the worldline integration $\tau_k$ that joins them; after the momentum-space reduction of Section~\ref{subsec:F_mom}, link $k$ carries exactly one inverse power of the $v$-symbol, $\vts{(k-1)}{k}$, cf.~\eqref{eq:tau_is_vtilde}.

\paragraph{Derivative monomials $\{m_k\}$ and $\{l_k\}$.} We assign a non-negative integer $m_k$ to each slot $k$, such that $\sum_k m_k = 4$ for $n\geq 4$. Each $m_k$ corresponds to a number of $\tau$ derivatives $\partial_{\tau_k}$ acting on the corresponding field at slot $k$, so that this field has helicity $h_k=\tfrac12(m_k+m_{k+1})-1$. As discussed in the preceding sections, the $\tau$ derivatives must be accompanied by a definite number of light-cone derivatives. We write them as $\partial_-^{\,l_k+1}$ acting on the field at slot $k$ and call $l\equiv\{l_k\}$ the second monomial. The two monomials are not independent:
\begin{equation}
 l_k=-\Big\lfloor \frac{m_k-m_{k+1}}{2}\Big\rfloor\,,
 \label{eq:l_rule}
\end{equation}
where a negative value means an inverse power of $\partial_-$, as is common in the light-cone gauge. The rule is derived after eq.~\eqref{eq:family_mom} below by requiring that the output of the functional is boost invariant leg by leg and that its numerator is a pure product of spinor brackets.

\paragraph{The functional.} We are ready to define the \emph{off-shell} generating functional, which, in momentum space has the Parke-Taylor denominator and a monomial of order 4 in the numerator. The functional, in general, depends on scalar components of physical fields of various helicities, $\phi^{(h)}$, with $h=-1,-1/2,0,+1/2,+1$. For a given derivative monomial $m\equiv\{m_i\}$ we define
\begin{equation}
  \mathcal F^{\{m\}}\left[\phi^{(h)}\right]
  =\int\!d^4x\,\sum_{n\ge3}\,
  \Tr\prod_{i\in\mathbb{Z}_n}
  \int\!ds_i\,d\alpha_i\;\int_{-\infty}^0\!d\tau_i\;
  \partial^{\,l_i+1}_-
  \partial^{\,m_i}_{\tau_{i}}  
  \hat\phi^{\left(\frac{m_i}{2}+\frac{m_{i+1}}{2}-1\right)}
  \big(x+\tau_i\varepsilon^+_{\alpha_{i}}+s_{i+1}\varepsilon^+_{\alpha_{i+1}}\big)\,,
  \label{eq:family}
\end{equation}

\paragraph{Momentum space}
In momentum space, upon reduction of the external fields and taking the on-shell limit, the above functional generates the landscape of $\mathcal{N}=4$ super Yang-Mills theory, as we shall discuss later below. Without taking the on-shell limit, the above formula leads to an expression, formally, in terms of the $v$-symbols which provide a natural off-shell continuation of the spinor brackets. More concretely, repeating the steps of Section~\ref{subsec:F_mom}, slot $k$ contributes $(-ip_k^+)^{l_k+1}$ from the light-cone derivatives, $\sgn{p_k^+}/p_k^+$ from the $\alpha_{k+1}$ integral, $(-i\vts{(k-1)}{k})^{m_k}$ from the $\tau$ derivatives and $i/\vts{(k-1)}{k}$ from the $\tau_k$ integral, so that, up to an overall constant, for an $n$-th term of~\eqref{eq:family} in momentum space we have
\begin{multline}
  \mathcal F^{\{m\}}\left[\phi^{(h)}\right]\Big|_n
  \propto \Tr\!\left(t^{b_1}\cdots t^{b_n}\right)
    \prod_{k \in \mathbb{Z}_n}\int\frac{d^4p_k}{(2\pi)^4}\;
    \delta^4\!\!\left(\textstyle\sum_k p_k\right)\sgn{p_k^+}
    \frac{1}{\vts{(k-1)}{k}}\; \left( \sqrt{\frac{p_{k-1}^{+}}{p_{k}^+}}\,\vts{(k-1)}{k}\right)^{m_k} \\
   \left(p^+_k\right)^{\frac{\epsilon_k}{2}}\, \widetilde\phi^{\,(h_k)}_{b_k}(p_k)\,,\qquad \epsilon_k=\begin{cases}1 & h_k\ \text{half-integer}\,,\\ 0 & h_k\ \text{integer}\,.\end{cases}
  \label{eq:family_mom}
\end{multline}
Above, the $p^+$ factors accompanying the fields assure their boost invariance for various spins, while the $p^+$ factors accompanying the $v$-symbols in the numerator assure they become pure spinor brackets, as required by our twistor space considerations. All of that is guaranteed by the proper choice of the $\{l_k\}$ monomial, eq.~\eqref{eq:l_rule}, as we now show.

The rule \eqref{eq:l_rule} follows from \eqref{eq:family_mom} by counting the powers of $p^+$ attached to each leg. The $v$-symbol of slot $k$ is not a pure spinor bracket. By \eqref{eq:vij_to_spinor} it reads
\begin{equation}
 \vts{(k-1)}{k}=-\frac{1}{\sqrt2}\sqrt{\frac{p_k^+}{p_{k-1}^+}}\;\sqb{k-1}{k}\,,
\end{equation}
so that the factor $\vts{(k-1)}{k}^{\,m_k-1}$ produced by slot $k$ carries $(p_k^+)^{(m_k-1)/2}$ and $(p_{k-1}^+)^{-(m_k-1)/2}$ on top of a product of brackets. The momentum $p_k^+$ therefore appears with the following powers: $l_k$ from the light-cone derivatives of slot $k$ (the power $l_k+1$ from $\partial_-^{\,l_k+1}$ minus one from the $\alpha_{k+1}$ integral), $(m_k-1)/2$ from the $v$-symbol of slot $k$ and $-(m_{k+1}-1)/2$ from the $v$-symbol of slot $k+1$. In total the field of leg $k$ is multiplied by
\begin{equation}
 \left(p_k^+\right)^{\,l_k+\frac{m_k-m_{k+1}}{2}}
\end{equation}
times brackets. The brackets are boost invariant. The field $\widetilde\phi^{(h_k)}(p_k)$ is boost invariant for integer $h_k$ and acquires the factor $\sqrt{\bar\kappa}$ for half-integer $h_k$, so that the invariant combination is $(p_k^+)^{\epsilon_k/2}\,\widetilde\phi^{(h_k)}(p_k)$ with $\epsilon_k$ as in \eqref{eq:family_mom}. Requiring the output to be boost invariant leg by leg thus fixes
\begin{equation}
 l_k=\frac{\epsilon_k}{2}-\frac{m_k-m_{k+1}}{2}\,.
\end{equation}
Finally $h_k=\tfrac12(m_k+m_{k+1})-1$ is half-integer exactly when $m_k-m_{k+1}$ is odd. For even $m_k-m_{k+1}$ the last equation gives $l_k=-(m_k-m_{k+1})/2$ and for odd $m_k-m_{k+1}$ it gives $l_k=-(m_k-m_{k+1}-1)/2$. Both cases are summarized by the floor function in \eqref{eq:l_rule}. For a single $\tau$ derivative at slot $m$ the rule gives $l_m=0$ and $l_{m-1}=1$, which is the assignment of \eqref{eq:tau_derivat_2}. For the gluonic monomial $m=(4,0,\dots,0)$ it gives $l_1=-2$ and $l_n=2$, so that inverse powers of $\partial_-$ cannot be avoided for general monomials.

\subsection{Three points}
\label{sub:3p}

Before turning to the scattering amplitudes with $n\geq 4$, let us discuss the central off-shell result of the calculus, which is at three points, i.e. for $n=3$ above.

\paragraph{Off-shell. } With the monomial $m=(m_1,m_2,m_3)$ on the links carrying $\vts{3}{1},\vts{1}{2},\vts{2}{3}$ the structure collapses on (off-shell) momentum conservation. Using the three-point identity~\eqref{eq:v_threept} we get (upon stripping the fields and their corresponding boost factors)
\begin{multline}
  \mathcal F^{\{m\}}\Big|_3
  \;\longrightarrow\;
  \vts{3}{1}^{\,m_1-1}\,\vts{1}{2}^{\,m_2-1}\,\vts{2}{3}^{\,m_3-1} \left(p_1^+\right)^{\frac{m_2-m_1}{2}} \left(p_2^+\right)^{\frac{m_3-m_2}{2}} \left(p_3^+\right)^{\frac{m_1-m_3}{2}} \\
  =  (p_1^+ v_{12} )^{h_1+h_2+h_3} \left(p_1^+\right)^{-h_1} \left(p_2^+\right)^{-h_2} \left(p_3^+\right)^{-h_3}   \,,\qquad h_k=\frac{m_k+m_{k+1}}{2}-1\,,
  \label{eq:metsaev_v}
\end{multline}
 Equation~\eqref{eq:metsaev_v} corresponds to  the  \emph{kinematic form} of the light-cone cubic vertices of chiral higher-spin theory \cite{Metsaev:1991mt,Metsaev:1991nb,Ponomarev:2016lrm,Ponomarev:2017nrr}, off shell, with the invariant $p_1^+\vts{1}{2}$ playing the role of the antisymmetric light-front momentum of that literature.
Keep in mind, that the chiral higher-spin \emph{theory} is defined not by this kinematic factor alone but by the coupling assignment $C_{\lambda_1\lambda_2\lambda_3}=\ell^{\,\Sigma\lambda-1}/\Gamma(\Sigma\lambda)$, fixed by closure of the light-cone Poincar\'e algebra \cite{Metsaev:1991mt,Ponomarev:2016lrm}. The functional in its current form does not include any relative couplings.

\paragraph{On-shell. }
In the on-shell limit the $v$-symbols, together with the corresponding $\sqrt{p^+}$ factors,  become the spinor brackets
\begin{equation}
  \mathcal F^{\{m\}}\Big|^{\rm on\text{-}shell}_3
  \;\longrightarrow\;
  \sqb{3}{1}^{\,m_1-1}\,\sqb{1}{2}^{\,m_2-1}\,\sqb{2}{3}^{\,m_3-1}\,,
  \label{eq:threept_br}
\end{equation}
which is the holomorphic three-point amplitude on complex kinematics, the on-shell face of the off-shell kinematic form~\eqref{eq:metsaev_v}, coinciding vertex by vertex with the cubic amplitudes of chiral higher-spin theory \cite{Metsaev:1991mt,Metsaev:1991nb,Ponomarev:2016lrm,Ponomarev:2017nrr} once that theory's coupling assignment is supplied  with the $U(N)$ Chan--Paton dressings under which chiral theories remain consistent \cite{Skvortsov:2020wtf,Skvortsov:2020gpn} and with supersymmetric extensions populating the half-integer entries \cite{Metsaev:2019dqt,Metsaev:2019aig,Tsulaia:2022csz}. The interesting monomials, their corresponding helicity content and the on-shell output can be summarized in the following table:
\begin{center}
\begin{tabular}{llll}
$(m_1,m_2,m_3)$ & $(h_1,h_2,h_3)$ & output & identification \\
\hline
$(4,0,0)$ & $(+1,-1,+1)$ & $\sqb{3}{1}^{\,3}/(\sqb{1}{2}\sqb{2}{3})$ &
$\MHVb$ three-gluon amplitude \cite{Parke:1986gb} \\
$(2,2,2)$ & $(+1,+1,+1)$ & $\sqb{3}{1}\,\sqb{1}{2}\,\sqb{2}{3}$ &
all-plus amplitude of $\Tr F_{\rm ASD}^3$ \cite{Brandhuber:2018kqb,AccettulliHuber:2019mxn} \\
$(3,3,3)$ & $(+2,+2,+2)$ & $\sqb{3}{1}^{2}\sqb{1}{2}^{2}\sqb{2}{3}^{2}$ &
chiral-graviton vertex \cite{Metsaev:1991mt,Ponomarev:2016lrm} \\
$(2,1,1)$ & $(+\tfrac12,0,+\tfrac12)$ & $\sqb{3}{1}$ &
fermionic chiral vertex \cite{Metsaev:2019dqt,Tsulaia:2022csz} \\
$(3,2,1)$ & $(+\tfrac32,+\tfrac12,+1)$ & $\sqb{3}{1}^{2}\,\sqb{1}{2}$ &
mixed-spin supersymmetric vertex \cite{Metsaev:2019dqt,Metsaev:2019aig} \\
$(4,1,0)$ & $(+\tfrac32,-\tfrac12,+1)$ & $\sqb{3}{1}^{\,3}/\sqb{2}{3}$ &
one negative light-cone exponent \cite{Metsaev:2019dqt} \\
\end{tabular}
\end{center}

Since $m_k\ge0$, the exponents $m_k-1$ of the three brackets in \eqref{eq:threept_br} are bounded from below by $-1$, which corresponds to a bare link. The chiral theory also contains vertices with an exponent $\le-2$, e.g. helicities $(+3,+3,-2)$ with exponents $(8,-2,-2)$ or the $0$-$0$-$s$ couplings with $s\ge2$; these are outside the derivative calculus, see Section~\ref{sub:boundary}.

\subsection{\texorpdfstring{$\mathcal{N}=4$}{N=4} dictionary}
\label{sub:N4_dictionary}

We now proceed to a discussion of scattering amplitudes generated by the  functional with the monomial $m$ of order 4. We shall assume the on-shell reduction of the functional and express the results in terms of the spinor brackets, instead of their off-shell continuations -- the $v$-symbols.

\paragraph{Nair's superamplitude.}
The most simple way of organizing the $\MHVb$ sector is via the superamplitude of $\Ncal=4$ super Yang--Mills \cite{Nair:1988bq}
\begin{equation}
  \mathcal A_n^{\MHVb}
  =\frac{\delta^4\big(\sum_ip_i\big)\,\delta^8\big(\sum_i\tlam_i\tilde\eta_i\big)}{\sqb{1}{2}\sqb{2}{3}\cdots\sqb{n}{1}}\,,
  \qquad
  \delta^8=\prod_{A=1}^4\sum_{i<j}\sqb{i}{j}\,\tilde\eta_i^A\tilde\eta_j^A\,,
  \label{eq:nair}
\end{equation}
written in the conventions of~\eqref{eq:Fsd_final}: the Grassmann-degree-zero component is the all-minus configuration and each $\tilde\eta$ raises a helicity by $+\tfrac12$. Comparing with \eqref{eq:family_mom} on shell: $\partial_{\tau_k}$ supplies one factor of $\sqb{(k-1)}{k}$ and one half-unit on each of legs $k-1,k$; it is the  counterpart of the \emph{adjacent} bilinear $\sqb{(k-1)}{k}\,\tilde\eta_{k-1}\!\cdot\!\tilde\eta_k$ of $\delta^8$. Distributing four derivatives among the links is the same combinatorics as extracting a component of $\delta^8$ whose four flavor pairings all connect neighbors.

\paragraph{The four-point atlas.}
With $s=\ab{1}{2}\sqb{2}{1}$, $t=\ab{1}{4}\sqb{4}{1}$, $s+t+u=0$  
 the extracted on-shell outputs are:
\begin{align}
  m=(4,0,0,0):&\quad
  \frac{\sqb{4}{1}^{\,4}}{\sqb{1}{2}\sqb{2}{3}\sqb{3}{4}\sqb{4}{1}}
  =A\big(1^{+},2^{-},3^{-},4^{+}\big)
  \label{eq:atlas_gluons}\\
  m=(3,1,0,0):&\quad
  \frac{\sqb{4}{1}^{\,3}\sqb{1}{2}}{\sqb{1}{2}\sqb{2}{3}\sqb{3}{4}\sqb{4}{1}}
  =\frac{\sqb{4}{1}^{\,2}}{\sqb{2}{3}\sqb{3}{4}}
  =A\big(1^{+1},2^{-1/2},3^{-1},4^{+1/2}\big)
  \label{eq:atlas_gluino}\\
  m=(2,2,0,0):&\quad
  \frac{\sqb{4}{1}\,\sqb{1}{2}}{\sqb{2}{3}\,\sqb{3}{4}}\,,
  \quad\text{scalars at }2,4\,;\;\;
  \sqb{4}{1}\sqb{1}{2}\,\ab{1}{2}\ab{4}{1}=ts=\ab{2}{3}\ab{3}{4}\,\sqb{2}{3}\sqb{3}{4}\,,
  \label{eq:atlas_two_scalars}\\
  m=(2,0,2,0):&\quad
  \frac{\sqb{4}{1}\,\sqb{2}{3}}{\sqb{1}{2}\,\sqb{3}{4}}=-\frac{t}{s}\,,
  \quad\text{four scalars: $s$-channel exchange $+$ contact, no $t$-channel pole}\,,
  \label{eq:atlas_ts}\\
  m=(2,1,1,0):&\quad
  \frac{\sqb{4}{1}^{\,2}\sqb{1}{2}\sqb{2}{3}}{\sqb{1}{2}\sqb{2}{3}\sqb{3}{4}\sqb{4}{1}}
  =\frac{\sqb{4}{1}}{\sqb{3}{4}}\,,
  \quad\text{gluinos at }1,3\,,\;\text{scalars at }2,4\,,
  \label{eq:atlas_mixed}\\
  m=(1,1,1,1):&\quad 1\,,
  \quad\text{the quartic scalar contact vertex.}
  \label{eq:atlas_contact}
\end{align}
Every nontrivial entry agrees with known results, in particular the first one is the  Parke-Taylor amplitude \cite{Parke:1986gb,Berends:1987me} and the second agrees with \cite{Grisaru:1976vm,Parke:1985pn,Gunion:1985vca}. For the four-scalar entry~\eqref{eq:atlas_ts} the flavor routing behind the pole structure is instructive: the monomial $m=(2,0,2,0)$ corresponds, through the $\delta^8$ dictionary above, to the component $A\big(1_{\phi^{AB}},2_{\phi^{CD}},3_{\phi^{CD}},4_{\phi^{AB}}\big)$ with $\{A,B\}\cap\{C,D\}=\emptyset$; legs $1,2$ carry conjugate scalars (since $\phi_{AB}=\tfrac12\epsilon_{ABCD}\phi^{CD}$), so a gluon can be exchanged in the planar $s=(p_1{+}p_2)^2$ channel, while legs $2,3$ carry identical scalars and no gluon exchange is possible in the planar $t$ channel, in agreement with the single pole of $-t/s$.

\paragraph{Different monomials with the same helicities.}
A monomial $m$ fixes more than the helicities of the legs. In the dictionary above each $\tau$ derivative at slot $k$ corresponds to one factor $\sqb{k-1}{k}\,\tilde\eta_{k-1}^A\tilde\eta_k^A$ of $\delta^8$, i.e. to one $SU(4)$ index $A$ shared by the legs $k-1$ and $k$. Two monomials can therefore give the same helicities to all legs and still distribute the four indices differently. At four points this happens for $m=(0,3,0,1)$ and $m=(1,2,1,0)$. Both give helicities $(+\tfrac12,+\tfrac12,-\tfrac12,-\tfrac12)$. In the first case three indices are shared by legs $1,2$ and one index by legs $3,4$, so that all four gluinos carry the same $SU(4)$ label. In the second case the four indices are shared by the pairs of legs $(4,1)$, $(1,2)$, $(1,2)$ and $(2,3)$; as a result legs $1$ and $3$ carry one label and legs $2$ and $4$ carry another one. These are different components of the superamplitude. The functional gives $\sqb{1}{2}^2/(\sqb{2}{3}\sqb{4}{1})$ for the first and $\sqb{1}{2}/\sqb{3}{4}$ for the second, and the two differ by the factor $-s/t$.

\paragraph{Higher-point amplitudes.}
At five points $m=(4,0,0,0,0)$ gives 
$\sqb{5}{1}^{\,4}/(\sqb{1}{2}\sqb{2}{3}\sqb{3}{4}\sqb{4}{5}\sqb{5}{1})=A(1^+,2^-,3^-,4^-,5^+)$, 
 and the same monomial at any $n$ yields the adjacent-plus $\MHVb$ gluon amplitude, i.e.\ the extracted output is, on shell, the conjugate Parke--Taylor amplitude. The single-fermion-pair components are, function by function, the QCD amplitudes with a quark pair \cite{Gunion:1985vca,Kunszt:1985mg}, the color factor being the only difference; $\Ncal<4$ truncations select the corresponding sub-atlas.

\subsection{Collinear limit}
\label{sub:collinear}

In the holomorphic collinear limit $\sqb{1}{2}\to0$, the light-front collinearity $p_1^\bul/p_1^+=p_2^\bul/p_2^+$ of Section~\ref{sec:Basics}, the canonical spinors~\eqref{eq:spinor_standard} of the pair and of $P=p_1+p_2$ obey exactly $\tlam_{1,2}=c_{1,2}\,\tlam_P$ with $c_i=\sqrt{p_i^+/P^+}$, so that $c_1^2+c_2^2=1$. 
 The pole comes entirely from the $\tau_2$-integral and its residue fuses the chain,
\begin{equation}
  \mathrm{Res}_{\sqb{1}{2}\to0}\,
  \frac{\Tr\big(t^{b_1}t^{b_2}t^{b_3}\!\cdots\big)}
       {\sqb{n}{1}\sqb{1}{2}\sqb{2}{3}\cdots}
  =\frac{1}{c_1c_2}\,
  \frac{\Tr\big(t^{b_1}t^{b_2}\,t^{b_3}\!\cdots\big)}
       {\sqb{n}{P}\,\sqb{P}{3}\cdots}\,,
  \label{eq:collinear}
\end{equation}
with the momentum-fraction factor $1/(c_1c_2)=P^+\!/\sqrt{p_1^+p_2^+}$ of the holomorphic splitting function. 
 The soft towers organized on such collinear data are the $w_{1+\infty}$ symmetry of celestial holography \cite{Guevara:2021abz,Strominger:2021mtt}, and the sharpened statement that self-dual observables are correlators of a two-dimensional chiral algebra is the chiral-algebra bootstrap \cite{Costello:2021bah,Costello:2022wso}.

\subsection{Off-shell continuation of the spinor brackets}
\label{sub:offshell_brackets}

The spinor map~\eqref{eq:vij_to_spinor}, read from right to left, extends the square bracket off shell through the $v$-symbols: we \emph{define}, for arbitrary momenta,
\begin{equation}
  \sqb{i}{j}\;\equiv\;-\sqrt2\,\sqrt{\frac{p_i^+}{p_j^+}}\;\vts{i}{j}
  \qquad(\textit{off shell})\,,
  \label{eq:offshell_bracket}
\end{equation}
which reduces to the spinor product on shell and introduces no new kinematic object: it is the bracket itself, continued. This continuation coincides with the off-shell brackets of the light-front MHV-action literature \cite{Mansfield:2005yd,Ettle:2006bw,Kotko2017}. It differs from the CSW prescription, the unnormalized $\lambda_\alpha\propto P_{\alpha\dot\alpha}\widetilde\eta^{\dot\alpha}$ \cite{Cachazo:2004kj} (where the normalization is immaterial as it cancels in assembled amplitudes).

\subsection{The boundary of the landscape}
\label{sub:boundary}

It is rather obvious that one can build a more general functional by symmetrizing over non-identical fields and promoting the spin-1/2 fields to Grassman-valued functions. This would allows to reconstruct the color-dressed amplitudes of the (anti-)MHV sectors. More precisely, such extended functional therefore computes exactly: (1) the color-dressed $\MHVb$ superamplitude of $\Ncal=4$ SYM \cite{Nair:1988bq,Elvang:2013cua}, with its QCD and $\Ncal<4$ truncations \cite{Gunion:1985vca,Parke:1985pn,Kunszt:1985mg}; (2) the three-point $\MHVb$ vertex of self-dual Yang--Mills; (3) holomorphic three-point amplitudes with all light-cone exponents $\ge-1$, and off shell the kinematic forms~\eqref{eq:metsaev_v} of the light-cone chiral vertices; (4) the current-algebra collinear structure.

Let us summarize the limitations of the functional calculus in its current form.  \emph{${\rm N}^k\MHVb$ amplitudes}: for the monomial $m$ order $p>4$ distributions with helicities $|h_j|\le1$ (e.g.\ $n=6$, $m=(4,0,0,4,0,0)$, output $\sqb{6}{1}^4\sqb{3}{4}^4\big/\prod_k\sqb{(k-1)}{k}$) are well-defined holomorphic objects but not amplitudes, since beyond (anti-)MHV amplitudes require multiparticle poles \cite{Benincasa:2007xk,McGady:2013sga}; assembling them from the outputs is the (conjugate) CSW expansion \cite{Witten:2003nn,Cachazo:2004kj,Mansfield:2005yd,Gorsky:2005sf}, external to the functional, although the \emph{off-shell} $p>4$ outputs connect to the ${\rm N}^k\MHVb$ vertices of the Wilson-line-based actions \cite{Kakkad:2021uhv,Kakkad_2022}. \emph{Loop level}: the all-same-helicity one-loop family, e.g.\ $A_4^{\rm 1\text{-}loop}\propto\ab{1}{2}\ab{3}{4}/(\sqb{1}{2}\sqb{3}{4})$ in our conventions, exits holomorphy; it was conjectured in \cite{Bern:1993qk}, proven in \cite{Mahlon:1993fi}, tied to self-duality in \cite{Bern:1996ja} and to an anomaly in \cite{Bardeen:1995gk}; the incompleteness of naive MHV rules at one loop, and its cure by matter completions or careful regularization, is documented in \cite{Brandhuber:2007vm,Ettle:2007qc}, and the light-front quantum completions of the MHV and Wilson-line actions \cite{Kakkad_2022,Kakkad:2024ase} address precisely the pieces the tree-level functional misses. \emph{Gravity}: MHV gravity is not a cyclic chain \cite{Berends:1988zp,Mason:2008jy,Hodges:2012ym}; self-dual gravity trades the color trace for the kinematic algebra \cite{Plebanski:1975wn,Monteiro2011,Krasnov:2016emc}, a different functional. \emph{Massive states} break the $\lambda\tlam$ factorization; \emph{chiral vertices with an exponent $\le-2$} are outside the derivative calculus.

\section{Conclusions and outlook}
\label{sec:conclusions}

We have studied a position-space object, a kernel-free functional of algebra-valued fields, defined on points that belong to a velocity-correlated piece-wise path on the anti-self-dual plane. The path has a built-in "time" arrow, by demanding that one of the path parameter $\tau$ is bounded from above, $-\infty<\tau \leq 0$;  the exact definition of the path was provided in Eq.~(\eqref{eq:piecewise_path}. We have shown that it encodes a substantial amount of gauge-theory structure. Its momentum-space reduction produces the cyclic Parke--Taylor factor, off shell in the light-front $v$-symbols and on shell in spinor brackets, the only two kinematic objects used in this work. Reorganizing the same reduction, twistor space emerges: the functional is a closed cyclic chain of Penrose transforms on $\mathbb{CP}^3$, with the twistor functions defined through certain integral transforms of the fields, eg.~Eq~\eqref{eq:twistor_f_def}. We also formulate the piece-wise correlated path directly in twistor space.
In twistor space the picture is geometric: at fixed $x$ the $n$-th term lives in the projective plane $\Pi_x\subset\mathbb{CP}^3$ dual to the plane $\Pi^+(x)$; each twistor function dual to a field is carried by a twistor which, under the $\tau_k$ integration, slides along a line through the reference twistor $Z_*$ of the light-cone gauge; the colour ordering closes these motions into a $2n$-gon with alternate vertices on the line $L_x$, and the Parke--Taylor denominators are the boundary terms of the $\tau_k$ integrals. Setting all $\tau_k=0$ collapses the polygon onto $L_x$, the configuration of $n$ points on a single line familiar from the MHV vertex.

Recognizing next that derivatives over the path parameter $\tau$ create an anti-holomorphic spinor brackets on-shell, we developed a family of functionals defined on algebra-valued fields of different spins that exhaust the $\MHVb$ amplitude super multiplet. The fields in question are either scalar, $\pm 1$ helicity fields corresponding to physical components of the vector gauge fields, or the "good" spin $+1/2$ and $-1/2$ components of the Dirac field. The character of the fields is simply a consequence of the Lorentz boost invariance and complex rotational invariance of the functional with $\tau$ derivatives.

Several directions remain open. The off-shell outputs at total monomial order $p>4$ are natural candidates for the ${\rm N}^k\MHVb$ vertices of the Wilson-line-based actions, and matching their normalizations and composite-momentum arguments would make direct contact with the light-front Lagrangian program. 
On the other hand, it is natural to try to search for paths that exit the "holomorphy", i.e. that wander off the anti-self-dual plane  in complexified Minkowski space, and generate the other amplitude sectors. 
On the mathematical side, the functional is built on concatenations of half-lines with correlated directions, i.e.\ non-differentiable paths, and we expect that its definition and the derivative calculus admit a formulation within rough-path theory
\cite{Lyons:1998rough}.

\section*{Acknowledgements}

We are grateful to A.~Stasto for motivation and  support in exploring the twistor space formulation of the functional. HK thanks T.~Adamo, Y.-t.~Huang, A.~Laddha, L.~Mason, A.~Ochirov and L.~Smith for discussions.  HK is supported by the National Natural Science Foundation of China via NSFC grant No. 12447147 and by the Science and Technology Commission of the Shanghai Municipality via STCSM grant No. 24ZR1450600. PK was supported by the National Science Centre, Poland, grant no.~\mbox{DEC-2024/54/E/ST2/00369}. 

\newpage
\appendix

\section{Dynamical scalar components and their Lorentz transformations}
\label{app:Lorentz}

In our framework the functional family is built out of the physical components of the covariant fields. Below we summarize the residual Lorentz transformation properties for fields of different spin. Throughout, a field transforms as $\Phi'(x')=D(\Lambda)\Phi(x)$ with $x'=\Lambda x$, and the \emph{weight} of a component is the factor $D(\Lambda)$ multiplying it; in momentum space the components of $p$ transform like those of $x$, so that $x\cdot p$ is invariant, and the Fourier transforms of the fields have the same weights as the fields themselves.

\paragraph{Lorentz residual transformations.} Three residual transformations play central role in our work:\\ (i) longitudinal boost
\begin{equation}
 x^+ \rightarrow \kappa x^+ \,, \quad x^- \rightarrow \bar{\kappa} x^- \,, \quad \text{with $\kappa\bar\kappa =1$} \,,
\end{equation}
(ii) transverse complex rotation
\begin{equation}
 x^{\bullet} \rightarrow \gamma x^{\bullet} \,, \quad x^{\star} \rightarrow \bar\gamma x^{\star} \,, \quad \text{with $\gamma\bar\gamma = 1$ } \,,
\end{equation}
 (iii) combined complex scaling 
\begin{equation}
 x^+ \rightarrow \lambda x^+ \,, \quad x^- \rightarrow \bar{\lambda} x^- \,,\quad
 x^{\bullet} \rightarrow \bar\lambda x^{\bullet} \,, \quad x^{\star} \rightarrow \lambda x^{\star} \,, \quad \text{with $\lambda\bar\lambda = 1$ } \,,
\end{equation}
i.e. (i) and (ii) with $\kappa=\lambda$, $\gamma=\bar\lambda$. This is the transformation used in Subsection~\ref{sub:tau_spin}; it is induced on the momentum-space arguments of the fields by the scaling $Z^I\to\bar\lambda Z^I$ of the twistor variables, see the proof of the homogeneity of \eqref{eq:twistor_f_def}, and it controls the homogeneities of the twistor functions. Under (i) and (ii) the measures and derivatives of the functional carry the weights $ds,d\tau\to\bar\gamma$, $d\alpha\to\bar\kappa\gamma$, $\partial_-\to\kappa$, $\partial_\tau\to\gamma$.

\paragraph{Spin-1 fields.} For a vector gauge field $A^{\mu}$ the physical components are $A^{\bullet}$ and $A^{\star}$. One way to see this is to choose the light-cone gauge $A^+=0$ and notice that the action becomes quadratic in $A^-$, meaning, it can be integrated out of the path integral, see eg.~\cite{Scherk1975}. Equivalently, one sees that $A^-$ is a non-dynamical component that can be eliminated through equation of motion. In our work, we identify
\begin{equation}
  \phi^{(-1)}\equiv A^{\bullet} \,, \quad   \phi^{(+1)}\equiv A^{\star} \,.
\end{equation}
Therefore, being components of a vector, they transform like the corresponding coordinates:
\begin{equation}
 \phi^{(-1)}\rightarrow  \phi^{(-1)} \,, \quad \phi^{(+1)}\rightarrow  \phi^{(+1)}\,, \quad \text{(longitudinal boost)} \,,
\end{equation}
\begin{equation}
 \phi^{(-1)}\rightarrow \gamma \phi^{(-1)} \,, \quad \phi^{(+1)}\rightarrow \bar\gamma \phi^{(+1)}\,, \quad \text{(transverse rotation)} \,,
\end{equation}
\begin{equation}
 \phi^{(-1)}\rightarrow \bar\lambda \phi^{(-1)} \,, \quad \phi^{(+1)}\rightarrow \lambda\phi^{(+1)}\,, \quad \text{(combined scaling)} \,.
\end{equation}

\paragraph{Spin-1/2 fields.} We describe spin-1/2 particles in terms of two scalar fields $\phi^{(-\frac{1}{2})}$ and $\phi^{(+\frac{1}{2})}$. These fields correspond to components of the "good" Dirac field $\Psi_+$, which is the only dynamical component of the field. More precisely, the light cone coordinates allow to define the nilpotent Dirac gamma matrices $\boldsymbol{\gamma}^{\pm}=(\boldsymbol{\gamma}^0\pm\boldsymbol{\gamma}^3)/\sqrt2$, $(\boldsymbol{\gamma}^{\pm})^2=0$, satisfying $\{\boldsymbol{\gamma}^{+}, \boldsymbol{\gamma}^{-}\}=2$, which naturally form light-cone projections operators $\Lambda_{\pm}=\frac{1}{2}\boldsymbol{\gamma}^{\mp}\boldsymbol{\gamma}^{\pm}=\frac12\left(1\pm\boldsymbol{\gamma}^0\boldsymbol{\gamma}^3\right)$, such that $\Lambda_+ + \Lambda_- = 1$ and $\Lambda_{\pm}^2 = \Lambda_{\pm}$, $\Lambda_{+}\Lambda_{-}=\Lambda_{-}\Lambda_{+} = 0$. The full Dirac field is decomposed in that basis as $\Psi=\Psi_+ + \Psi_-$. Similar to the vector field case, treating the $x^+$ as light-cone time one finds that $\Psi_-$ is non-dynamical and can be eliminated through equations of motion. To explicitly find the remaining "good" component (in Weyl basis), let us take the four-component Dirac field and project it onto $\Lambda_+$:
\begin{equation}
 \Psi = \begin{pmatrix}
  \xi \\
  \chi
 \end{pmatrix}
 \,, \quad \Rightarrow \quad
 \Psi_+ = \Lambda_+\Psi = \begin{pmatrix}
 \xi_2 \\
 \chi_1
 \end{pmatrix}
  \equiv \partial_-\begin{pmatrix} \phi^{(-\frac{1}{2})} \\ \phi^{(+\frac{1}{2})} \end{pmatrix} \,.
\end{equation}
The fields $\phi^{(\pm\frac12)}$ used in the functional are therefore the good components with one inverse light-cone derivative, $\phi^{(\pm\frac12)}=\partial_-^{-1}\Psi_+^{(\pm\frac12)}$. This is a natural object in the light-cone gauge, where the non-dynamical components are expressed through $1/\partial_-$ as well.

Consider first the longitudinal boost. The generator is $J^{03}=\frac{1}{2}i\gamma^0\gamma^3$ and the action of the corresponding boost matrix on $\Psi_+$ reads
\begin{equation}
 S_{\mathrm{boost}}\Psi_+ = \exp \left( \frac{1}{2} \ln\kappa\; \boldsymbol{\gamma}^0\boldsymbol{\gamma}^3 \right) \Lambda_+ \Psi 
 = e^{\frac{\ln\kappa}{2}} \Psi_+ =  \sqrt{\kappa}\, \Psi_+ \,,
 \end{equation}
 where $\ln\kappa$ is the rapidity and we have used the fact that $\boldsymbol{\gamma}^0\boldsymbol{\gamma}^3 \Lambda_+ = \Lambda_+$. The good components themselves thus scale as $\sqrt\kappa$, which is the familiar statement that the light-cone spinors $u_+(p)$ are proportional to $\sqrt{p^+}$. Since $\partial_-^{-1}$ carries the weight $\bar\kappa$, both fields $\phi^{(\pm\frac12)}=\partial_-^{-1}\Psi_+^{(\pm\frac12)}$ scale with the same factor $\bar\kappa\sqrt\kappa=\sqrt{\bar\kappa}$:
 \begin{equation}
 \phi^{(-\frac{1}{2})}\rightarrow \sqrt{\bar\kappa} \phi^{(-\frac{1}{2})} \,, \quad 
 \phi^{(+\frac{1}{2})}\rightarrow \sqrt{\bar\kappa} \phi^{(+\frac{1}{2})}\,, \quad \text{(longitudinal boost)} \,.
\end{equation}
Consider next the transverse rotation with the parameter $\gamma$, i.e. $x^{\bullet}\to\gamma x^{\bullet}$ with $\gamma=e^{i\theta}$. The generator is $J^{12}=\frac{1}{2}\left(\begin{matrix}
\sigma_3 && 0 \\
0 && \sigma_3
\end{matrix}\right)$  and the spinor matrix of the rotation is $S_{\mathrm{rot}}=\exp(-i\theta J^{12})=\exp(-\ln\gamma\,J^{12})$, as fixed by the condition $S^{-1}\boldsymbol{\gamma}^{\mu}S=\Lambda^{\mu}{}_{\nu}\boldsymbol{\gamma}^{\nu}$ that makes the Dirac equation covariant. It commutes with $\boldsymbol{\gamma}^0\boldsymbol{\gamma}^3$ and therefore with $\Lambda_+$, so
\begin{equation}
 S_{\mathrm{rot}}\Psi_+ = \exp\left\{ -\frac{1}{2} \ln\gamma \; \left(\begin{matrix}
\sigma_3 && 0 \\
0 && \sigma_3
\end{matrix}\right) \right\} \begin{pmatrix}
 \xi_2 \\
 \chi_1 
 \end{pmatrix}
 = \begin{pmatrix}
 e^{+\frac{1}{2}\ln\gamma}\, \xi_2 \\
e^{-\frac{1}{2}\ln\gamma}\, \chi_1 
 \end{pmatrix}
 =  \begin{pmatrix}
 \sqrt{\gamma}\, \xi_2 \\
 \sqrt{\bar\gamma}\, \chi_1 
 \end{pmatrix} \,,
\end{equation}
since $\xi_2$ has $\sigma_3=-1$ and $\chi_1$ has $\sigma_3=+1$. The factor $\sqrt\gamma$ for $\xi_2$ is the expected one: $\xi$ is the left-handed spinor, its massless modes have helicity $-1/2$, and a state with $J_z=-1/2$ picks up $e^{-iJ_z\theta}=e^{i\theta/2}=\sqrt\gamma$, the same factor as $\sqrt{x^{\bullet}}$. The derivative $\partial_-$ is not affected by a rotation in the transverse plane, so the fields $\phi^{(\pm\frac12)}=\partial_-^{-1}\Psi_+^{(\pm\frac12)}$ transform in the same way as $\xi_2$ and $\chi_1$.
 Therefore
 \begin{equation}
 \phi^{(-\frac{1}{2})}\rightarrow \sqrt{\gamma} \phi^{(-\frac{1}{2})} \,, \quad 
 \phi^{(+\frac{1}{2})}\rightarrow \sqrt{\bar\gamma} \phi^{(+\frac{1}{2})}\,, \quad \text{(transverse rotation)} \,.
\end{equation}
Finally, for the combined scaling (iii), i.e. $\kappa=\lambda$ and $\gamma=\bar\lambda$, we find
\begin{equation}
 \phi^{(-\frac{1}{2})}\rightarrow \bar\lambda \phi^{(-\frac{1}{2})} \,, \quad 
 \phi^{(+\frac{1}{2})}\rightarrow \phi^{(+\frac{1}{2})}\,, \quad \text{(combined scaling)} \,,
\end{equation}
where the factors combine as follows: the good components $\Psi_+^{(-\frac12)}$ and $\Psi_+^{(+\frac12)}$ get $\sqrt\kappa=\sqrt\lambda$ from the boost and $\sqrt\gamma=\sqrt{\bar\lambda}$ and $\sqrt{\bar\gamma}=\sqrt\lambda$ from the rotation, i.e. $1$ and $\lambda$ in total, and $\partial_-^{-1}$ contributes a further $\bar\kappa=\bar\lambda$. Hence the $+1/2$ field is invariant while the $-1/2$ field transforms as the bullet component $\phi^{(-1)}$, exactly as used in Subsection~\ref{sub:tau_spin}. 

\newpage

\bibliographystyle{JHEP}
\bibliography{library}

\providecommand{\href}[2]{#2}\begingroup\raggedright\begin{thebibliography}{10}

\bibitem{Parke:1986gb}
S.~J. Parke and T.~R. Taylor, \emph{{Amplitude for n-Gluon Scattering}},
  \href{http://dx.doi.org/10.1103/PhysRevLett.56.2459}{\emph{Phys. Rev. Lett.}
  {\bf 56} (jun, 1986) 2459--2460}.

\bibitem{Berends:1987me}
F.~Berends and W.~Giele, \emph{{Recursive calculations for processes with n
  gluons}}, \href{http://dx.doi.org/10.1016/0550-3213(88)90442-7}{\emph{Nucl.
  Phys. B} {\bf 306} (sep, 1988) 759--808}.

\bibitem{Grisaru:1976vm}
M.~T. Grisaru, H.~N. Pendleton and P.~van Nieuwenhuizen, \emph{{Supergravity
  and the S Matrix}},
  \href{http://dx.doi.org/10.1103/PhysRevD.15.996}{\emph{Phys. Rev. D} {\bf 15}
  (1977) 996}.

\bibitem{Grisaru:1977px}
M.~T. Grisaru and H.~N. Pendleton, \emph{{Some Properties of Scattering
  Amplitudes in Supersymmetric Theories}},
  \href{http://dx.doi.org/10.1016/0550-3213(77)90277-2}{\emph{Nucl. Phys. B}
  {\bf 124} (1977) 81--92}.

\bibitem{Parke:1985pn}
S.~J. Parke and T.~R. Taylor, \emph{{Perturbative QCD Utilizing Extended
  Supersymmetry}},
  \href{http://dx.doi.org/10.1016/0370-2693(85)91216-X}{\emph{Phys. Lett. B}
  {\bf 157} (1985) 81}.

\bibitem{Gunion:1985vca}
J.~F. Gunion and Z.~Kunszt, \emph{{Improved Analytic Techniques for Tree Graph
  Calculations and the G g q anti-q Lepton anti-Lepton Subprocess}},
  \href{http://dx.doi.org/10.1016/0370-2693(85)90774-9}{\emph{Phys. Lett. B}
  {\bf 161} (1985) 333}.

\bibitem{Kunszt:1985mg}
Z.~Kunszt, \emph{{Combined use of the CALKUL method and N = 1 supersymmetry to
  calculate QCD six-parton processes}},
  \href{http://dx.doi.org/10.1016/S0550-3213(86)80013-X}{\emph{Nucl. Phys. B}
  {\bf 271} (1986) 333--348}.

\bibitem{Mangano:1990by}
M.~L. Mangano and S.~J. Parke, \emph{{Multi-Parton Amplitudes in Gauge
  Theories}}, \href{http://dx.doi.org/10.1016/0370-1573(91)90091-Y}{\emph{Phys.
  Rep.} {\bf 200} (feb, 2005) 84}, [\href{http://arxiv.org/abs/0509223}{{\tt
  0509223}}].

\bibitem{Elvang:2013cua}
H.~Elvang and Y.-t. Huang, \emph{{Scattering Amplitudes}},
  \href{http://arxiv.org/abs/1308.1697}{{\tt 1308.1697}}.

\bibitem{Travaglini:2022uwo}
G.~Travaglini et~al., \emph{{The SAGEX review on scattering amplitudes}},
  \href{http://dx.doi.org/10.1088/1751-8121/ac8380}{\emph{J. Phys. A} {\bf 55}
  (2022) 443001}, [\href{http://arxiv.org/abs/2203.13011}{{\tt 2203.13011}}].

\bibitem{Nair:1988bq}
V.~P. Nair, \emph{{A Current Algebra for Some Gauge Theory Amplitudes}},
  \href{http://dx.doi.org/10.1016/0370-2693(88)91471-2}{\emph{Phys. Lett. B}
  {\bf 214} (1988) 215--218}.

\bibitem{Kleiss:1988ne}
R.~Kleiss and H.~Kuijf, \emph{{Multi - Gluon Cross-sections and Five Jet
  Production at Hadron Colliders}},
  \href{http://dx.doi.org/10.1016/0550-3213(89)90574-9}{\emph{Nucl. Phys. B}
  {\bf 312} (1989) 616--644}.

\bibitem{DelDuca:1999rs}
V.~Del~Duca, L.~J. Dixon and F.~Maltoni, \emph{{New color decompositions for
  gauge amplitudes at tree and loop level}},
  \href{http://dx.doi.org/10.1016/S0550-3213(99)00809-3}{\emph{Nucl. Phys. B}
  {\bf 571} (2000) 51--70}, [\href{http://arxiv.org/abs/hep-ph/9910563}{{\tt
  hep-ph/9910563}}].

\bibitem{Cachazo:2013hca}
F.~Cachazo, S.~He and E.~Y. Yuan, \emph{{Scattering of Massless Particles in
  Arbitrary Dimensions}},
  \href{http://dx.doi.org/10.1103/PhysRevLett.113.171601}{\emph{Phys. Rev.
  Lett.} {\bf 113} (2014) 171601}, [\href{http://arxiv.org/abs/1307.2199}{{\tt
  1307.2199}}].

\bibitem{Mason:2013sva}
L.~Mason and D.~Skinner, \emph{{Ambitwistor strings and the scattering
  equations}}, \href{http://dx.doi.org/10.1007/JHEP07(2014)048}{\emph{JHEP}
  {\bf 07} (2014) 048}, [\href{http://arxiv.org/abs/1311.2564}{{\tt
  1311.2564}}].

\bibitem{Cachazo:2004kj}
F.~Cachazo, P.~Svrcek and E.~Witten, \emph{{MHV Vertices And Tree Amplitudes In
  Gauge Theory}},
  \href{http://dx.doi.org/10.1088/1126-6708/2004/09/006}{\emph{J. High Energy
  Phys.} {\bf 2004} (sep, 2004) 006--006}.

\bibitem{Witten:2003nn}
E.~Witten, \emph{{Perturbative Gauge Theory as a String Theory in Twistor
  Space}}, \href{http://dx.doi.org/10.1007/s00220-004-1187-3}{\emph{Commun.
  Math. Phys.} {\bf 252} (dec, 2004) 189--258},
  [\href{http://arxiv.org/abs/0312171}{{\tt 0312171}}].

\bibitem{Roiban:2004yf}
R.~Roiban, M.~Spradlin and A.~Volovich, \emph{{On the tree level S matrix of
  Yang-Mills theory}},
  \href{http://dx.doi.org/10.1103/PhysRevD.70.026009}{\emph{Phys. Rev. D} {\bf
  70} (2004) 026009}, [\href{http://arxiv.org/abs/hep-th/0403190}{{\tt
  hep-th/0403190}}].

\bibitem{Britto:2004ap}
R.~Britto, F.~Cachazo and B.~Feng, \emph{{New recursion relations for tree
  amplitudes of gluons}},
  \href{http://dx.doi.org/10.1016/j.nuclphysb.2005.02.030}{\emph{Nucl. Phys. B}
  {\bf 715} (may, 2005) 499--522},
  [\href{http://arxiv.org/abs/hep-th/0412308}{{\tt hep-th/0412308}}].

\bibitem{Britto:2005fq}
R.~Britto, F.~Cachazo, B.~Feng and E.~Witten, \emph{{Direct Proof of the
  Tree-Level Scattering Amplitude Recursion Relation in Yang-Mills Theory}},
  \href{http://dx.doi.org/10.1103/PhysRevLett.94.181602}{\emph{Phys. Rev.
  Lett.} {\bf 94} (may, 2005) 181602},
  [\href{http://arxiv.org/abs/hep-th/0501052}{{\tt hep-th/0501052}}].

\bibitem{Berends:1988zp}
F.~A. Berends, W.~T. Giele and H.~Kuijf, \emph{{On relations between multi -
  gluon and multigraviton scattering}},
  \href{http://dx.doi.org/10.1016/0370-2693(88)90813-1}{\emph{Phys. Lett. B}
  {\bf 211} (1988) 91--94}.

\bibitem{Mason:2008jy}
L.~J. Mason and D.~Skinner, \emph{{Gravity, Twistors and the MHV Formalism}},
  \href{http://dx.doi.org/10.1007/s00220-009-0972-4}{\emph{Commun. Math. Phys.}
  {\bf 294} (2010) 827--862}, [\href{http://arxiv.org/abs/0808.3907}{{\tt
  0808.3907}}].

\bibitem{Hodges:2012ym}
A.~Hodges, \emph{{A simple formula for gravitational MHV amplitudes}},
  \href{http://arxiv.org/abs/1204.1930}{{\tt 1204.1930}}.

\bibitem{Dirac:1949cp}
P.~A.~M. Dirac, \emph{{Forms of Relativistic Dynamics}},
  \href{http://dx.doi.org/10.1103/RevModPhys.21.392}{\emph{Rev. Mod. Phys.}
  {\bf 21} (1949) 392--399}.

\bibitem{Mansfield:2005yd}
P.~Mansfield, \emph{{The lagrangian origin of MHV rules}},
  \href{http://dx.doi.org/10.1088/1126-6708/2006/03/037}{\emph{J. High Energy
  Phys.} {\bf 2006} (mar, 2006) 037--037}.

\bibitem{Gorsky:2005sf}
A.~Gorsky and A.~Rosly, \emph{{From Yang-Mills lagrangian to MHV diagrams}},
  \href{http://dx.doi.org/10.1088/1126-6708/2006/01/101}{\emph{J. High Energy
  Phys.} {\bf 2006} (jan, 2006) 101--101}.

\bibitem{Ettle:2006bw}
J.~H. Ettle and T.~R. Morris, \emph{{Structure of the MHV-rules lagrangian}},
  \href{http://dx.doi.org/10.1088/1126-6708/2006/08/003}{\emph{J. High Energy
  Phys.} {\bf 2006} (aug, 2006) 003--003}.

\bibitem{Motyka2009}
L.~Motyka and A.~M. Sta{\'{s}}to, \emph{{Exact kinematics in the small-x
  evolution of the color dipole and gluon cascade}},
  \href{http://dx.doi.org/10.1103/PhysRevD.79.085016}{\emph{Phys. Rev. D} {\bf
  79} (apr, 2009) 085016}, [\href{http://arxiv.org/abs/0901.4949}{{\tt
  0901.4949}}].

\bibitem{Brandhuber:2004yw}
A.~Brandhuber, B.~Spence and G.~Travaglini, \emph{One-loop gauge theory
  amplitudes in super yang–mills from mhv vertices},
  \href{http://dx.doi.org/10.1016/j.nuclphysb.2004.11.023}{\emph{Nuclear
  Physics B} {\bf 706} (Jan, 2005) 150–180}.

\bibitem{Brandhuber:2007vm}
A.~Brandhuber, B.~Spence, G.~Travaglini and K.~Zoubos, \emph{{One-loop MHV
  rules and pure Yang-Mills}},
  \href{http://dx.doi.org/10.1088/1126-6708/2007/07/002}{\emph{J. High Energy
  Phys.} {\bf 2007} (jul, 2007) 002--002}.

\bibitem{Ettle:2007qc}
J.~H. Ettle, C.-H. Fu, J.~P. Fudger, P.~R. Mansfield and T.~R. Morris,
  \emph{{S-matrix equivalence theorem evasion and dimensional regularisation
  with the canonical MHV lagrangian}},
  \href{http://dx.doi.org/10.1088/1126-6708/2007/05/011}{\emph{J. High Energy
  Phys.} {\bf 2007} (may, 2007) 011--011}.

\bibitem{Chalmers:1996rq}
G.~Chalmers and W.~Siegel, \emph{{Self-dual sector of QCD amplitudes}},
  \href{http://dx.doi.org/10.1103/PhysRevD.54.7628}{\emph{Phys. Rev. D} {\bf
  54} (dec, 1996) 7628--7633}, [\href{http://arxiv.org/abs/hep-th/9606061}{{\tt
  hep-th/9606061}}].

\bibitem{Bardeen:1995gk}
W.~Bardeen, \emph{{Self-dual Yang-Mills theory, integrability and multiparton
  amplitudes}}, {\emph{Prog. Theor. Phys. Suppl.} (1996) 1--8}.

\bibitem{Cangemi:1996rx}
D.~Cangemi, \emph{{Self-dual Yang-Mills theory and one-loop maximally helicity
  violating multi-gluon amplitudes}},
  \href{http://dx.doi.org/10.1016/S0550-3213(96)00586-X}{\emph{Nucl. Phys. B}
  {\bf 484} (jan, 1997) 521--537},
  [\href{http://arxiv.org/abs/hep-th/9605208}{{\tt hep-th/9605208}}].

\bibitem{Bern:1993qk}
Z.~Bern, G.~Chalmers, L.~Dixon and D.~A. Kosower, \emph{One-loopn-gluon
  amplitudes with maximal helicity violation via collinear limits},
  \href{http://dx.doi.org/10.1103/physrevlett.72.2134}{\emph{Physical Review
  Letters} {\bf 72} (Apr, 1994) 2134–2137}.

\bibitem{Mahlon:1993fi}
G.~Mahlon, \emph{{Multi - gluon helicity amplitudes involving a quark loop}},
  \href{http://dx.doi.org/10.1103/PhysRevD.49.4438}{\emph{Phys. Rev. D} {\bf
  49} (1994) 4438--4453}, [\href{http://arxiv.org/abs/hep-ph/9312276}{{\tt
  hep-ph/9312276}}].

\bibitem{Bern:1996ja}
Z.~Bern, L.~J. Dixon, D.~C. Dunbar and D.~A. Kosower, \emph{{One loop selfdual
  and N=4 superYang-Mills}},
  \href{http://dx.doi.org/10.1016/S0370-2693(96)01676-0}{\emph{Phys. Lett. B}
  {\bf 394} (1997) 105--115}, [\href{http://arxiv.org/abs/hep-th/9611127}{{\tt
  hep-th/9611127}}].

\bibitem{Rosly:1996vr}
A.~A. Rosly and K.~G. Selivanov, \emph{{On amplitudes in the self-dual sector
  of Yang-Mills theory}}, {\emph{Phys. Lett. B} {\bf 399} (1997) 135--140}.

\bibitem{Selivanov:1997aq}
K.~G. Selivanov, \emph{{On tree form-factors in (supersymmetric) Yang-Mills
  theory}}, \href{http://dx.doi.org/10.1007/s002200050006}{\emph{Commun. Math.
  Phys.} {\bf 208} (2000) 671--687},
  [\href{http://arxiv.org/abs/hep-th/9809046}{{\tt hep-th/9809046}}].

\bibitem{Dixon:2004za}
L.~J. Dixon, E.~W.~N. Glover and V.~V. Khoze, \emph{{MHV rules for Higgs plus
  multi-gluon amplitudes}},
  \href{http://dx.doi.org/10.1088/1126-6708/2004/12/015}{\emph{JHEP} {\bf 12}
  (2004) 015}, [\href{http://arxiv.org/abs/hep-th/0411092}{{\tt
  hep-th/0411092}}].

\bibitem{Boels:2013bi}
R.~H. Boels, R.~S. Isermann, R.~Monteiro and D.~O'Connell,
  \emph{{Colour-Kinematics Duality for One-Loop Rational Amplitudes}},
  \href{http://dx.doi.org/10.1007/JHEP04(2013)107}{\emph{JHEP} {\bf 04} (2013)
  107}, [\href{http://arxiv.org/abs/1301.4165}{{\tt 1301.4165}}].

\bibitem{Brandhuber:2018kqb}
A.~Brandhuber, M.~Kostacinska, B.~Penante and G.~Travaglini,
  \emph{{$\text{Tr}(F^3)$ supersymmetric form factors and maximal
  transcendentality Part I: $\mathcal{N}=4$ super Yang-Mills}},
  \href{http://dx.doi.org/10.1007/JHEP12(2018)076}{\emph{JHEP} {\bf 12} (2018)
  076}, [\href{http://arxiv.org/abs/1804.05703}{{\tt 1804.05703}}].

\bibitem{AccettulliHuber:2019mxn}
M.~Accettulli~Huber, A.~Brandhuber, S.~De~Angelis and G.~Travaglini,
  \emph{{Complete Form Factors in Yang-Mills from Unitarity and Spinor Helicity
  in Six Dimensions}},
  \href{http://dx.doi.org/10.1103/PhysRevD.101.026004}{\emph{Phys. Rev. D} {\bf
  101} (2020) 026004}, [\href{http://arxiv.org/abs/1910.04772}{{\tt
  1910.04772}}].

\bibitem{Brandhuber:2010ad}
A.~Brandhuber, B.~Spence, G.~Travaglini and G.~Yang, \emph{{Form Factors in N=4
  Super Yang-Mills and Periodic Wilson Loops}},
  \href{http://dx.doi.org/10.1007/JHEP01(2011)134}{\emph{JHEP} {\bf 01} (2011)
  134}, [\href{http://arxiv.org/abs/1011.1899}{{\tt 1011.1899}}].

\bibitem{Brandhuber:2011tv}
A.~Brandhuber, O.~Gurdogan, R.~Mooney, G.~Travaglini and G.~Yang,
  \emph{{Harmony of Super Form Factors}},
  \href{http://dx.doi.org/10.1007/JHEP10(2011)046}{\emph{JHEP} {\bf 10} (2011)
  046}, [\href{http://arxiv.org/abs/1107.5067}{{\tt 1107.5067}}].

\bibitem{Monteiro2011}
R.~Monteiro and D.~O'Connell, \emph{{The kinematic algebra from the self-dual
  sector}}, \href{http://dx.doi.org/10.1007/JHEP07(2011)007}{\emph{J. High
  Energy Phys.} {\bf 2011} (2011) 1--29},
  [\href{http://arxiv.org/abs/arXiv:1105.2565v2}{{\tt arXiv:1105.2565v2}}].

\bibitem{Bern:2008qj}
Z.~Bern, J.~J.~M. Carrasco and H.~Johansson, \emph{{New Relations for
  Gauge-Theory Amplitudes}},
  \href{http://dx.doi.org/10.1103/PhysRevD.78.085011}{\emph{Phys. Rev. D} {\bf
  78} (2008) 085011}, [\href{http://arxiv.org/abs/0805.3993}{{\tt 0805.3993}}].

\bibitem{Guevara:2021abz}
A.~Guevara, E.~Himwich, M.~Pate and A.~Strominger, \emph{{Holographic symmetry
  algebras for gauge theory and gravity}},
  \href{http://dx.doi.org/10.1007/JHEP11(2021)152}{\emph{JHEP} {\bf 11} (2021)
  152}, [\href{http://arxiv.org/abs/2103.03961}{{\tt 2103.03961}}].

\bibitem{Strominger:2021mtt}
A.~Strominger, \emph{{$w_{1+\infty}$ Algebra and the Celestial Sphere: Infinite
  Towers of Soft Graviton, Photon, and Gluon Symmetries}},
  \href{http://dx.doi.org/10.1103/PhysRevLett.127.221601}{\emph{Phys. Rev.
  Lett.} {\bf 127} (2021) 221601}, [\href{http://arxiv.org/abs/2105.14346}{{\tt
  2105.14346}}].

\bibitem{Costello:2021bah}
K.~Costello and N.~M. Paquette, \emph{{Celestial holography meets twisted
  holography: 4d amplitudes from chiral correlators}},
  \href{http://dx.doi.org/10.1007/JHEP10(2022)193}{\emph{JHEP} {\bf 10} (2022)
  193}, [\href{http://arxiv.org/abs/2201.02595}{{\tt 2201.02595}}].

\bibitem{Costello:2022wso}
K.~Costello, N.~M. Paquette and A.~Sharma, \emph{{Top-Down Holography in an
  Asymptotically Flat Spacetime}},
  \href{http://dx.doi.org/10.1103/PhysRevLett.130.061602}{\emph{Phys. Rev.
  Lett.} {\bf 130} (2023) 061602}, [\href{http://arxiv.org/abs/2208.14233}{{\tt
  2208.14233}}].

\bibitem{Plebanski:1975wn}
J.~F. Plebanski, \emph{{Some solutions of complex Einstein equations}},
  \href{http://dx.doi.org/10.1063/1.522505}{\emph{J. Math. Phys.} {\bf 16}
  (1975) 2395--2402}.

\bibitem{Krasnov:2016emc}
K.~Krasnov, \emph{{Self-Dual Gravity}},
  \href{http://dx.doi.org/10.1088/1361-6382/aa65e5}{\emph{Class. Quant. Grav.}
  {\bf 34} (2017) 095001}, [\href{http://arxiv.org/abs/1610.01457}{{\tt
  1610.01457}}].

\bibitem{Kotko2017}
P.~Kotko and A.~M. Stasto, \emph{{Wilson lines in the MHV action}},
  \href{http://dx.doi.org/10.1007/JHEP09(2017)047}{\emph{J. High Energy Phys.}
  {\bf 2017} (2017) 47}.

\bibitem{Kakkad2020}
H.~Kakkad, P.~Kotko and A.~Stasto, \emph{{Exploring straight infinite Wilson
  lines in the self-dual and the MHV Lagrangians}},
  \href{http://dx.doi.org/10.1103/PhysRevD.102.094026}{\emph{Phys. Rev. D} {\bf
  102} (nov, 2020) 094026}.

\bibitem{Kakkad:2021uhv}
H.~Kakkad, P.~Kotko and A.~Stasto, \emph{{A new Wilson line-based action for
  gluodynamics}}, \href{http://dx.doi.org/10.1007/JHEP07(2021)187}{\emph{JHEP}
  {\bf 07} (2021) 187}, [\href{http://arxiv.org/abs/2102.11371}{{\tt
  2102.11371}}].

\bibitem{Kakkad:2021dkw}
H.~Kakkad, P.~Kotko and A.~Stasto, \emph{{A new Wilson line-based classical
  action for gluodynamics}},
  \href{http://dx.doi.org/10.21468/SciPostPhysProc.7.011}{\emph{SciPost Phys.
  Proc.} {\bf 7} (2022) 011}, [\href{http://arxiv.org/abs/2110.13301}{{\tt
  2110.13301}}].

\bibitem{Kakkad_2022}
H.~Kakkad, P.~Kotko and A.~Stasto, \emph{One-loop effective action approach to
  quantum {MHV} theory},
  \href{http://dx.doi.org/10.1007/jhep11(2022)132}{\emph{Journal of High Energy
  Physics} {\bf 2022} (nov, 2022) }.

\bibitem{Kakkad:2024ase}
H.~Kakkad, P.~Kotko and A.~Stasto, \emph{{Wilson line-based action for
  gluodynamics at the loop level}},
  \href{http://dx.doi.org/10.1007/JHEP11(2024)073}{\emph{JHEP} {\bf 11} (2024)
  073}, [\href{http://arxiv.org/abs/2406.13314}{{\tt 2406.13314}}].

\bibitem{Kakkad:2023bnk}
H.~Kakkad, \emph{Scattering amplitudes in the yang-mills sector of quantum
  chromodynamics},  2023.

\bibitem{vanHameren:2012if}
A.~{Van Hameren}, P.~Kotko and K.~Kutak, \emph{{Helicity amplitudes for
  high-energy scattering}}, {\emph{J. High Energy Phys.} {\bf 2013} (2013) }.

\bibitem{Kotko:2014aba}
P.~Kotko, \emph{{Wilson lines and gauge invariant off-shell amplitudes}},
  \href{http://dx.doi.org/10.1007/JHEP07(2014)128}{\emph{J. High Energy Phys.}
  {\bf 2014} (jul, 2014) 128}, [\href{http://arxiv.org/abs/1403.4824}{{\tt
  1403.4824}}].

\bibitem{Alday:2007hr}
L.~F. Alday and J.~M. Maldacena, \emph{{Gluon scattering amplitudes at strong
  coupling}},
  \href{http://dx.doi.org/10.1088/1126-6708/2007/06/064}{\emph{JHEP} {\bf 06}
  (2007) 064}, [\href{http://arxiv.org/abs/0705.0303}{{\tt 0705.0303}}].

\bibitem{Metsaev:1991mt}
R.~R. Metsaev, \emph{{Poincare invariant dynamics of massless higher spins:
  Fourth order analysis on mass shell}},
  \href{http://dx.doi.org/10.1142/S0217732391000348}{\emph{Mod. Phys. Lett. A}
  {\bf 6} (1991) 359--367}.

\bibitem{Metsaev:1991nb}
R.~R. Metsaev, \emph{{S matrix approach to massless higher spins theory. 2: The
  Case of internal symmetry}},
  \href{http://dx.doi.org/10.1142/S0217732391002839}{\emph{Mod. Phys. Lett. A}
  {\bf 6} (1991) 2411--2421}.

\bibitem{Ponomarev:2016lrm}
D.~Ponomarev and E.~D. Skvortsov, \emph{{Light-Front Higher-Spin Theories in
  Flat Space}}, \href{http://dx.doi.org/10.1088/1751-8121/aa56e7}{\emph{J.
  Phys. A} {\bf 50} (2017) 095401},
  [\href{http://arxiv.org/abs/1609.04655}{{\tt 1609.04655}}].

\bibitem{Ponomarev:2017nrr}
D.~Ponomarev, \emph{{Chiral Higher Spin Theories and Self-Duality}},
  \href{http://dx.doi.org/10.1007/JHEP12(2017)141}{\emph{JHEP} {\bf 12} (2017)
  141}, [\href{http://arxiv.org/abs/1710.00270}{{\tt 1710.00270}}].

\bibitem{Skvortsov:2018jea}
E.~D. Skvortsov, T.~Tran and M.~Tsulaia, \emph{{Quantum Chiral Higher Spin
  Gravity}},
  \href{http://dx.doi.org/10.1103/PhysRevLett.121.031601}{\emph{Phys. Rev.
  Lett.} {\bf 121} (2018) 031601}, [\href{http://arxiv.org/abs/1805.00048}{{\tt
  1805.00048}}].

\bibitem{Skvortsov:2020wtf}
E.~Skvortsov, T.~Tran and M.~Tsulaia, \emph{{More on Quantum Chiral Higher Spin
  Gravity}}, \href{http://dx.doi.org/10.1103/PhysRevD.101.106001}{\emph{Phys.
  Rev. D} {\bf 101} (2020) 106001},
  [\href{http://arxiv.org/abs/2002.08487}{{\tt 2002.08487}}].

\bibitem{Skvortsov:2020gpn}
E.~Skvortsov and T.~Tran, \emph{{One-loop Finiteness of Chiral Higher Spin
  Gravity}}, \href{http://dx.doi.org/10.1007/JHEP07(2020)021}{\emph{JHEP} {\bf
  07} (2020) 021}, [\href{http://arxiv.org/abs/2004.10797}{{\tt 2004.10797}}].

\bibitem{Metsaev:2019dqt}
R.~R. Metsaev, \emph{{Cubic interaction vertices for N=1 arbitrary spin
  massless supermultiplets in flat space}},
  \href{http://dx.doi.org/10.1007/JHEP08(2019)130}{\emph{JHEP} {\bf 08} (2019)
  130}, [\href{http://arxiv.org/abs/1905.11357}{{\tt 1905.11357}}].

\bibitem{Metsaev:2019aig}
R.~R. Metsaev, \emph{{Cubic interactions for arbitrary spin $ \mathcal{N} $
  -extended massless supermultiplets in 4d flat space}},
  \href{http://dx.doi.org/10.1007/JHEP11(2019)084}{\emph{JHEP} {\bf 11} (2019)
  084}, [\href{http://arxiv.org/abs/1909.05241}{{\tt 1909.05241}}].

\bibitem{Tsulaia:2022csz}
M.~Tsulaia and D.~Weissman, \emph{{Supersymmetric quantum chiral higher spin
  gravity}}, \href{http://dx.doi.org/10.1007/JHEP12(2022)002}{\emph{JHEP} {\bf
  12} (2022) 002}, [\href{http://arxiv.org/abs/2209.13907}{{\tt 2209.13907}}].

\bibitem{Weinberg:1964ew}
S.~Weinberg, \emph{{Photons and Gravitons in $S$-Matrix Theory: Derivation of
  Charge Conservation and Equality of Gravitational and Inertial Mass}},
  \href{http://dx.doi.org/10.1103/PhysRev.135.B1049}{\emph{Phys. Rev.} {\bf
  135} (1964) B1049--B1056}.

\bibitem{Coleman:1967ad}
S.~R. Coleman and J.~Mandula, \emph{{All Possible Symmetries of the S Matrix}},
  \href{http://dx.doi.org/10.1103/PhysRev.159.1251}{\emph{Phys. Rev.} {\bf 159}
  (1967) 1251--1256}.

\bibitem{Benincasa:2007xk}
P.~Benincasa and F.~Cachazo, \emph{{Consistency Conditions on the S-Matrix of
  Massless Particles}},  \href{http://arxiv.org/abs/0705.4305}{{\tt
  0705.4305}}.

\bibitem{McGady:2013sga}
D.~A. McGady and L.~Rodina, \emph{{Higher-spin massless $S$-matrices in
  four-dimensions}},
  \href{http://dx.doi.org/10.1103/PhysRevD.90.084048}{\emph{Phys. Rev. D} {\bf
  90} (2014) 084048}, [\href{http://arxiv.org/abs/1311.2938}{{\tt 1311.2938}}].

\bibitem{Brink:1982wv}
L.~Brink, O.~Lindgren and B.~{E.W. Nilsson}, \emph{N = 4 yang-mills theory on
  the light cone},
  \href{http://dx.doi.org/https://doi.org/10.1016/0550-3213(83)90678-8}{\emph{Nuclear
  Physics B} {\bf 212} (1983) 401--412}.

\bibitem{Mandelstam:1982cb}
S.~Mandelstam, \emph{{Light Cone Superspace and the Ultraviolet Finiteness of
  the N=4 Model}},
  \href{http://dx.doi.org/10.1016/0550-3213(83)90179-7}{\emph{Nucl. Phys. B}
  {\bf 213} (1983) 149--168}.

\bibitem{Penrose:1967wn}
R.~Penrose, \emph{{Twistor algebra}},
  \href{http://dx.doi.org/10.1063/1.1705200}{\emph{J. Math. Phys.} {\bf 8}
  (1967) 345}.

\bibitem{Penrose:1969ae}
R.~Penrose, \emph{{Solutions of the zero-rest-mass equations}},
  \href{http://dx.doi.org/10.1063/1.1664756}{\emph{J. Math. Phys.} {\bf 10}
  (1969) 38--39}.

\bibitem{Bateman:1904xu}
H.~Bateman, \emph{{The Solution of partial differential equations by means of
  definite integrals}},
  \href{http://dx.doi.org/10.1112/plms/s2-1.1.451}{\emph{Proc. Lond. Math.
  Soc.} {\bf 1} (1904) 451--458}.

\bibitem{Ward:1977ta}
R.~S. Ward, \emph{{On Selfdual gauge fields}},
  \href{http://dx.doi.org/10.1016/0375-9601(77)90842-8}{\emph{Phys. Lett. A}
  {\bf 61} (1977) 81--82}.

\bibitem{Mason:2005zm}
L.~J. Mason, \emph{{Twistor actions for non-self-dual fields: A Derivation of
  twistor-string theory}},
  \href{http://dx.doi.org/10.1088/1126-6708/2005/10/009}{\emph{JHEP} {\bf 10}
  (2005) 009}, [\href{http://arxiv.org/abs/hep-th/0507269}{{\tt
  hep-th/0507269}}].

\bibitem{Boels:2006ir}
R.~Boels, L.~J. Mason and D.~Skinner, \emph{{Supersymmetric Gauge Theories in
  Twistor Space}},
  \href{http://dx.doi.org/10.1088/1126-6708/2007/02/014}{\emph{JHEP} {\bf 02}
  (2007) 014}, [\href{http://arxiv.org/abs/hep-th/0604040}{{\tt
  hep-th/0604040}}].

\bibitem{Penrose:1985bww}
R.~Penrose and W.~Rindler, \emph{{Spinors and Space-Time}}.
\newblock Cambridge Monographs on Mathematical Physics. Cambridge Univ. Press,
  Cambridge, UK, 4, 2011,
  \href{http://dx.doi.org/10.1017/CBO9780511564048}{10.1017/CBO9780511564048}.

\bibitem{Kogut:1969xa}
J.~B. Kogut and D.~E. Soper, \emph{{Quantum Electrodynamics in the Infinite
  Momentum Frame}},
  \href{http://dx.doi.org/10.1103/PhysRevD.1.2901}{\emph{Phys. Rev. D} {\bf 1}
  (1970) 2901--2913}.

\bibitem{Lyons:1998rough}
T.~J. Lyons, \emph{Differential equations driven by rough signals},
  {\emph{Revista Matem{\'a}tica Iberoamericana} {\bf 14} (1998) 215--310}.

\bibitem{Scherk1975}
J.~Scherk and J.~H. Schwarz, \emph{{Gravitation in the light cone gauge}},
  \href{http://dx.doi.org/10.1007/BF00761962}{\emph{Gen. Relativ. Gravit.} {\bf
  6} (1975) 537--550}.

\end{thebibliography}\endgroup
\end{document}